\documentclass[aps,twocolumn,preprintnumbers,nofootinbib,prd,superscriptaddress]{revtex4-2}

\usepackage[utf8]{inputenc}

\usepackage{graphicx,amssymb,amsmath,amsthm,amsfonts}
\usepackage{mathtools}
\usepackage[usenames,dvipsnames,svgnames,table]{xcolor}
\definecolor{darkred}{rgb}{0.5,0,0}
\usepackage{aas_macros}
\usepackage{bm}
\usepackage{dcolumn}
\usepackage{latexsym}
\usepackage{rotating}

\usepackage[normalem]{ulem}

\usepackage{enumerate}
\usepackage{tensor,multirow}
\usepackage{url}
\usepackage[linktocpage]{hyperref}

\newcommand{\OO}{\mathcal{O}}

\newcommand{\gint}{\int d^4x \, \sqrt{-g} \,}

\newcommand{\tder}{\partial_{t}}
\newcommand{\rder}{\partial_{r}}

\usepackage{braket}

\def\be{\begin{equation}}
\def\ee{\end{equation}}
\newcommand{\beq}{\begin{eqnarray}}
\newcommand{\eeq}{\end{eqnarray}}

\newcommand{\GG}{\mathcal G}
\newcommand{\RR}{\mathcal R}

\newcommand{\ea}{\textit{et al.} }

\def\ba{\begin{align}}
\def\ea{\end{align}}

\newcommand{\Sch}[1]{\mathfrak{#1}}

\begin{document}

\title{
Nonperturbative gedanken experiments in Einstein-dilaton-Gauss-Bonnet gravity:\\ nonlinear transitions and tests of the cosmic censorship beyond General Relativity 
}

\author{Fabrizio Corelli}
\email{fabrizio.corelli@uniroma1.it}
\affiliation{Dipartimento di Fisica, ``Sapienza" Universit\`a di Roma \& Sezione INFN Roma1, Piazzale Aldo Moro 
5, 00185, Roma, Italy}

\author{Marina de Amicis}
\affiliation{Dipartimento di Fisica, ``Sapienza" Universit\`a di Roma \& Sezione INFN Roma1, Piazzale Aldo Moro 
5, 00185, Roma, Italy}
\affiliation{Niels Bohr International Academy, Niels Bohr Institute, Blegdamsvej 17, 2100 Copenhagen, Denmark}

\author{Taishi Ikeda}
\affiliation{Dipartimento di Fisica, ``Sapienza" Universit\`a di Roma \& Sezione INFN Roma1, Piazzale Aldo Moro 
5, 00185, Roma, Italy}

\author{Paolo Pani}
\email{paolo.pani@uniroma1.it}
\affiliation{Dipartimento di Fisica, ``Sapienza" Universit\`a di Roma \& Sezione INFN Roma1, Piazzale Aldo Moro 
5, 00185, Roma, Italy}

\begin{abstract}
As the only gravity theory with quadratic curvature terms and second-order field equations, Einstein-dilaton-Gauss-Bonnet gravity is a natural testbed to probe the high-curvature regime beyond General Relativity in a fully nonperturbative way. 
Due to nonperturbative effects of the dilatonic coupling, black holes in this theory have a minimum mass which separates a stable branch from an unstable one. 
The minimum mass solution is a double point in the phase diagram of the theory, wherein the critical black hole and a wormhole solution coexist.
We perform extensive nonlinear simulations of the spherical collapse onto black holes with scalar hair in this theory, especially focusing on the region near the minimum mass.
We study the nonlinear transition from the unstable to the stable branch and assess the nonlinear stability of the latter.
Furthermore, motivated by modeling the mass loss induced by Hawking radiation near the minimum mass at the classical level, we study the collapse of a phantom field onto the black hole. When the black-hole mass decreases past the critical value, the apparent horizon shrinks significantly, eventually unveiling a high-curvature elliptic region.
We argue that evaporation in this theory is bound to either violate the weak cosmic censorship or produce horizonless remnants. Addressing the end-state might require a different evolution scheme.
\end{abstract}

\maketitle

\section{Introduction \& Motivations}
%
Penrose's weak cosmic censorship conjecture~\cite{1969NCimR...1..252P} posits that --~within Einstein's General Relativity~(GR)~-- naked singularities cannot form from typical regular initial data (see~\cite{Wald:1997wa} for an overview). 
Lacking a rigorous proof of this conjecture, great effort has been devoted to devise \emph{gedanken experiments}~\cite{gedankenexperiments} aimed at supporting or disproving it. This has been done by trying to overcharge/overspin a black hole~(BH) past extremality in order to destroy the BH horizon and unveil the curvature singularity concealed in its interior (see~\cite{1974AnPhy..82..548W,Hubeny:1998ga,Jacobson:2009kt,Saa:2011wq,Isoyama:2011ea,Natario:2016bay,Siahaan:2021bzc,Aniceto:2015klq,Semiz:2005gs,Duztas:2013wua,Duztas:2021kuj,Siahaan:2021bzc} for various different attempts).

While most attempts have focused on the dynamics of test particles/fields onto a fixed BH geometry, this regime is insufficient to test the conjecture, since backreaction and finite-size effects can be key to avoid naked-singularity formation (see, e.g.,~\cite{Barausse:2010ka,Barausse:2011vx}).
Therefore, gedanken experiments relying on the fully nonlinear dynamics of a theory are particularly important~\cite{Corelli:2021ikv}.

In this paper (a companion of the letter~\cite{letter}) we perform extensive nonlinear numerical simulations of the spherical collapse of scalar fields onto BHs in a theory of gravity with quadratic curvature terms. Our testbed is Einstein-dilaton-Gauss-Bonnet~(EdGB) gravity~\cite{Kanti:1995vq}, a theory that stands out within those containing curvature-squared terms as the only one featuring second-order field equations. This avoids the Ostrograski's instability~\cite{Woodard:2006nt}, and allows studying the theory at the fully nonperturbative level~\cite{Ripley:2019hxt,Ripley:2019irj,Kovacs:2020pns,Kovacs:2020ywu,East:2020hgw,East:2021bqk,Kuan:2021lol,Kuan:2021yih}, i.e. beyond an effective field theory (see~\cite{Witek:2018dmd,Okounkova:2019zjf,Okounkova:2020rqw,Silva:2020omi,Doneva:2022byd,Elley:2022ept} for simulations in the perturbative regime). Thus, one of the questions we wish to explore here is whether naked singularities can form dynamically in the high-curvature regime when this theory dramatically differs from GR.

Another broad motivation for our study is an intriguing aspect of BHs in this theory that is often overlooked. By simple dimensional arguments, any theory with ultraviolet curvature-squared terms has a natural length scale $\ell$ below which GR deviations become dominant. Indeed, due to nonperturbative effects, in this theory BHs may have a \emph{minimum radius} and a \emph{minimum mass}, both of ${\cal O}(\ell)$~\cite{Kanti:1995vq,Torii:1996yi,Alexeev:1996vs,Pani:2009wy}. This is a striking difference with respect to GR, where the BH mass is an unconstrained free parameter, so in GR BHs can have any size.
As we shall discuss in details, in EdGB gravity the minimum-radius solution and the minimum-mass solution exist but do \emph{not} coincide~\cite{Torii:1996yi, Guo:2008hf, MarinaThesis, Blazquez-Salcedo:2017txk}. Furthermore, the minimum-mass solution --~like all BH solutions in this theory~-- actually corresponds to a 
\emph{double point} in the phase space in which the BH solution and a regular wormhole solution~\cite{Kanti:2011jz} co-exist~\cite{MarinaThesis}.

Remarkably, as we shall show, the Hawking temperature~\cite{Hawking:1975vcx} and graybody factor of BHs in this theory are finite and nonvanishing, even at the critical (i.e., minimum mass) solution (see also~\cite{Konoplya:2019hml}). This unveils a conundrum: \emph{What is the final fate of Hawking evaporation in this theory since a BH cannot evaporate completely?}~\cite{Torii:1996yi,Alexeyev:2002tg}\footnote{Note that a similar question emerges in other scenarios with a new fundamental length scale, e.g. in the context of the generalized uncertainty principle~\cite{Adler:2001vs}.} We shall argue that, due to nonperturbative high-curvature effects, EdGB gravity is bound to either violate the weak cosmic censorship\footnote{Note that one might consider a version of the weak cosmic censorship that requires matter fields to satisfy some energy conditions~\cite{Wald:1997wa}, in which case Hawking radiation (and the phantom field we shall use to mimic it at the classical level) would be excluded as a possible dynamical process to test this conjecture. Here we shall adopt a more agnostic viewpoint and define the violation of the weak cosmic censorship as the formation of naked singularity from typical regular initial data.} or produce horizonless remnants. 

One might argue that Hawking evaporation is irrelevant for real BHs and that also higher-curvature corrections are negligible if the fundamental length scale $\ell$ is much smaller than the typical size of an astrophysical BH. However, the problem has potentially deep implications, as put forward by the following gedanken experiment. 
Imagine a BH with radius (and mass) much bigger than $\ell$ (we shall use natural units henceforth). In this regime higher-curvature corrections are negligible and EdGB gravity reduces to GR. Due to Hawking evaporation, the BH mass (and size) decreases, and inevitably reaches the length scale $\ell$. In that regime, nonperturbative EdGB effects become important as testified by the fact that there is a critical mass, $M_{\rm min}\propto \ell$, below which no static BH solutions exist. Since Hawking emission is not halted at the critical point, something dramatic must happen to the system.
Note that this conclusion holds no matter how small $\ell$ is: Hawking radiation will dynamically bring the system toward the nonperturbative regime. 

With the above motivations in mind, we wish to perform a gedanken experiment which is similar (in spirit) to Hawking evaporation, by studying the dynamics of a nearly-critical BH in EdGB gravity past the minimum mass. In order to mimic the mass loss due to Hawking evaporation at the classical level, we shall use a massless ``phantom'' scalar field with the ``wrong'' sign of the kinetic term. In this setup, a BH would reduce its mass after absorbing a phantom perturbation.

The rest of this paper is organized as follows. 
In Sec.~\ref{sec:framework} we present the theory and field equations in covariant form, as well as the sets of coordinates used in different parts of the analysis. 
In Sec.~\ref{sec:staticBHs} we discuss static BH solutions in this theory, compute their temperature and graybody factors, and their interior. We also discuss the phase space of static solutions in EdGB gravity, which includes wormholes and singular solitons.
Section~\ref{sec:numsetup} presents our numerical setup, whereas Sec.~\ref{sec:gedanken} is devoted to our numerical simulations using both dilaton and phantom perturbations.
We conclude with a discussion of the results in Sec.~\ref{sec:conclusion}.
The paper is supplemented by several appendices: Appendix~\ref{app:Eqs} gives the set of field equations to be solved for the static solutions and for the initial-value problem; Appendix~\ref{app:Phases} provides details on the static wormholes and soliton solutions; finally, Appendix~\ref{app:Tests} presents some details and convergence tests of our code.
%

%
\section{Framework}\label{sec:framework}
We consider the action of Einstein-scalar-Gauss-Bonnet gravity~\cite{Kanti:1995vq} with an additional (real) phantom scalar field:
\begin{align}
	S &= \frac{1}{16\pi}\gint \biggl\{ \RR - \bigl(\nabla_\mu \phi\bigr)\bigl(\nabla^\mu \phi\bigr) \notag \\ 
	&+ \bigl(\nabla_\mu \xi\bigr)\bigl(\nabla^\mu \xi\bigr) + 2 F[\phi] \GG\biggr\},
	\label{eq:Action}
\end{align}
where $\RR$ is the scalar curvature, $\phi$ is the dilatonic field, $\xi$ is the phantom field, $F[\phi]$ is the coupling function, and $\GG = \frac{1}{4} \delta^{\mu\nu\alpha\beta}_{\rho\sigma\lambda\omega} \tensor{R}{^{\rho\sigma}_{\mu\nu}} \tensor{R}{^{\lambda\omega}_{\alpha\beta}}$ is the Gauss-Bonnet invariant, $\delta^{\mu\nu\alpha\beta}_{\rho\sigma\lambda\omega} = \epsilon^{\mu\nu\alpha\beta} \epsilon_{\rho\sigma\lambda\omega}$ is the generalized Kronecker delta, with $\epsilon_{\mu\nu\alpha\beta} = \epsilon^{\mu\nu\alpha\beta}$ being the Levi-Civita symbol.

From this action we obtain the following field equations
\begin{align}
	R_{\mu\nu}-\frac{1}{2}g_{\mu\nu}R &= 8\pi T_{\mu\nu}, \label{eq:field_grav} \\
	\Box \phi &= -\frac{\delta F[\phi]}{\delta \phi} \GG, \label{eq:field_dilaton}\\
	\Box \xi &= 0, \label{eq:field_phantom}
\end{align}
where $\Box = \nabla_\mu \nabla^\mu$ and 
\begin{align}
	T_{\mu\nu} &= \frac{1}{8\pi} \biggl[ \bigl( \nabla_\mu \phi \bigr) \bigl( \nabla_\nu \phi) - \frac{1}{2} \bigl(\nabla_\alpha \phi\bigr)\bigl(\nabla^\alpha \phi\bigr) g_{\mu\nu} \notag \\
		&- \bigl( \nabla_\mu \xi \bigr) \bigl( \nabla_\nu \xi) + \frac{1}{2} \bigl(\nabla_\alpha \xi\bigr)\bigl(\nabla^\alpha \xi\bigr) g_{\mu\nu} + \notag \\
		   &- 2\bigl( \nabla_\gamma\nabla^\alpha F[\phi] \bigr) \delta^{\gamma\delta\kappa\lambda}_{\alpha\beta\rho\sigma} \tensor{R}{^{\rho\sigma}_{\kappa\lambda}} \tensor{\delta}{^\beta_{(\mu}} \tensor{g}{_{\nu) \delta}}\biggr]
		   \label{eq:TSF}
\end{align} 
is the effective stress-energy tensor. 
For concreteness, we will consider a dilatonic coupling function of the form~\cite{Gross:1986mw} 
\begin{equation}
	F[\phi] = \lambda e^{-\gamma \phi},
	\label{eq:Coupling}
\end{equation}
where $\lambda$ is the Gauss-Bonnet coupling constant and $\gamma$ is the dilaton coupling constant. We expect that several of the qualitative features discussed below hold also with different coupling functions, as long as the quadratic-curvature interactions are sufficiently strong.
Henceforth we will refer to this class of quadratic-gravity theories as EdGB gravity.

Note that, in term of the generic length scale discussed in the introduction, $\lambda\simeq\ell^2$ since the coupling is dimensionally the inverse of a curvature.

We shall construct static BH solutions in this theory and compute their Hawking temperature and graybody factor.
We shall also study their nonlinear stability by performing numerical simulations in full-fledged EdGB gravity.
We use different coordinate systems for these studies. To compute the Hawking temperature and graybody factors, we use Schwarzschild-like coordinates $(\Sch{t},\Sch{r},\theta,\varphi)$ and assume the following ansatz for the metric
\begin{equation}
ds^2=-e^{\Gamma(\Sch{r})}d\Sch{t}^2+e^{\Lambda(\Sch{r})}d\Sch{r}^2+\Sch{r}^2d\Omega^2\,,
\label{eq:Sch coord}
\end{equation}
where $\Gamma(\Sch{r})$ and $\Lambda(\Sch{r})$ are functions of the areal radius $\Sch{r}$.
On the other hand, when performing nonlinear simulations of wave packets absorbed by dilatonic BHs, we use Painlev\'e-Gullstrand~(PG)-like coordinates $(t,R,\theta,\varphi)$ that penetrate the BH horizon, since in this case we are also interested in monitoring the BH interior. The line element in this case reads
\begin{equation}
	ds^2 = -\alpha(t,R)^2 dt^2 + (dR + \alpha(t,R) \zeta(t,R) \, dt)^2 + R^{2} d\Omega^2,
	\label{eq:PG coord}
\end{equation}
where $R$ is the areal radius. These two coordinates are connected by
\begin{align}
d\Sch{r}=dR\,,\qquad d\Sch{t}=dt-\frac{\zeta}{\alpha(1-\zeta^{2})}dR\,.
\end{align}
In some selected cases, we checked that the solutions obtained with different coordinates are consistent with each other.

\section{Static dilatonic BHs and other horizonless solutions in EdGB gravity} \label{sec:staticBHs}
In this section, we construct static dilatonic BH solutions in EdGB gravity and discuss the minimum BH mass and the Hawking emission (Sec.~\ref{sec:BH1}), as well as construct the BH interior (Sec.~\ref{sec:BH2}) which would be needed for the initial data of the simulations performed in the next sections. 
In Sec.~\ref{sec:phase}, we shall also discuss the phase space of static objects in this theory and present other horizonless solutions.
In this section we switch off the phantom field, thus dealing with pure EdGB gravity in vacuum.

\subsection{Static dilatonic BH solutions in Schwarzschild-like coordinates} \label{sec:BH1}

We consider static and spherically symmetric solutions to the field equations~\eqref{eq:field_grav} and~\eqref{eq:field_dilaton}, when the phantom field vanishes. In particular, we are interested in BH solutions with a dilaton hair that vanishes at spatial infinity~\cite{Kanti:1995vq}. 
In Schwarzschild-like coordinates, Eq.~\eqref{eq:Sch coord},
we obtain a set of differential equations for the metric functions and the dilaton, which are given in Appendix~\ref{app:Eqs}.

The metric functions and dilaton near the BH horizon ($\Sch{r}\sim \Sch{r}_{\rm H}$) read
\begin{equation}
    \begin{cases}
        e^{\Gamma(\Sch{r})}&\simeq \Gamma_1(\Sch{r}-\Sch{r}_H)+\mathcal{O}\left[(\Sch{r}-\Sch{r}_H)^2\right]\\
        e^{-\Lambda(\Sch{r})}&\simeq \lambda_1(\Sch{r}-\Sch{r}_H)+\mathcal{O}\left[(\Sch{r}-\Sch{r}_H)^2\right]\\
        \phi(\Sch{r})&\simeq\phi_H+\phi'_H(\Sch{r}-\Sch{r}_{\rm H})+\mathcal{O}\left[(\Sch{r}-\Sch{r}_{\rm H})^2\right]
    \end{cases}
\end{equation}
where $\Gamma_1$  is related to a time rescaling and can be set by requiring $e^{\Gamma(\Sch{r})}\to1$ at infinity, whereas $\lambda_1$ and $\phi'_H$ can be written in terms of $\phi_H$ and $\Sch{r}_H$ through the field equations~\cite{Kanti:1995vq}
\begin{equation}
    \begin{split}
        \phi_H'&=\frac{\Sch{r}_{\rm H}}{8\gamma\lambda}e^{\gamma\phi_H}\left(1-\sqrt{1-\frac{192\gamma^2\lambda^2}{\Sch{r}_{\rm H}^4}e^{-2\gamma\phi_H}}\right)\\
        \lambda_1&=\frac{1}{\Sch{r}_{\rm H}-4\gamma\lambda e^{-\gamma\phi_H}\phi_H'}
        \label{eq:MetricDilatonExpansionInfSchw}
    \end{split}
\end{equation}
Thus, for a fixed coupling function and choosing units such as $\Sch{r}_H$ is fixed, the near-horizon solution depends on a single parameter, $\phi_H$. 
Near spatial infinity,
\begin{eqnarray}
    e^{\Gamma(\Sch{r})}&\simeq& e^{-\Lambda(\Sch{r})}\simeq 1-\frac{2M_{\rm BH}}{\Sch{r}}+\mathcal{O}\left(\Sch{r}^{-2}\right) \,\\
    \phi(\Sch{r})&\simeq& C-\frac{D}{\Sch{r}}+\mathcal{O}\left(\Sch{r}^{-2}\right)\,, 
\end{eqnarray}
where $M_{\rm BH}$ is the BH mass and $D$ is the dilaton charge.
We integrate the field equations from the horizon outward and find a family of asymptotically-flat BH solutions by adjusting $\phi_H$ in order to impose $C=0$ at spatial infinity. We do so with two different procedures, details are given in Sec.~\ref{sec:BH2}.

\begin{figure}[ht]
	\centering
	\includegraphics[width = \columnwidth]{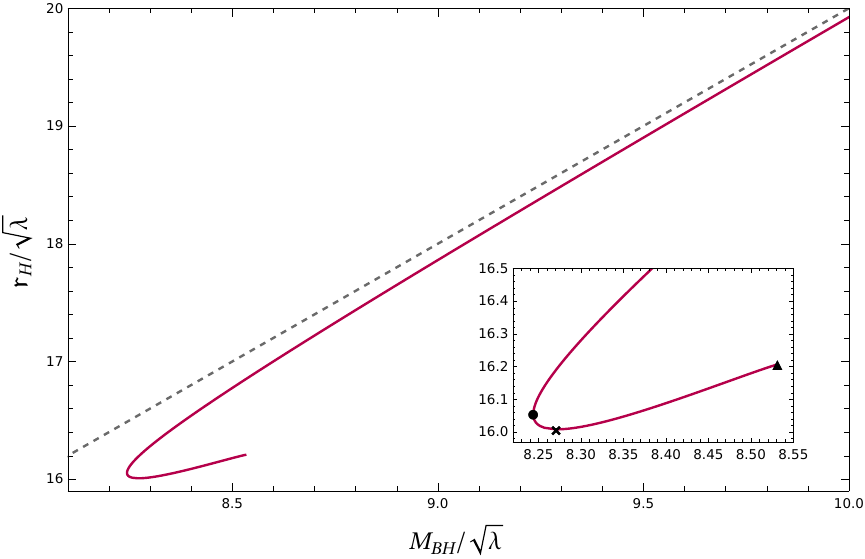}
	\caption{Areal radius of the event horizon as a function of the BH mass for static BH solutions in EdGB gravity with coupling $F[\phi]=\lambda e^{-4\phi}$. 
	 The gray dashed line is the Schwarzschild limit $\Sch{r}_H=2M_{\rm BH}$, reached when $M\gg M_{\rm crit}\approx 8.244 \sqrt{\lambda}$.
	 The inset is a zoom-in around the minimum-mass solution, which separates a stable branch from an unstable branch.
	 The minimum-mass, minimum-radius, and singular BH solutions are denoted by a circle, cross, and triangle, respectively.
	}
	\label{fig:RHvsMstatic}
\end{figure}

For concreteness, we shall now focus on the $\gamma=4$ case; different couplings are discussed later and give qualitatively similar results, including $\gamma=\sqrt{2}$ which is motivated by string theory~\cite{Gross:1986mw}.
In Fig.~\ref{fig:RHvsMstatic} we show the areal radius of the event horizon as a function of the BH mass $M_{\rm BH}$ in this theory. 
When $\lambda/\Sch{r}_H^2\ll1$, there exists only one asymptotically-flat solution for given BH mass, which reduces to the GR Schwarzschild BH in the $\lambda\to0$ limit. In this limit one gets $\Sch{r}_H\approx 2M_{\rm BH}$ as in GR.
However, for any finite $\lambda$ there exists a minimum-mass\footnote{As later discussed, other values of $\gamma\gtrsim1$ change the proportionality factor of the minimum mass but in general $M_{\rm crit}\propto \sqrt{\lambda}$.} BH solution~\cite{Kanti:1995vq,Torii:1996yi,Alexeev:1996vs,Pani:2009wy}, $M_{\rm BH}\geq M_{\rm crit}\simeq 8.244\sqrt{\lambda}$. 
The critical BH divides two branches of solutions with the same mass and different radii. The upper branch (i.e., larger radii) is linearly stable, whereas the lower branch (i.e., smaller radii) is linearly unstable~\cite{Torii:1998gm,MarinaThesis}.
As later discussed, the details (and existence) of the second branch depends on the specific values of $\gamma$.
In our context it is important to highlight that, just as the Schwarzschild solution, these metrics have a curvature singularity inside the horizon~\cite{Alexeev:1996vs}, except for the solution at the end of the unstable branch in which such singularity coincides with the horizon and becomes naked (see, e.g.,~\cite{Sotiriou:2013qea,Sotiriou:2014pfa} for BHs in shift-symmetric theories with $F[\phi]\propto \phi$). 
Since for $\gamma\gtrsim1$ the singular solution does not coincide with the minimum-mass solution, the latter is regular on and outside the horizon, just as in the GR case. On the other hand, the singular solution is unphysical as it is part of the unstable branch. 

\subsubsection{BH temperature and graybody factor}

\begin{figure}
	\centering
	\includegraphics[width = \columnwidth]{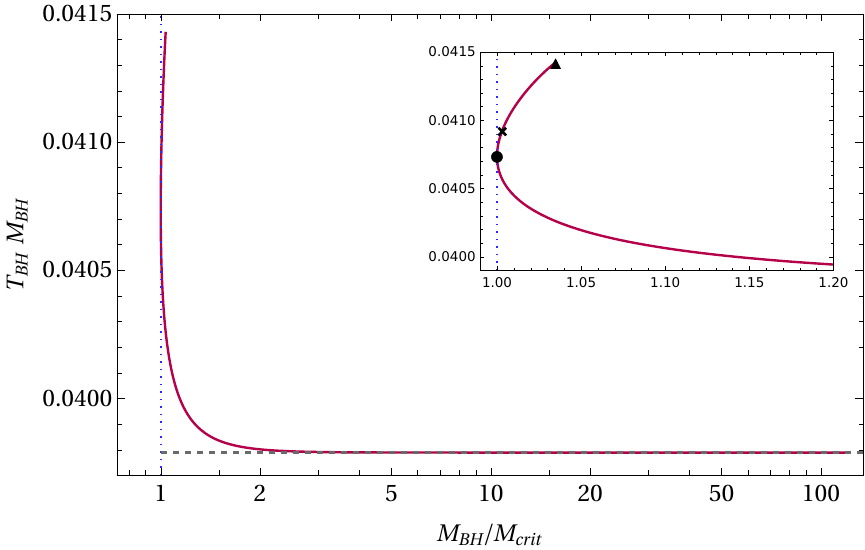}
	\caption{Hawking temperature of a dilatonic BH in EdGB gravity as a function of the BH mass. 
	The horizontal dashed line denotes the temperature of a Schwarzschild BH, $T_{\rm BH}^{\rm GR}=1/(8\pi M_{\rm BH})\approx 0.0398/M_{\rm BH}$. The inset is a zoom-in around the minimum-mass solution.
	The minimum-mass, minimum-radius, and singular BH solutions are denoted by a circle, cross, and triangle, respectively.
	}
	\label{fig:TemperatureStatic}
\end{figure}

We are interested in how these modified BH solutions emit Hawking radiation. 
Thus, we first compute their Hawking temperature~\cite{gibbons1993action}
\begin{equation}
    T_{\rm BH}=\frac{1}{4\pi}\lim_{\Sch{r}\to \Sch{r}_H}\frac{{dg_{\Sch{tt}}}/{d\Sch{r}}}{\sqrt{g_{\Sch{tt}}g_{\Sch{rr}}}}\,.
\end{equation}
As shown in Fig.~\ref{fig:TemperatureStatic}, the temperature of a dilatonic BH in EdGB gravity is always higher than that of the corresponding Schwarzschild BH with same mass. This suggests that a BH evaporates faster in EdGB gravity than in GR. Furthermore, we note that the temperature is always nonvanishing also for the minimum-mass solution. This suggests that the BH continues emitting energy once it reaches the minimum mass configuration. 

However, the BH  mass loss depends also on its graybody factor ${G}_{lm}(\omega)$, which is the fraction of energy flux at frequency $\omega$ coming from spatial infinity that is captured by the horizon.
Specifically:
\begin{equation}
    \frac{dM}{d\Sch{t}}=-\frac{1}{2\pi}\sum_{lm}\int d\omega\frac{\omega {G}_{lm}(\omega)}{e^{\omega/T_{\rm BH}}\pm 1} \,,\label{massloss}
\end{equation}
where the sum is over the $(l,m)$ angular mode of the radiation and, at the denominator, the plus/minus applies to the emission of fermions/bosons.
Thus, in order to study the BH evaporation, it is not sufficient to compute its temperature, we also need the behavior of the graybody factors relative to the emitted modes.
We compute these quantities for minimally-coupled scalar massless particles and for photons\footnote{Of course also gravitons would be radiated, and in EdGB theory the gravitational sector is coupled to the dilaton. The computation of the graybody factor for gravitons and dilatons is technically more involved but does not change the qualitative picture.} (see also~\cite{Konoplya:2019hml}). In particular, we consider the lowest angular modes, i.e. $l=0$ and $l=1$, for the scalar and vector emission, respectively, which give the leading contribution to the mass loss in this case.

The scalar $\Psi$ and electromagnetic $A_{\mu}$ fields satisfy the following field equations:
\begin{equation}
    \begin{split}
        &\nabla_{\mu}\partial^{\mu}\Psi=0\,,\\
        &\nabla_{\mu}(\partial^{\mu}A^{\nu}-\partial^{\nu}A^{\mu})=0\,,
    \end{split}
    \label{eq:ScalarMasslessEMmodesStaticBG}
\end{equation}
on the background metric described by the dilatonic BH solution. 
Since the background metric~\eqref{eq:Sch coord} is spherically symmetric, it is possible to decompose the scalar field in spherical harmonics $Y_{lm}(\theta,\varphi)$ and the electromagnetic field in vector harmonics~\cite{Regge:1957td}:
\begin{equation}
    \begin{split}
        &\Psi(\Sch{t},\Sch{r},\theta,\varphi)=\sum_{lm}\frac{R_{lm}(\Sch{t},\Sch{r})}{\Sch{r}} Y_{lm}\,,\\
        &
        \begin{split}
             A_{\mu}(\Sch{t},\Sch{r},\theta,\varphi)=&\sum_{lm}
       \begin{pmatrix}
        &f_{lm}(\Sch{t,r})\\
        &h_{lm}(\Sch{t,r})\\
        &a_{lm}(\Sch{t,r})\frac{1}{\sin\theta}\partial_{\varphi}+k_{lm}(\Sch{t,r})\partial_{\theta}\\
        &a_{lm}(\Sch{t,r})\sin\theta\partial_{\theta}-k_{lm}(\Sch{t,r})\partial_{\varphi}
        \end{pmatrix}Y_{lm} \nonumber\,.
        \end{split}
    \end{split}
\end{equation}
Substituting these expansions in the field equations~\eqref{eq:ScalarMasslessEMmodesStaticBG} and assuming a time dependence $e^{-i\omega \Sch{t}}$, the radial part of the equations separates and takes the form of a Schr\"{o}edinger-like equation:
\begin{equation}
    \frac{d^2}{d\Sch{r}_{\ast}^{2}}\Theta_{lm}(\Sch{r})+\left[\omega^2-V_{slm}(\Sch{r})\right]\Theta_{lm}(\Sch{r})=0\,,
    \label{eq:MasterEqStaticSection}
\end{equation}
where $\Theta_{lm}$ collectively denotes the master function for the scalar or the electromagnetic field, and
\begin{equation}
    \begin{split}
        &V_{s=0}(\Sch{r})=\frac{l(l+1)}{\Sch{r}^2}e^{\Gamma(\Sch{r})}+e^{\frac{\Gamma(r)-\Lambda(\Sch{r})}{2}}\frac{1}{\Sch{r}}\frac{d}{d\Sch{r}}e^{\frac{\Gamma(\Sch{r})-\Lambda(\Sch{r})}{2}}\,,\\
        &V_{s=1}(\Sch{r})=\frac{l(l+1)}{\Sch{r}^2}e^{\Gamma(\Sch{r})}\,,
    \end{split}
    \label{eq:EffectivePotentialsScalarMasslessEMmodes}
\end{equation}
for the scalar ($s=0$) and electromagnetic ($s=1$) cases, respectively.
In the above equations, $\Sch{r}_{\ast}$ is the generalized tortoise coordinate defined through
\begin{equation}
    \frac{d\Sch{r}_{\ast}}{d\Sch{r}}=e^{\frac{\Lambda(\Sch{r})-\Gamma(\Sch{r})}{2}}\,.
\end{equation}

\begin{figure}[h!]
	\centering
	\includegraphics[width = \columnwidth]{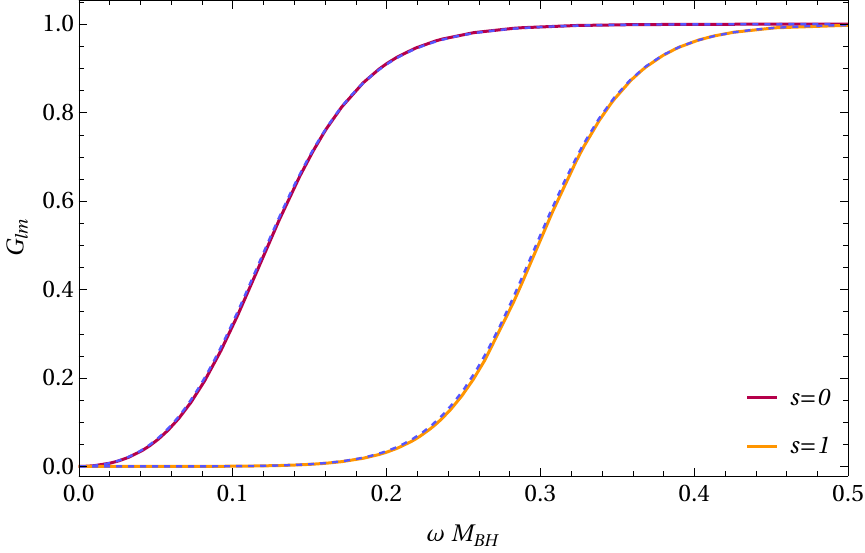}
	\caption{Graybody factors of the dilatonic BH with minimum mass ($\lambda\approx0.01552$ in units such that $\Sch{r}_h = 2$, as we shall fix from now on), for the emission of massless scalar particles (purple) and photons (orange) in their lowest angular modes ($l=0,1$, respectively). We compare each curve with the corresponding graybody factors of a Schwarzschild BH with same mass (dashed blue lines).}
	\label{fig:GrayBodyFactors}
\end{figure}

The potentials in Eq.~\eqref{eq:EffectivePotentialsScalarMasslessEMmodes} vanish both at the horizon and at spatial infinity and their radial profile is in fact qualitatively very similar to the case of a Schwarzschild BH. The asymptotic solutions are ingoing/outgoing waves in tortoise coordinates, $\Theta_{lm}\sim e^{\pm i \omega \Sch{r}_{\ast}}$.
If we normalize the flux coming from infinity, the graybody factor is simply related to the transmission coefficient of the master function,
\begin{equation}
    \begin{cases}
    &\Theta_{lm}=e^{-i\omega \Sch{r}_{\ast}}+\mathcal{R}_{lm}e^{i\omega \Sch{r}_{\ast}} \ \ \ \ \ \ \Sch{r}_{\ast}\rightarrow\infty\\
    &\Theta_{lm}=G_{lm}e^{-i\omega \Sch{r}_{\ast}} \ \ \ \ \ \ \ \ \ \ \ \ \ \ \ \ \Sch{r}_{\ast}\rightarrow-\infty
    \end{cases}\,.
\end{equation}

We have studied this scattering problem for the lowest angular modes of the massless scalar and the electromagnetic field, for different values of the coupling constant $\lambda$. In Fig.~\eqref{fig:GrayBodyFactors}, we show the graybody factors of the dilatonic BH with minimum mass, compared with those of a Schwarzschild BH of equal mass. Overall these two quantities are very similar to each other for any value of the coupling (of course the agreement further improves for smaller values of the coupling than that shown in Fig.~\eqref{fig:GrayBodyFactors}). 
This is consistent with the fact that the graybody factor is mainly governed by the BH photon-sphere, which is slightly outside the horizon, where the higher-curvature corrections are already smaller relative to their value at and inside the horizon.

Therefore, the main difference between the spectrum of a dilatonic and a Schwarzschild BH comes from the (slightly) different temperature. Since the temperature of a dilatonic BH is (slightly) higher than that of a Schwarzschild BH of the same mass, the former evaporates (slightly) faster than the latter.
Using Eq.~\eqref{massloss}, we estimate that near the minimum mass a dilatonic BH evaporates $\approx 7\%$ and $\approx 14\%$ faster than in GR for scalar and vector modes, respectively.

Intriguingly, when the dilatonic BH reaches the minimum mass configuration, the graybody factor and temperature are finite and nonvanishing. In other words, the BH should continue evaporating, but since there are no static BH solutions with lower mass, it is natural to ask toward which state the BH evolves.

 \begin{figure}[h!]
	\centering
	\includegraphics[width = \columnwidth]{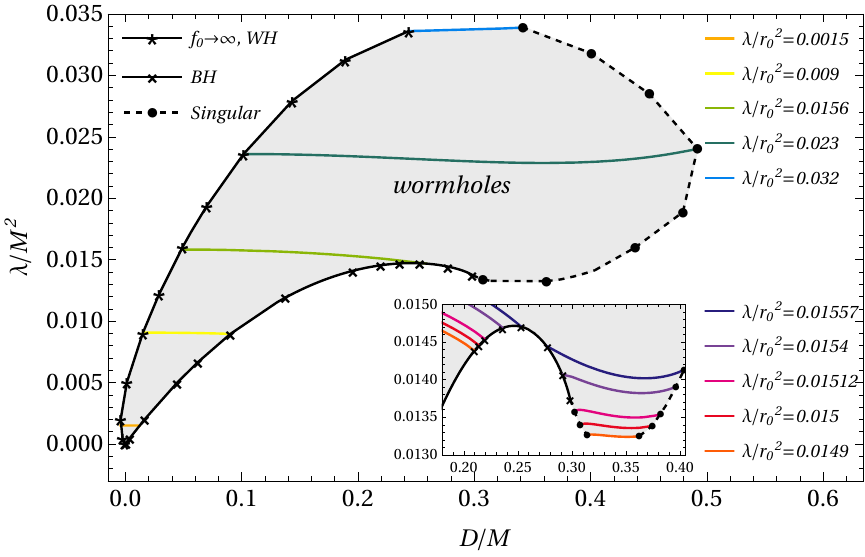}
	\caption{Families of asymptotically flat solutions to EdGB gravity as plotted in the phase space $(D/M,\lambda/M^2)$, where $D$ is the dilatonic charge and $M$ is the mass of the object measured by an observer at spatial infinity (see also~\cite{Kanti:2011jz,Kanti:2011yv,Kleihaus:2019rbg,Kleihaus:2020qwo}).
	The gray region represents the domain of existence of the wormhole solutions, each colored line represents a family of wormhole solutions characterized by a specific value of $\lambda$ (in units where the wormhole throat is $r_0=2$): $\lambda=0.0015$ (orange), $\lambda=0.009$ (yellow), $\lambda=0.0156$ (green), $\lambda=0.023$ (dark green) and $\lambda=0.032$ (light blue). The upper black line (asterisks) corresponds to regular wormhole solutions in the $f_0\to\infty$ limit (see Appendix~\ref{app:Phases}). The lower black line (crosses) corresponds to BH solutions, whereas the dashed line (dots) corresponds to solitonic solutions with a singularity in the second derivative of the dilaton.
	}
	\label{fig:SolutionsEdGB}
\end{figure}

\subsubsection{Phase diagram in EdGB gravity} \label{sec:phase}

To start addressing the question related to the evolution of BHs past the minimum mass in EdGB gravity, it is useful to study in detail the parameter space of static and spherically-symmetric solutions in this theory.
In particular, one might entertain the idea of phase transitions from the critical BH toward some other solutions, should the parameter space allow for that.
Interestingly, EdGB gravity admits other, horizonless, asymptotically flat solutions: traversable wormholes~\cite{Kanti:2011jz,Kanti:2011yv} and particle-like (solitonic) solutions
characterized by a singularity in the second derivative of the dilaton field~\cite{Kleihaus:2019rbg,Kleihaus:2020qwo}. 
We have built these solutions following Refs.~\cite{Kanti:2011jz,Kanti:2011yv,Kleihaus:2019rbg,Kleihaus:2020qwo}. Details are presented in Appendix~\ref{app:Phases}.

In Fig.~\ref{fig:SolutionsEdGB}, we present the phase diagram $(D/M,\lambda/M^2)$, first computed in Ref.~\cite{Kleihaus:2019rbg,Kleihaus:2020qwo}. 
BHs and solitons form a one-parameter family of solutions, so they are represented by curves which encloses a two-dimensional surface. The latter is the domain of existence of the wormhole solutions.
An interesting feature of this phase diagram is that the BH solutions (including the minimum mass) correspond to \emph{double points} in the phase space, wherein the BH and the wormhole solution co-exist (see inset in Fig.~\ref{fig:SolutionsEdGB}).
Furthermore, the singular BH solution at the end of the unstable branch connects also to the solitonic solution which has a derivative singularity (i.e., a cusp), being therefore a \emph{triple point} in the phase space of the theory. Thus, even though the soliton solution is probably not a good candidate for the endpoint of a phase transition, the regular wormhole solution is more appealing.

\subsection{Static solutions in horizon-penetrating coordinates} \label{sec:BH2}

As discussed in Sec.~\ref{sec:numsetup}, for our nonlinear simulations we are interested also in the BH interior.
Therefore, we need to construct initial data using horizon-penetrating coordinates such as PG-like ones (Eq.~\eqref{eq:PG coord}). Since we are interested in simulating the BH evolution close to the critical configuration, for which the curvature singularity is close to the horizon, we also need small grid steps to resolve properly the BH region. In order to reduce the computational cost by increasing the resolution only in the central region, we define the areal radius $R(r)$ in terms of a radial coordinate $r$. As explained in Appendix~\ref{app:Tests}, the function $R(r)$ is accurately chosen as to achieve better resolution in high-curvature regions while keeping a uniform grid for the coordinate radius $r$.
The line element in PG-like coordinates can thus be written as
\begin{equation}
	ds^2 = -\alpha^2 dt^2 + (R'(r) dr + \alpha \zeta \, dt)^2 + R(r)^2 d\Omega^2,
	\label{eq:PGLineElementRT}
\end{equation}
where $\alpha$ and $\zeta$ depend in general on $(r, t)$. In the following equations we shall often leave the $r$ dependence of $R$ implicit.

\subsubsection{Equations and boundary conditions} 
Replacing the static line element~\eqref{eq:PGLineElementRT} into the field equations~\eqref{eq:field_grav}-\eqref{eq:field_dilaton} and performing algebraic operations, we obtain two first-order equations for $\alpha$ and $\zeta$, and a second-order equation for $\phi$, which are reported in Appendix~\ref{app:Eqs}.

The expansion of the future-directed outgoing null geodesics normal to the 2-spheres $S_R$ of (areal) radius $R$ is given by
\begin{equation}
	\theta_{(l)} = \frac{2}{R} (1 - \zeta),
	\label{eq:Expansion}
\end{equation}
where $l^\mu = \bigl(\frac{1}{\alpha}, \frac{1 - \zeta}{R'}, 0, 0\bigr)$ is the future-directed null vector normal to $S_R$. Thus, the horizon $r_h$ is located where $\zeta = 1$. 

The denominator of the right-hand side of the equation for the dilaton (Eq.~\eqref{eq:PhiPrimePrimeDenominator}) goes to zero at the horizon, and imposing that the singular terms in $\phi_h'' := \phi''(r_h)$ vanish, we recover the regularity condition~\cite{Kanti:1995vq}: 
\begin{equation}
	\phi'_h = \frac{R_h' \left(-R_h^2 + \sqrt{R_h^4-192 F'[\phi_h]^2}\right)}{8 R_h F'[\phi_h]},
	\label{eq:HorizonPhiDer}
\end{equation}
where the subscript $h$ indicates that the quantities are evaluated at the horizon, and $F'[\phi] = \frac{\delta F[\phi]}{\delta \phi}$. This expression, together with the regularity condition $\zeta_h = 1$, are the analog of Eq.~\eqref{eq:MetricDilatonExpansionInfSchw} in different coordinates.

In PG-like coordinates the spatial 3-metric is flat, and thus the Arnowitt-Deser-Misner mass identically vanishes. Following~\cite{Ripley:2019aqj}, we use the asymptotic value of the Misner-Sharp mass function $m_{\rm MS}(r)$ as a definition of the total mass of the spacetime:
\begin{equation}
	M_{\rm MS} := \lim_{r \to +\infty} m_{\rm MS}(r) = \lim_{r \to +\infty} \frac{R}{2} \zeta^2.
	\label{eq:MSmass}
\end{equation}
We can now write the asymptotic behaviors of $\phi$, $\alpha$ and $\zeta$ in the asymptotically flat case as
\begin{align}
	\phi &= -\frac{D}{R} + \OO\Bigl(\frac{1}{R^2}\bigr), \label{eq:AsymptoticPhi} \\
	\zeta &= \sqrt{2 \frac{M_{\rm MS}}{R}} + \OO\Bigl(\frac{1}{R^{5/2}}\Bigr), \label{eq:AsymptoticZeta} \\
	\alpha &= A + \OO\Bigl( \frac{1}{R^2} \Bigr), \label{eq:AsymptoticAlpha}
\end{align}
where the constant $A$ in Eq.~\eqref{eq:AsymptoticAlpha} is a free parameter, since $\alpha$ can be arbitrarily rescaled by a constant with a redefinition of the coordinate time.

\subsubsection{Numerical procedures}
We used two procedures for constructing the static dilatonic BH solutions.

The first is a standard shooting, wherein (for fixed values of the coupling constant $\lambda$ and the horizon radius $R_h$) we integrate the equations from the horizon outward, using  Newton's method to find the value of the only free parameter $\phi_h$ for which the asymptotic boundary conditions~\eqref{eq:AsymptoticPhi}-\eqref{eq:AsymptoticAlpha} are satisfied. We finally obtain the static dilatonic solution by performing an integration both outside and inside the BH region. Note that since the equations for $\phi$ and $\zeta$ do not depend on $\alpha$, we do not integrate the equation for this metric function. 

The second procedure is based on the invariance of the theory under the transformation
\begin{equation}
	\phi \to \phi + C  \qquad \qquad \lambda \to \lambda e^{\gamma C},
	\label{eq:ExponentialCouplingSymmetry}
\end{equation}
where $C$ is a real constant. The strategy is similar to the one outlined in Ref.~\cite{Kokkotas:2017ymc}. Namely, we start by fixing the horizon radius and setting the coupling constant to a generic value. We initialize $\phi_h$, and then $\zeta_h$ and $\phi_h'$ with the conditions at the horizon. We then integrate equations~\eqref{eq:ZetaPrime}-\eqref{eq:PhiPrimePrime}, obtaining the generic asymptotic behavior for $\phi \sim {\rm cost} - \frac{D}{R}$.
Finally, we perform a symmetry transformation~\eqref{eq:ExponentialCouplingSymmetry} to impose \eqref{eq:AsymptoticPhi}. 
This second procedure has the advantage of being faster, since it does not require solving the field equations multiple times to construct a single solution. Furthermore, it simplifies finding multiple solutions for the same coupling constant, when they exist. On the other hand, since it takes advantage of a symmetry of the theory, it can only be used with couplings such that the action is invariant under~\eqref{eq:ExponentialCouplingSymmetry}. 

In both cases, we perform the numerical integration using the fourth-order accurate Runge-Kutta method, starting from the horizon and moving both inward and outward. Even though from an analytical point of view the conditions at the horizon guarantee the regularity of the field equations, the presence of $(1 - \zeta^2)$ at the denominator of the equation for the dilaton can cause instabilities when used in a numerical integration algorithm. To overcome this issue we use the following strategy. First we integrate the field equations with the fourth-order accurate Runge-Kutta method for a single step from $r_h$ to $r_h + \frac{\Delta r}{2}$, where $\Delta r$ is the required grid step. We use the analytic expression of $\phi_h''$ and $\zeta_h'$ (Eqs.~\eqref{eq:HorizonPhiDder}-\eqref{eq:HorizonZetaDer}) as the right-hand sides of the equations at the horizon, while we use Eqs.~\eqref{eq:PhiPrimePrime}-\eqref{eq:ZetaPrime} in the intermediate steps. Then, we continue the numerical integration up to the outer boundary using $\Delta r$ as integration step. We repeat the same procedure inside the BH region and we obtain that in the final numerical data the horizon is staggered between two grid points.
We have found that, when the static solution is used to initialize the evolution code described in the next section, this strategy produces a better behaved constraint violation with respect to the standard Taylor's expansion at the horizon.

Let us stress that the BH solutions have a curvature singularity inside the horizon~\cite{Alexeev:1996vs}, so we can only integrate the equations from the horizon inward up to the radius of such singularity. The position of the singularity inside the horizon depends on the specific value of the coupling constant, which motivates the discussion presented in the next subsection.

\begin{figure}
	\centering
	\includegraphics[width = \columnwidth]{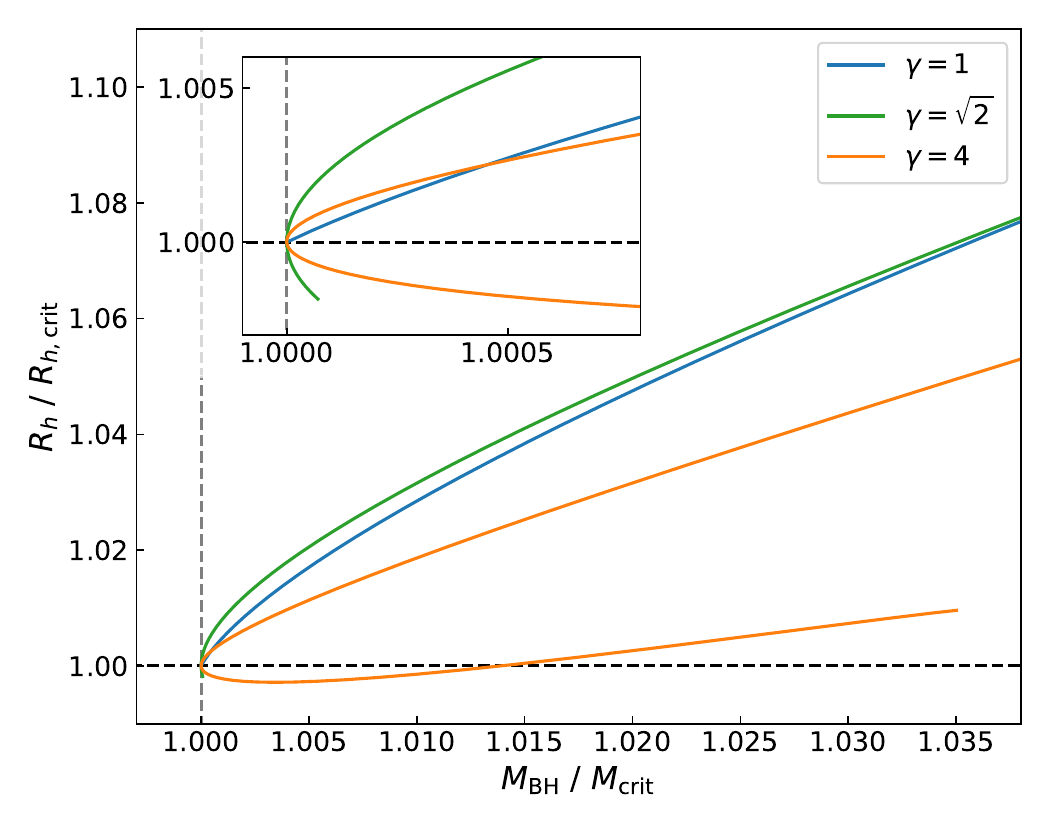}
	\caption{Dilatonic solutions in the $R_h-M_{\rm BH}$ plane for different values of $\gamma$. For $\gamma = 1$ the singular configuration has minimum mass, for $\gamma = \sqrt{2}$ a second branch forms and the singular configuration has minimum radius, while for $\gamma = 4$ both the minimum-mass and minimum-radius solutions are regular at the horizon.
	}
	\label{fig:DilatonicExponentialG4}
\end{figure}

\subsubsection{Properties of the solutions for different $\gamma$'s}
In Fig.~\ref{fig:DilatonicExponentialG4} we show the usual $R_h-M_{\rm BH}$ plane for some representative values of $\gamma$. For $\gamma=1$, there is only one branch of solutions and no local minimum of the BH mass. In this case the minimum-mass solution is also singular at the horizon, as in the shift-symmetric case~\cite{Sotiriou:2013qea,Sotiriou:2014pfa}. For slightly larger values of $\gamma$ (e.g. $\gamma=\sqrt{2}$ in the plot), there is a critical (minimum-mass) BH which is regular in and outside the horizon. This solution separates two branches, with the lower one terminating at the minimum-radius solution, which is singular at the horizon \cite{Torii:1996yi, Guo:2008hf}.
Finally, for even larger values of $\gamma$ (e.g., $\gamma=4$ in the plot), also the minimum-radius solution is regular in the BH exterior~\cite{Blazquez-Salcedo:2017txk}. In this case the second branch terminates at a different solution which is not the minimum-mass nor the minimum-radius one.
Note, however, that the lower branch is linearly unstable~\cite{Torii:1998gm}, as we shall also find at the fully nonlinear level in Sec.~\ref{sec:gedanken}. Therefore, the physically interesting solutions are those on the upper branch, and we are particularly interested in the critical (minimum-mass) BH in those cases in which it is regular.

\begin{figure}
	\centering
	\includegraphics[width = \columnwidth]{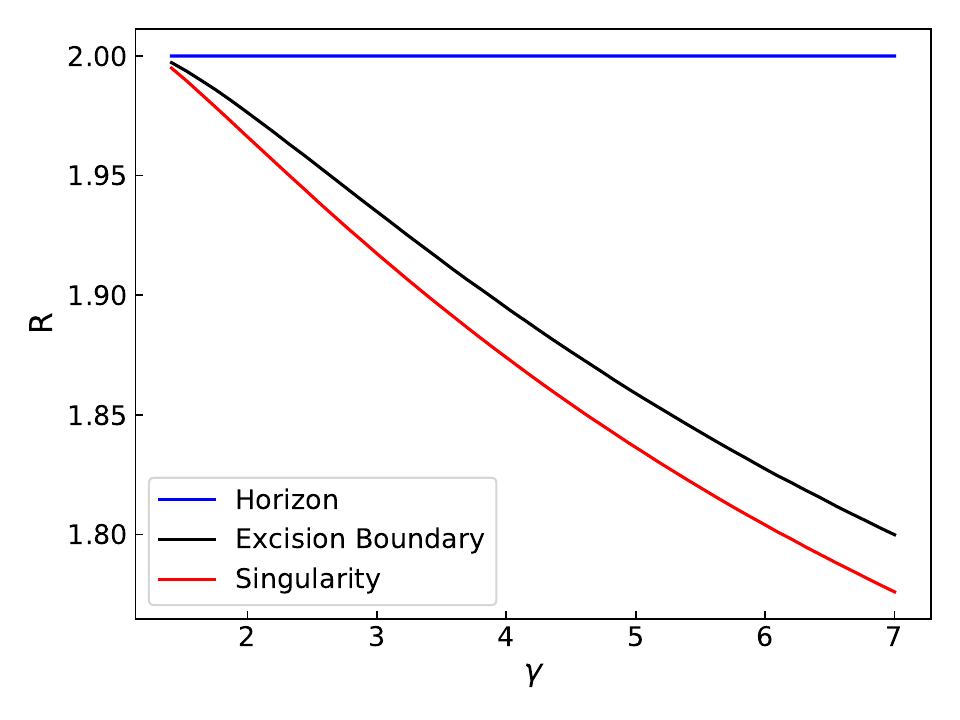}
	\caption{Position of the excision boundary and curvature singularity at the critical configuration for different choices of the parameter $\gamma$. Given a fixed value of the horizon radius, the radii of the excision boundary and the singularity at the critical configuration decrease as $\gamma$ increases. For this reason, using larger values of $\gamma$ allows us to use larger grid steps and reduce the computational cost.}
	\label{fig:StaticCriticalExcisionSingularity}
\end{figure}

It is also interesting to investigate in more details the location $R_s$ of the curvature singularity inside the horizon as a function of the dilaton coupling. To identify the singularity, we considered the numerical data obtained from the integration in the BH region, which starts from the horizon and proceeds inward.
At the singularity the denominator $D_\phi$ in the right-hand side of the equation for the dilaton (Eq.~\eqref{eq:PhiPrimePrime}) vanishes, thus the algorithm fails and the numerical data become less smooth, featuring spurious jumps. We determine $R_s$ as the radius where this happens, imposing numerical conditions that detect changes of sign or discontinuities in $D_\phi$ and its derivatives near the root.
In Fig.~\ref{fig:StaticCriticalExcisionSingularity}, we compare the location of the singularity with the horizon radius at the critical BH solution for different values of $\gamma$. The units are fixed in such a way that $R_h = 2$.
Overall, the smaller the $\gamma$ the smaller the areal distance between the singularity and the horizon, which also requires higher resolution to resolve the region around the horizon.
Thus, in order to reduce the computational cost of the nonlinear time evolution presented in the next section, in addition to using the radial transformation $R(r)$ we decided to set $\gamma = 4$. We also checked different values of $\gamma$, finding a qualitatively similar behavior.
Note that in Fig.~\ref{fig:StaticCriticalExcisionSingularity} we also show the radius of the excised region, $R_e$, obtained by initializing the evolution algorithm presented in Sec.~\ref{sec:numsetup}. Details on the excision are given later on.

\begin{figure}
	\centering
	\includegraphics[width = \columnwidth]{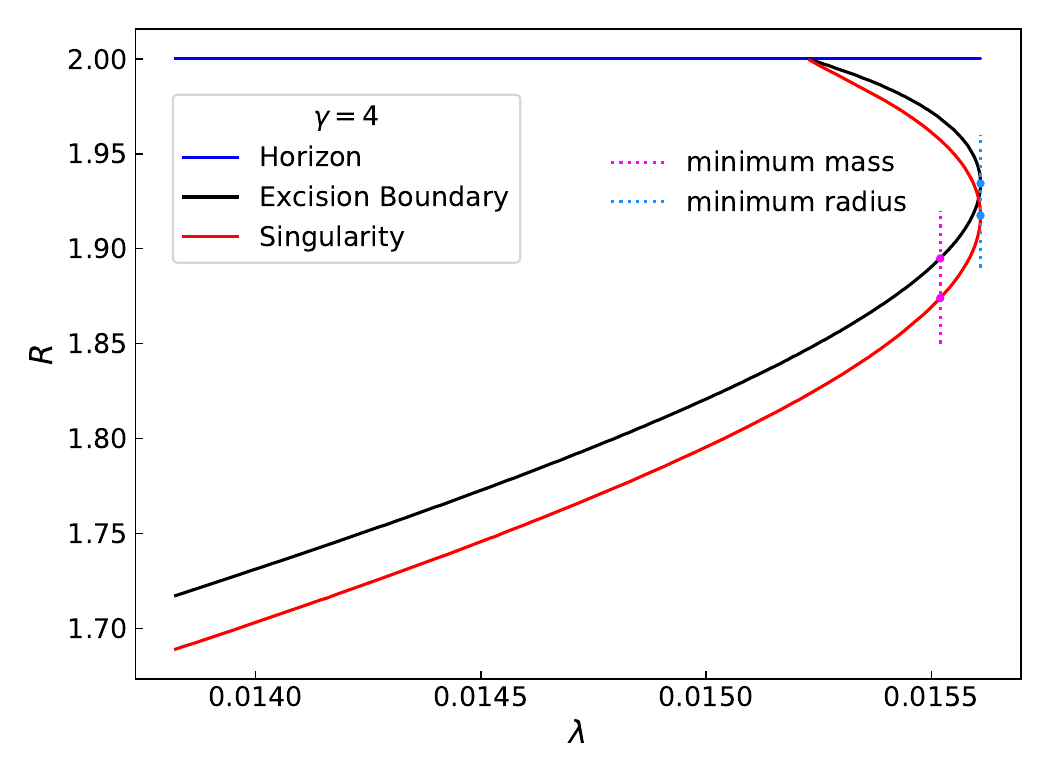}
	\caption{Position of the excision boundary (black) and the curvature singularity (red) for the static dilatonic solutions in the case of $\gamma = 4$. Both curves reach the horizon (blue line) at the singular configuration.
	}
	\label{fig:StaticExcisionSingularityG4}
\end{figure}

Finally, in Fig.~\ref{fig:StaticExcisionSingularityG4} we show the behavior of the excision radius (black curve) and of the singularity (red curve) with respect to the coupling constant $\lambda$ when $\gamma = 4$. 
As anticipated, for the minimum-mass solution the singularity is well within the horizon, whereas near the singular configuration both the excision and the singularity approach the horizon radius. 
Moreover, since these solutions are computed at fixed horizon areal radius $R_h = 2$ the coupling constant starts decreasing after the configuration that minimizes $\frac{R_h}{\sqrt{\lambda}}$.

\section{Numerical setup: initial value problem in EdGB gravity}\label{sec:numsetup}

In this section we discuss our numerical setup for the spherical collapse of fields onto a dilatonic BH in EdGB gravity. We mostly follow the formalism used in Ref.~\cite{Ripley:2019aqj} for shift-symmetric (i.e., $F[\phi]\propto\phi$) EdGB gravity.
We remind that we consider the collapse both of the dilatonic field $\phi$ directly coupled to the higher-curvature terms, and that of a phantom field $\xi$, which is needed to mimic BH evaporation at the classical level.

\subsection{System of equations and hyperbolicity}
To obtain the evolution equations for the system we start by defining the variables
\begin{align}
	Q = \rder \phi, \qquad 	\Theta = \rder \xi,
	\label{eq:QThetaDef}
\end{align}
and the conjugate momenta
\begin{align}
	P = \frac{1}{\alpha} \tder \phi - \frac{\zeta Q}{R'(r)}, \qquad 	\Pi = \frac{1}{\alpha} \tder \xi - \frac{\zeta \Theta}{R'(r)}.
	\label{eq:PPiDef}
\end{align}
We then substitute these definitions and the ansatz for the metric in the field equations and obtain a set of 7 evolution equations for $\phi$, $Q$, $P$, $\xi$, $\Theta$, $\Pi$, $\zeta$, plus 2 constraint equations for $\alpha$ and $\zeta$. All equations are reported in Appendix~\ref{app:Eqs}. The evolution equations for $\phi$ and $\xi$ are redundant, since the profiles of the scalar fields can be obtained using Eqs.~\eqref{eq:QThetaDef} as constraints. 

In PG-like coordinates the system of evolution equations and constraints is not everywhere hyperbolic~\cite{Ripley:2019aqj}. In order to identify elliptic regions during the numerical evolution we computed the discriminant of the characteristic equation following Ref.~\cite{Ripley:2019irj}. 

In particular, we consider the principal symbol of our system of equations
\begin{equation}
	\mathcal P_{IJ} = \frac{\delta E_{v^I}}{\delta \partial_\mu v^J} \eta_\mu,
	\label{eq:PrincipalSymbol}
\end{equation}
where $v^I = (\phi, Q, P, \xi, \Theta, \Pi, \alpha, \zeta)$ schematically denotes a variable of the system of equations, $E_{v^I}$ is the $I$-th field equation written in implicit form (6 evolution equations for $\phi$, $Q$, $P$, $\xi$, $\Theta$, $\Pi$, and 2 constraint for $\zeta$, $\alpha$), and $\eta^\mu$ is a 4-vector. 
The determinant of $\mathcal P$ has the form
\begin{align}
	\det{\mathcal P} &\propto \eta_t^2 \eta_r^2 \biggl[ \mathfrak{a}_\xi \biggl(\frac{\eta_t}{\eta_r}\biggr)^2 + \mathfrak{b}_\xi \biggl( \frac{\eta_t}{\eta_r} \biggr) + \mathfrak{c}_\xi \biggr] \notag \\
		& \times \biggl[ \mathfrak{a}_\phi \biggl(\frac{\eta_t}{\eta_r}\biggr)^2 + \mathfrak{b}_\phi \biggl( \frac{\eta_t}{\eta_r} \biggr) + \mathfrak{c}_\phi \biggr],
	\label{eq:DetP}
\end{align}
where $\mathfrak{a}_\phi$, $\mathfrak{b}_\phi$, $\mathfrak{c}_\phi$, $\mathfrak{a}_\xi$, $\mathfrak{b}_\xi$, and $\mathfrak{c}_\xi$ are lengthy expressions that depend on all the fields.
This determinant vanishes if $\eta_t^2 \eta_r^2 = 0$, $\mathfrak{a}_\phi \bigl(\frac{\eta_t}{\eta_r}\bigl)^2 + \mathfrak{b}_\phi \bigl( \frac{\eta_t}{\eta_r} \bigr) + \mathfrak{c}_\phi = 0$, or $\mathfrak{a}_\xi \bigl(\frac{\eta_t}{\eta_r}\bigl)^2 + \mathfrak{b}_\xi \bigl( \frac{\eta_t}{\eta_r} \bigr) + \mathfrak{c}_\xi = 0$. The first equation has two solutions $\eta_r = 0$, which come from the fact that $\alpha$ and $\zeta$ are constrained degrees of freedom, and two solutions $\eta_t = 0$, which come from the redundancy of the equations for $\tder \phi$ and $\tder \xi$. 

The second and the third equations have real solutions if the corresponding discriminants, $\Delta = \mathfrak{b}^2 - 4 \mathfrak{a} \mathfrak{c}$, are nonnegative. In this case the characteristic velocities $c_{\pm} = -\Bigl(\frac{\eta_t}{\eta_r}\Bigr)_{\pm}$ are given by
\begin{align}
	c_{\pm}^{(\phi)} = \frac{\mathfrak{b}_\phi \pm \sqrt{\Delta_\phi}}{2 \mathfrak{a}_\phi}, \qquad c_{\pm}^{(\xi)} = \frac{\mathfrak{b}_\xi \pm \sqrt{\Delta_\xi}}{2 \mathfrak{a}_\xi}.
	\label{eq:CharacteristicVelocities}
\end{align}
In order for the system to be hyperbolic we need to impose that both discriminants
\begin{align}
	\Delta_\phi = \mathfrak{b}_\phi^2 - 4 \mathfrak{a}_\phi \mathfrak{c}_\phi \,,\qquad 	\Delta_\xi = \mathfrak{b}_\xi^2 - 4 \mathfrak{a}_\xi \mathfrak{c}_\xi
	\label{eq:Discriminants}
\end{align}
are positive, so that there are 4 different real characteristic velocities.
As we shall later discuss, we use an excision procedure to exclude the spacetime region where the system is not hyperbolic.

\subsection{Initial data} \label{subsec:Initialization}
Our purpose is to simulate the evolution of small perturbations of scalar fields around initially static dilatonic BHs. To construct these initial configurations we first use the procedures described in Sec.~\ref{sec:staticBHs} to find the profiles $\phi_0(r)$, $Q_0(r)$, and $\zeta_0(r)$ corresponding to a static isolated BH. Next, we initialize the dilaton as
\begin{align}
	\phi(r, t=0) &= \phi_0(r) + \delta \phi(r), \notag \\
	Q(r, t=0) &= Q_0(r) + \delta Q(r), \notag \\
	P(r, t=0) &= P_0(r) + \delta P(r) = - \frac{\zeta_0(r) \, Q_0(r)}{R'(r)} + \delta P(r),
	\label{eq:InitialPhi}
\end{align}
where
\begin{align}
	\delta \phi(r) &= \frac{A_{0, \phi}}{R(r)} e^{-\frac{(R(r) - R_{0, \phi})^2}{\sigma_\phi^2}}, \notag \\
	\delta Q(r) &= \rder \delta \phi(r), \notag \\
	\delta P(r) &= \frac{\delta \phi(r)}{R(r)} + \partial_R \delta \phi(r) = \frac{\delta \phi(r)}{R(r)} + \frac{1}{R'} \delta Q(r)\,.
	\label{eq:InitialPhiPerturbation}
\end{align}
Similarly, since the phantom field vanishes in the background, we initialize its perturbation as
\begin{align}
	\xi(r, t=0) &= \delta \xi(r) = \frac{A_{0, \xi}}{R(r)} e^{-\frac{(R(r) - R_{0, \xi})^2}{\sigma_\xi^2}}, \notag \\
	\Theta(r, t=0) &= \delta \Theta(r) = \rder \delta \xi(r), \notag \\
	\Pi(r, t=0) &=	\delta \Pi(r) = \frac{\delta \xi(r)}{R(r)} + \partial_R \delta \xi(r) \notag \\
			&= \frac{\delta \xi(r)}{R(r)} + \frac{1}{R'} \delta \Theta(r).
	\label{eq:InitialXi}
\end{align}
In Eqs.~\eqref{eq:InitialPhiPerturbation}-\eqref{eq:InitialXi}, $A_{0, \phi}$ and $A_{0, \xi}$ represent the amplitudes of the dilaton and phantom perturbations, respectively, $R_{0, \phi}$ and $R_{0, \xi}$ represent the peak value of the Gaussian profiles, whereas $\sigma_\phi$ and $\sigma_\xi$ are the typical widths. The conjugate momenta of the perturbations are similar to Ref.\cite{Ripley:2019irj}. With this choice, the wave packets are approximately inward moving.

We then integrate the constraints with the fourth-order accurate Runge-Kutta method, starting from the first grid point outside the horizon and moving both outward and inward. We assume that the perturbations of both fields are far enough from the horizon that we can consider the metric to be initially unperturbed in that region, and we start the numerical integration using the value of $\zeta$ obtained from the shooting procedure. Initially we set $\alpha=1$, and at the end of the initialization process we rescale it in such a way that $\alpha(r_\infty) = 1$, where $r_\infty$ is the outermost grid point.

The fourth-order accurate Runge-Kutta method requires evaluating the right-hand side of the equations in intermediate grid points. In order to obtain the values of the dilatonic field in these points we construct the static BH solution using a double resolution compared to that required by the numerical evolution. Namely, if we want the grid step of the numerical evolution to be $\Delta r$, we perform the shooting procedure with $\frac{\Delta r}{2}$ as a grid step, and we use half of the grid points as intermediate values for the Runge-Kutta method. We then discard them at the end of the initialization procedure. We evaluate $\rder Q$ and $\rder P$ on the right-hand side of the constraints by applying the fourth-order accurate centered finite differences scheme on the data from the shooting procedure, i.e., using the profiles obtained with grid step equal to $\frac{\Delta r}{2}$.

\subsection{Numerical evolution algorithm}
We perform the numerical integration with the method of lines, using the fourth-order accurate Runge-Kutta method for the time integration, and the fourth-order accurate finite differences method for computing the radial derivatives. In particular, at each step of the time integration we use Eqs.~\eqref{eq:EqPhi}-\eqref{eq:EqZeta} to evaluate the intermediate profiles of $\phi$, $Q$, $P$, $\xi$, $\Theta$, $\Pi$, and $\zeta$ required by the Runge-Kutta method, and we perform a fourth-order accurate numerical integration of Eq.~\eqref{eq:ConstrAlpha} to obtain the profile of $\alpha$. 

This latter numerical integration cannot be performed using the Runge-Kutta method, as it requires the evaluation of the fields in intermediate grid steps, which cannot be done in our setup due to the fact that the fields are only defined on the grid points. Nevertheless the constraint for $\alpha$ can schematically be written as
\begin{equation}
	\frac{\rder \alpha}{\alpha} =  L[R, \phi, Q, P, \Theta, \Pi, \zeta],
	\label{eq:LogConstrAlpha}
\end{equation}
where $L$ does not depend on $\alpha$. The solution then reads
\begin{equation}
	\alpha(r) =  \exp\Bigl[ \ln \alpha(r_\infty) + \int_{r_\infty}^r L \, dr \Bigr],
	\label{eq:IntegratedAlpha}
\end{equation}
where $\alpha(r_\infty)$ is given by the boundary conditions on the outermost grid point. We compute the integral in the above equation using the trapezoidal rule when $r$ and $r_\infty$ are adjacent grid points, and with a combination of the Simpson's rules in the other cases. In this way we obtain an accuracy of order four in all the numerical grid except in the last grid step. 

We use an excision procedure to remove the region where the system is not hyperbolic. The strategy is similar to the one used in Ref.~\cite{Ripley:2019aqj}: at the end of each time step we compute the discriminants~\eqref{eq:Discriminants}, find the outermost radius in which at least one of the two is nonpositive, and then excise the region in the interior. The field equations are not evolved in the excised region, thus the radius of the excision boundary $R_e$ cannot decrease, but would at most remain constant if the elliptic region shrinks.  

We also monitor the evolution of the apparent horizon, which is located at the coordinate radius $r_h$ where the expansion vanishes, $\theta_{(l)}(r_h) = 0$. We estimate $r_h$ using a linear interpolation.
Since the results of the numerical integration lose physical meaning when an elliptic region appears outside the BH\footnote{Note, however, that since the apparent horizon does not coincide with the event horizon in dynamical situations, the emergence of an elliptic region outside the apparent horizon is not necessarily pathological. In other words, we cannot exclude in general that an elliptic region outside the apparent horizon would remain confined within the event horizon.}, we stop the simulation if the apparent horizon enters in the excision boundary.

Finally, we implemented a fifth-order Kreiss-Oliger dissipation scheme in order to stabilize the integration algorithm against high-frequency modes arising from the inner- and near-horizon region. The action of the dissipation term is restricted to the central region by means of a weighting function $\rho(r)$. Specifically, if we schematically denote a generic variable with $u$, we add to the right-hand side of each evolution equation the term $Q \, u$ contained in Appendix~C of Ref.~\cite{Babiuc:2007vr}, which we write as 
\begin{equation}
	Q \, u = \frac{\eta_{\rm KO}}{64 \, \Delta t}\bigl(\Delta r)^6 \bigl(D_{+}^3 \bigr) \rho \bigl(D_{-}^3 \bigr) u,
	\label{eq:KOTerm}
\end{equation}
where $\eta_{\rm KO}$ is a constant, $\Delta r$ is the grid step, $\Delta t$ is the time step, $\rho = \rho(r)$ is the weighting function, and $D_{\pm}$ are the operators of first-order numerical differentiation with the one-sided finite difference scheme. 
In particular, we use $\eta_{\rm KO} = 0.1$ and 
\begin{equation}
	\rho(r) = \frac{1}{1 + e^{5 \, (R(r) - 5)}}.
\end{equation} 

Since the computation of the numerical derivatives in Eq.~\eqref{eq:KOTerm} requires three grid points on each side, we do not use the dissipation term in the three grid points near each boundary of the domain of integration.

\subsection{Boundary conditions}

We do not impose conditions at the excision boundary. Since the elliptic region lies always inside the horizon (otherwise we stop the simulation) all the characteristics are ingoing. For this reason, we use the upwind differentiation scheme in the first two grid points outside the excision, while we use the centered scheme in the rest of the grid. 

At the outer boundary we impose $\alpha(r_\infty) = 1$, and we keep all the other variables constant in the outermost three grid points, which are used only for computing the numerical derivatives. This can be done as long as we use a numerical grid large enough that the signals coming from the outer boundary do not reach the horizon region we are interested in. Actually, in the code the condition $\alpha = 1$ is imposed at the first point in which the time integration is performed (the fourth outermost grid point), however the errors introduced in $\alpha$ are of order $\frac{1}{R^2}$ and do not affect the accuracy of the code at late times, as we can see from the results of the test simulations reported in Appendix~\ref{app:Tests}.

We tested our implementation of the integration algorithm by checking the scaling of the violation of the constraint for $\zeta$. Our code appears to be accurate and reliable for the evolution of a static dilatonic BH and for the collapsing scenarios that we will discuss in the next sections. The results of the convergence tests are presented in Appendix~\ref{app:Tests}.

\section{Nonperturbative gedanken experiments with dilatonic BHs in EdGB gravity} \label{sec:gedanken}

We now turn to describe our simulations of the spherical collapse of wave packets on static dilatonic BHs in EdGB gravity. In Secs.~\ref{subsec:dilaton1} and~\ref{subsec:dilaton2} we consider the case of dilatonic perturbations onto BHs in the upper and lower branch, respectively. 
In Secs.~\ref{subsec:phantom}, \ref{subsec:naked}, and~\ref{subsec:pair} we consider different setups of phantom perturbations that reduce the BH mass, thus mimicking BH evaporation at the classical level.
We remind that we use units such that the horizon areal radius of the initial BH is $R_h(t = 0) = 2$, which corresponds to setting the initial BH mass to unity in the GR limit. 

\subsection{Collapse of a dilaton field on a BH in the upper branch} \label{subsec:dilaton1}
Let us first discuss the case of an initial dilatonic BH in the upper branch. We set the coupling constant to $\lambda = 0.01536$, and we construct the initial data using the procedure described in Sec.~\ref{subsec:Initialization}. The parameters $A_\phi$, $R_\phi$, and $\sigma_\phi$ are set to
\begin{equation}
	A_{0, \phi} = 0.02, \qquad R_{0, \phi} = 15, \qquad \sigma_\phi = 0.5,
	\label{eq:DilatonPerturbationParametersUpper}
\end{equation}
while $A_\xi=0$, which implies that the phantom field is always zero in this case.
The outer boundary is at $R_\infty = 520$, the final simulation time is $T = 500 $, and the grid step is $\Delta r = 0.01$, with a Courant-Friedrichs-Lewy factor ${\rm CFL} =\frac{\Delta t}{\Delta r}= 0.025$. \footnote{This small CFL factor is required by the fact that near and inside the horizon the areal radius step $\Delta R$ is approximately 20 times smaller than $\Delta r$.} 

\begin{figure}
	\centering
	\includegraphics[width = \columnwidth]{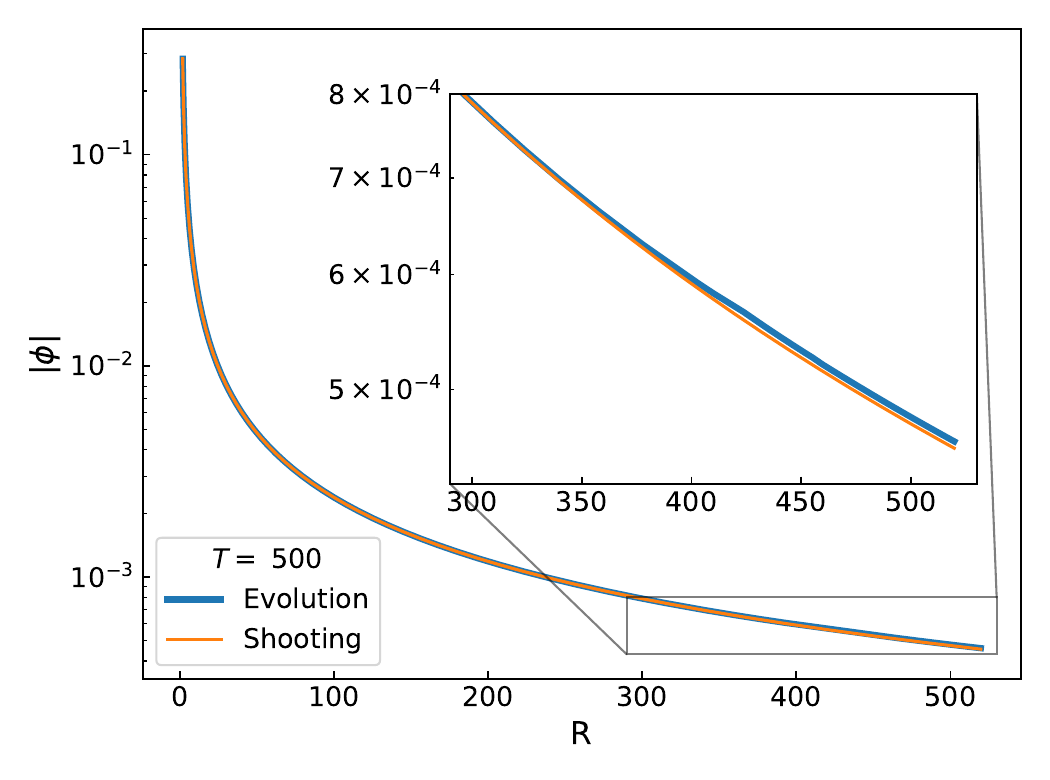}
	\caption{Profile of the dilaton $\phi$ at the end of the simulation of the collapse of a dilaton wave packet on a static BH in the upper branch. The blue curve is obtained at the end of the numerical evolution while the orange curve is obtained from the shooting procedure initialized with the horizon data at $t = T$.}
	\label{fig:DilatonFieldComparison}
\end{figure}

Since the upper branch is expected to be linearly stable~\cite{Torii:1998gm}, after the dilaton wave packet is absorbed the BH mass should increase, and the end-state of the numerical simulation should be approximated by a (slightly heavier) static dilatonic configuration in the upper branch. In order to check this we initialized the shooting algorithm described in Sec.~\ref{sec:BH2} with the horizon data at the end of our simulation ($t = T$), and constructed a static dilatonic BH solution. We then compared it with the profile of the dilaton at the end of the simulation, see Fig.~\ref{fig:DilatonFieldComparison}. The profile obtained by the shooting procedure (orange curve) is in excellent agreement with that obtained at the end of the numerical evolution (blue curve), except in the outer region. This is consistent with the fact that the information of the absorption of the pulse has not yet reached the outer boundary. 

\begin{figure}
	\centering
	\includegraphics[width = \columnwidth]{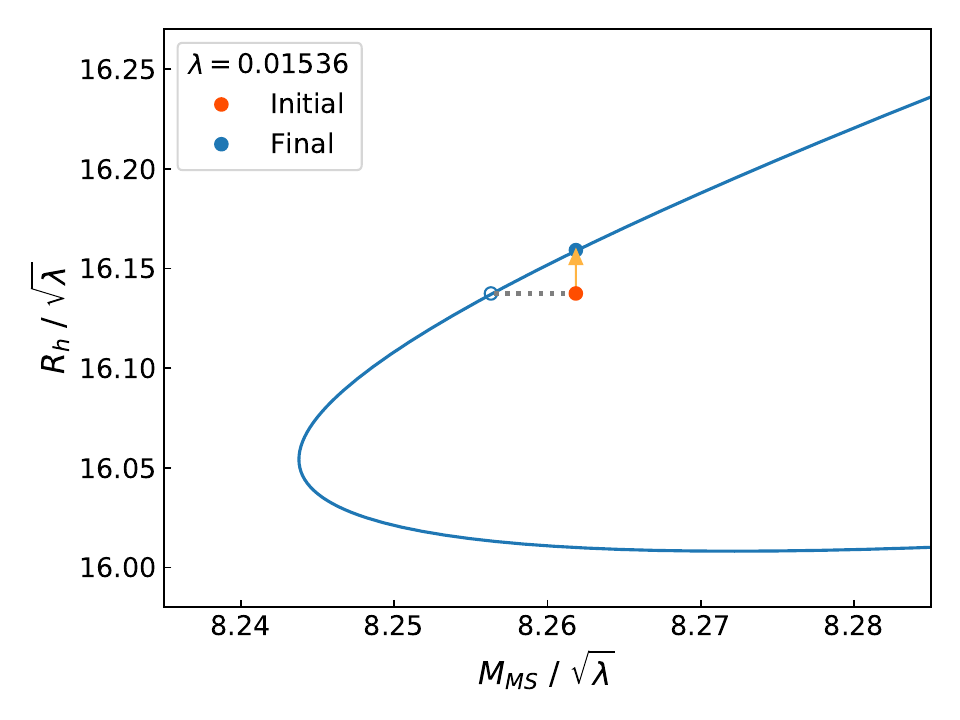}
	\caption{Evolution in the $R_h-M_{\rm MS}$ plane for the collapse of a dilaton wave packet on a BH in the upper branch. The blue curve is the domain of existence of static dilatonic solutions, while the blue point represents the static BH configuration that approximates the end-state of the numerical evolution.}
	\label{fig:DilatonUpperParameterSpace}
\end{figure}

In Fig.~\ref{fig:DilatonUpperParameterSpace} we show the evolution of the system during the simulation in the $R_h-M_{\rm MS}$ plane. The point corresponding to the initial configuration (red circle) is on the right of the domain of existence of static dilatonic BH solutions (blue curve), since the wave packet of the dilaton adds a positive contribution to the total Misner-Sharp mass. 
The initial (isolated) BH solution is marked by an empty circle, connected to the red one by a horizontal dotted line.
The blue full circle represents the static configuration that approximates the end-state of the numerical integration. It is clear that the final state of the evolution is in the upper branch, providing a first numerical confirmation of the stability of this family of solutions at the fully nonlinear level.

\subsection{Collapse of a dilaton field on a BH in the lower branch} \label{subsec:dilaton2}

We now perform a set of four simulations of the same type with different values of the coupling constant $\lambda$ in the range $[0.01554, 0.0156]$.
In this regime, there are two BH solutions for each mass, and we consider those in the lower branch (i.e., with smaller radii) as initial configurations.
These solutions should be linearly unstable~\cite{Torii:1998gm}.

We consider a dilaton wave packet with parameters
\begin{equation}
	A_{0, \phi} = 0.01 , \qquad R_{0, \phi} = 15 , \qquad \sigma_\phi = 2.5 .
	\label{eq:DilatonPerturbationParametersLowerBranch}
\end{equation}
The outer boundary is at $R_\infty = 2850$, the grid step is $\Delta r = 0.02$, and the total integration time is $T = 2800$. 

\begin{figure}
	\centering
	\includegraphics[width = \columnwidth]{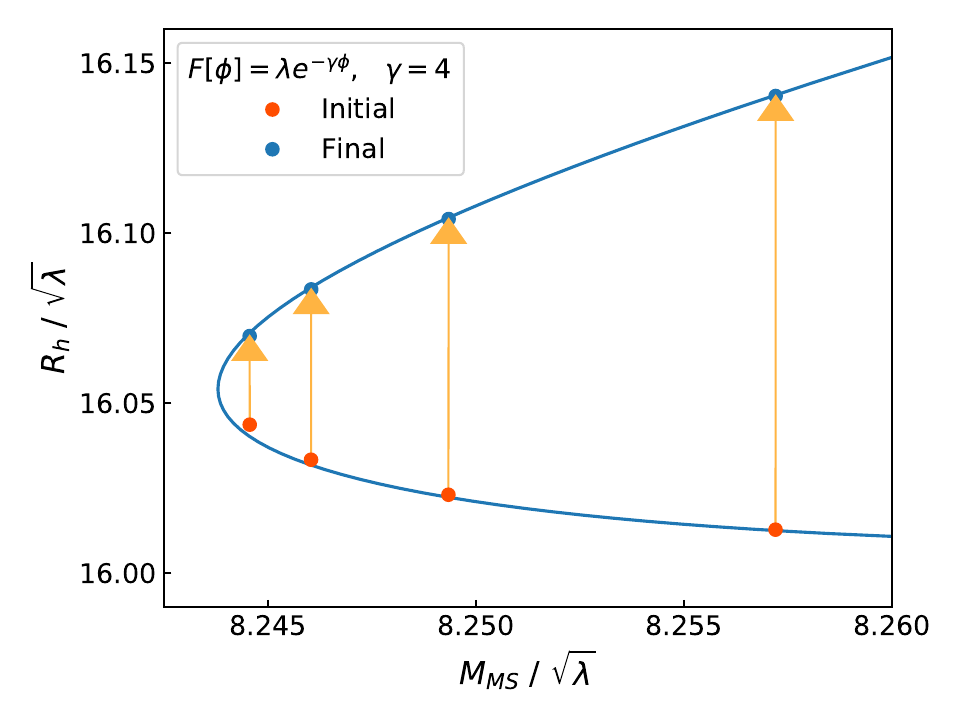}
	\caption{Same as in Fig.~\ref{fig:DilatonUpperParameterSpace} but for the simulations starting from dilatonic BHs in the lower branch. These solutions are unstable and migrate toward stable static BH configurations in the upper branch.}
	\label{fig:LowerBranchDilatonParameterSpace}
\end{figure}
\begin{figure}
	\centering
	\includegraphics[width = \columnwidth]{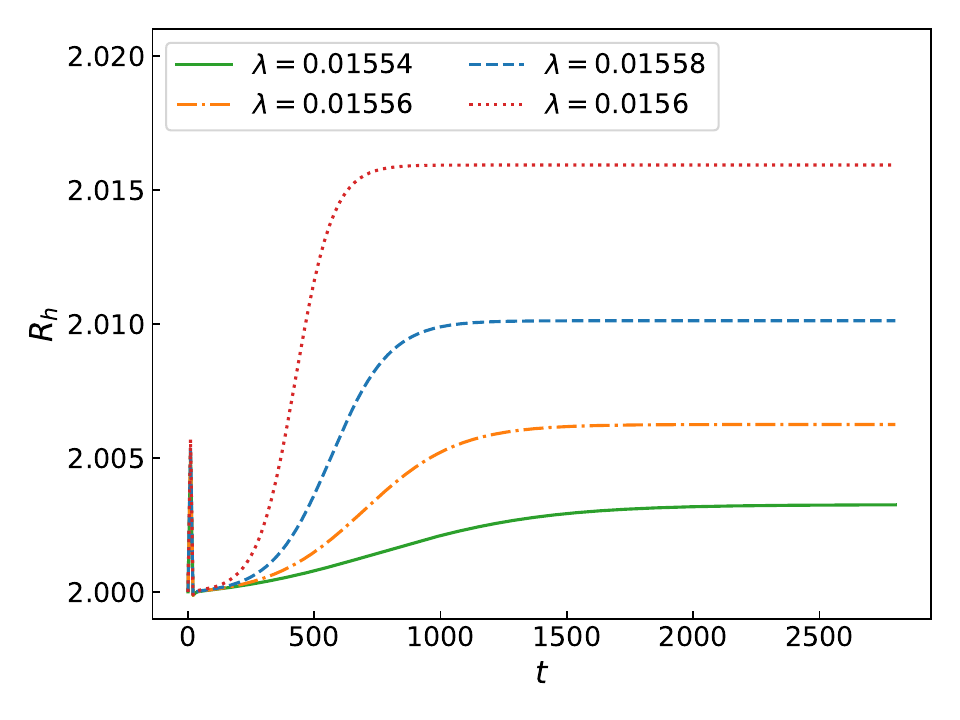}
	\caption{Evolution of the apparent horizon for a dilatonic perturbation on an initially static BH in the lower branch, showing the dynamics of the transition from the lower (unstable) to the upper (stable) branch.}
	\label{fig:LowerDilatonTransitionDynamics}
\end{figure}

In Fig~\ref{fig:LowerBranchDilatonParameterSpace} we show the evolution of the systems in the $R_h-M_{\rm MS}$ plane. In this case the BHs in the lower branch migrate toward the upper branch, hinting at the instability of the former and stability of the latter at the fully nonlinear level.  
In order to show the dynamics of the transition, we plot in Fig.~\ref{fig:LowerDilatonTransitionDynamics} the evolution of the apparent horizon areal radius. After the absorption of the wave packet, $R_h$ increases with time and approaches a constant value, which corresponds to the horizon radius of the final stable BH configuration.

\subsection{Collapse of a phantom field on a dilatonic BH} \label{subsec:phantom}
In the previous section we discussed the evolution of a BH when it absorbs a wave packet of the dilaton. However with this setup we are not able to test the behavior of the system when the BH mass falls below the critical value, since the pulse of the dilaton adds a positive contribution to the total mass and the initial setup is always supercritical.
We now move to investigate the dynamics of dilatonic BHs under a mass loss due to absorption of the phantom field, i.e. a scalar field whose kinetic term has the opposite, ``wrong'' sign.
We stress that the role of the phantom field is solely to mimic the mass loss due to BH evaporation at the classical level, but after the absorbption of the initial perturbation the evolution is governed only by the nonlinear dynamics of the theory, and the Hawking radiation is not taken into account anymore during the simulation. This allows us to dynamically reduce the BH mass below the critical value, and investigate the intrinsic behavior of the classical theory in this peculiar regime.

One might be concerned by the fact that a phantom field can lead to pathological dynamics, but this is not the case in spherical symmetry.
Indeed, in this case the phantom field does not induce runaway instabilities due to the absence of gravitational-wave emission.
We have checked this point by performing test simulations of the spherical collapse of a phantom field onto a Schwarzschild BH in GR (see Appendix~\ref{app:Tests}). As we are going to discuss, in this case the phantom perturbation is simply absorbed by the BH, which settles down to a stable Schwarzschild solution with a slightly smaller mass (and smaller horizon).
Note that here the second law of BH thermodynamics is violated even in GR, since the phantom field does not satisfy the null energy condition.

We performed different simulations of this process choosing the coupling constant $\lambda=\{0.01543 , 0.01545 , 0.01547 , 0.01549 , 0.01551 \}$, which correspond to $R_{h}/\sqrt{\lambda}=\{16.10,16.09,16.08,16.07,16.06\}$. The parameters of the initial phantom perturbation (see Eq.~\eqref{eq:InitialXi}) are 
\begin{equation}
	A_{0, \xi} = 0.01, \qquad R_{0, \xi} = 15, \qquad \sigma_\xi = 2.5.
	\label{eq:FixedAVaryingLParameters}
\end{equation}
The initial BH is always in the upper branch, and when $\lambda = 0.01551$ the total Misner-Sharp mass at the beginning of the simulation is slightly \emph{smaller} than the critical mass. The outer boundary is at $R_\infty = 2850$, the grid step is $\Delta r = 0.02$, and the final time of integration is $T = 2800$. The CFL factor is again set to $0.025$.

\begin{figure}
	\centering
	\includegraphics[width = \columnwidth]{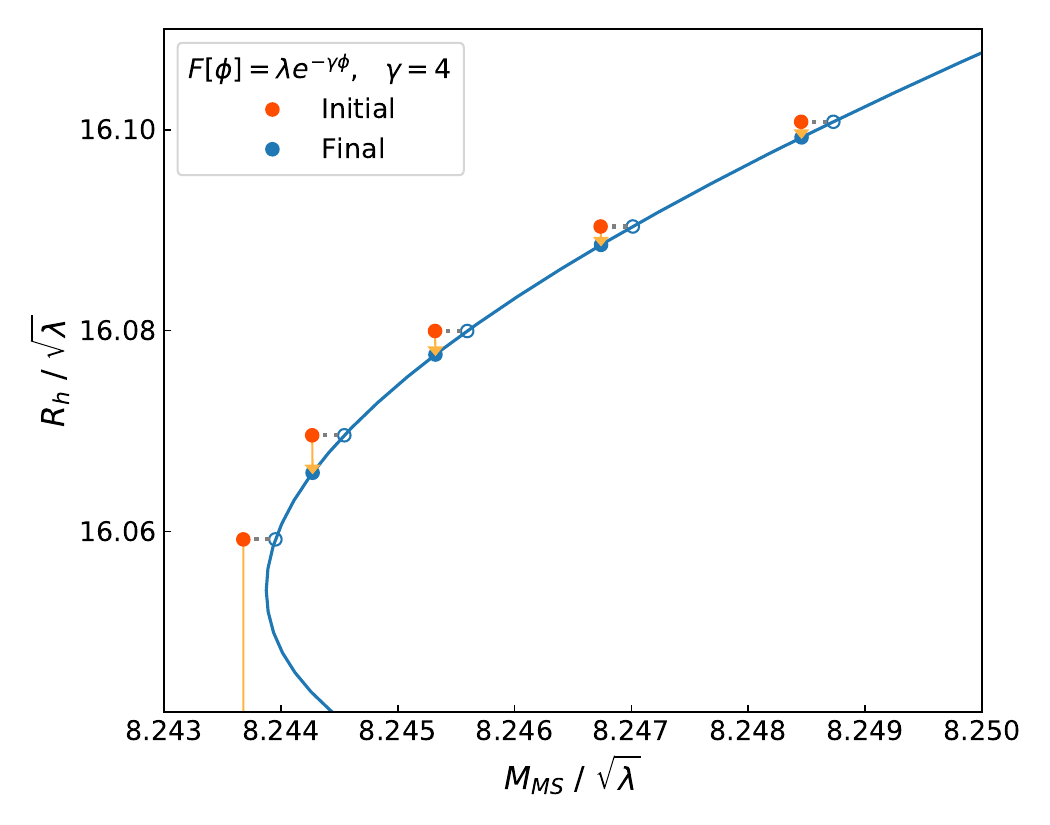}
	\caption{Collapse of a wave packet of the phantom field on different dilatonic BH configurations in the upper branch. The BH reaches a final stable configuration as long as the total mass in the spacetime at $t = 0$ is above the critical value $M_{\rm crit}\sim 8.244\sqrt{\lambda}$. Instead, when $\lambda = 0.01551$ the apparent horizon shrinks significantly and the excised region expands, until it emerges out of the apparent horizon and the simulation is stopped, see Fig.~\ref{fig:UpperGhostSubcriticalHorizon}.
	}
	\label{fig:UpperBranchGhostLAMBDA}
\end{figure}
The results of the simulations are shown in Fig.~\ref{fig:UpperBranchGhostLAMBDA}, in which we can see that the BH reaches a final stable configuration as long as the total mass in the spacetime at $t = 0$ is larger than the critical value. 
For $\lambda = 0.01551$ the situation changes dramatically. In this case the apparent horizon shrinks significantly until it crosses the excision boundary and the simulation is stopped.

\begin{figure}
	\centering
	\includegraphics[width = \columnwidth]{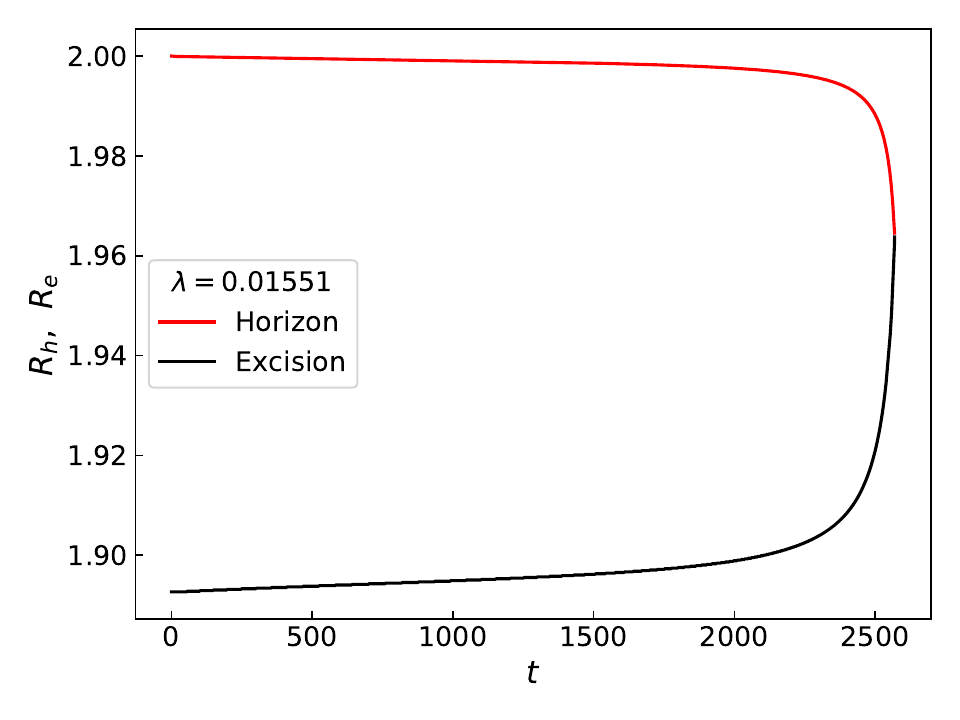}
	\caption{Evolution of the apparent horizon and excision boundary for the accretion of a phantom wave packet on a dilatonic BH in the subcritical case. After the initial absorption of the wave packet, on a much longer time scale the apparent horizon shrinks and the excised region expands, until they cross each other. When this happens the simulation is stopped, due to the presence of elliptic regions outside the BH.
	}
	\label{fig:UpperGhostSubcriticalHorizon}
\end{figure}

In this specific case we have repeated the numerical integration at different resolutions: $\Delta r = \{0.01 , 0.005 , 0.0025\}$, see Sec.~\ref{subsec:naked}.
In Fig.~\ref{fig:UpperGhostSubcriticalHorizon} we show the dynamics of the apparent horizon and of the excision boundary using the highest resolution. During the last stages of the simulation, the horizon shrinks increasingly fast\footnote{Note that the small phantom field is accreted in $\approx 10$ (in our units). Therefore, as discussed in more detail below, the dramatic shrink shown in Fig.~\ref{fig:UpperGhostSubcriticalHorizon} at much later times is entirely due to the intrinsic (nonperturbative but classical) dynamics of the theory past criticality, regardless of the details of the phantom-field accretion.}, and at the same time, the excised region expands at a similar pace. Eventually, they cross each other, and the simulation stops.

\subsection{Naked singularity formation in EdGB gravity?}  \label{subsec:naked}
\begin{figure*}[th]
	\centering
	\includegraphics[width = 0.32\textwidth]{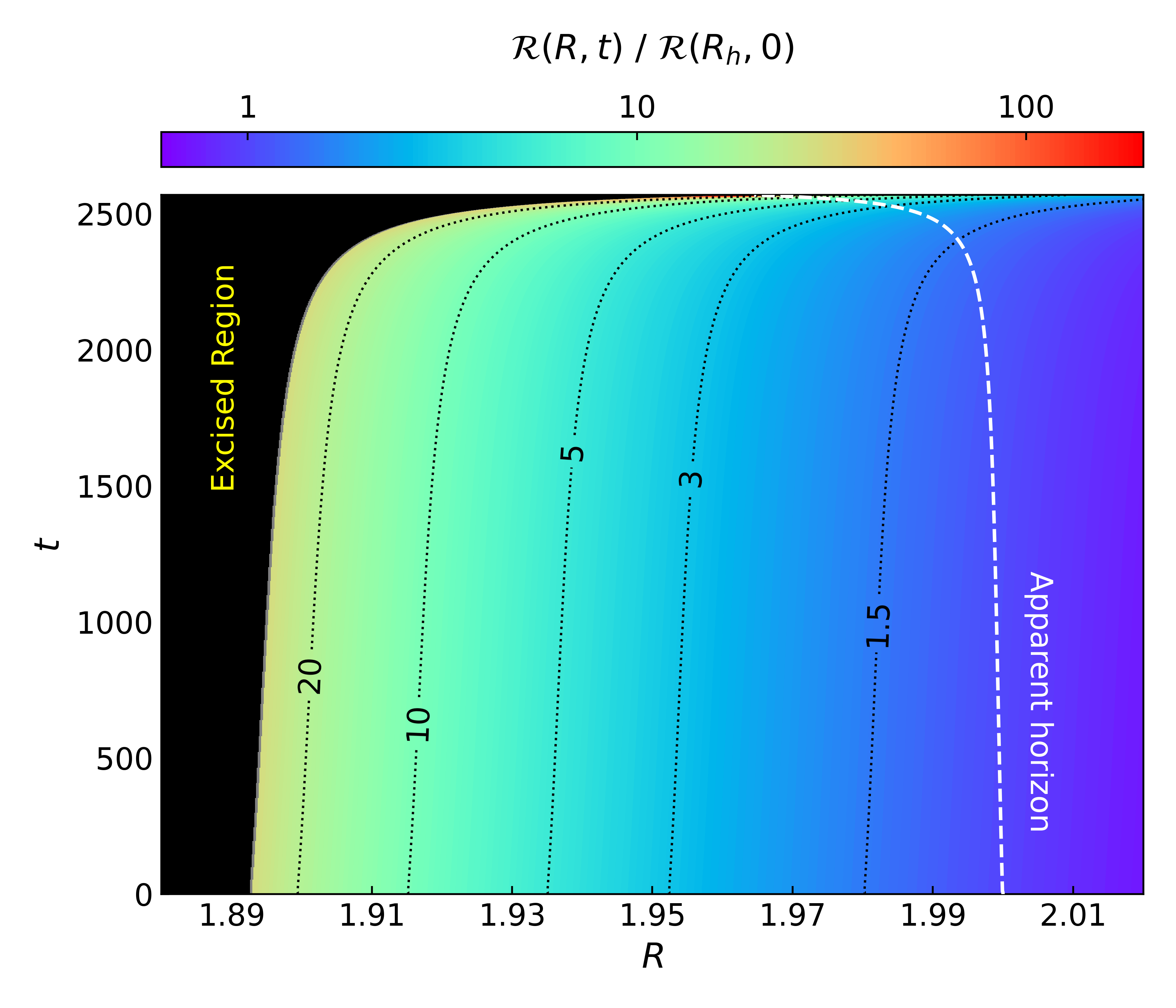}
	\includegraphics[width = 0.32\textwidth]{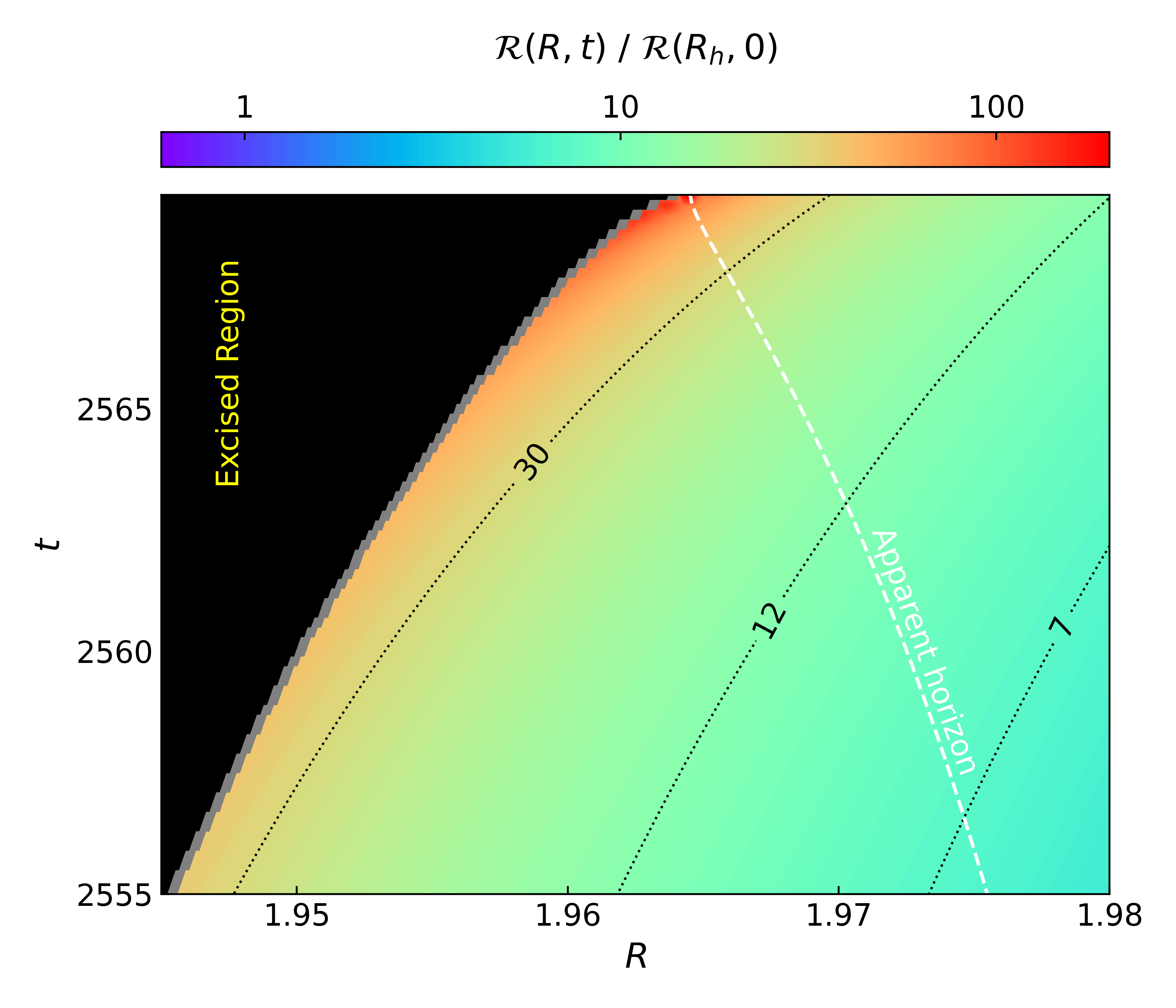}
	\includegraphics[width = 0.32\textwidth]{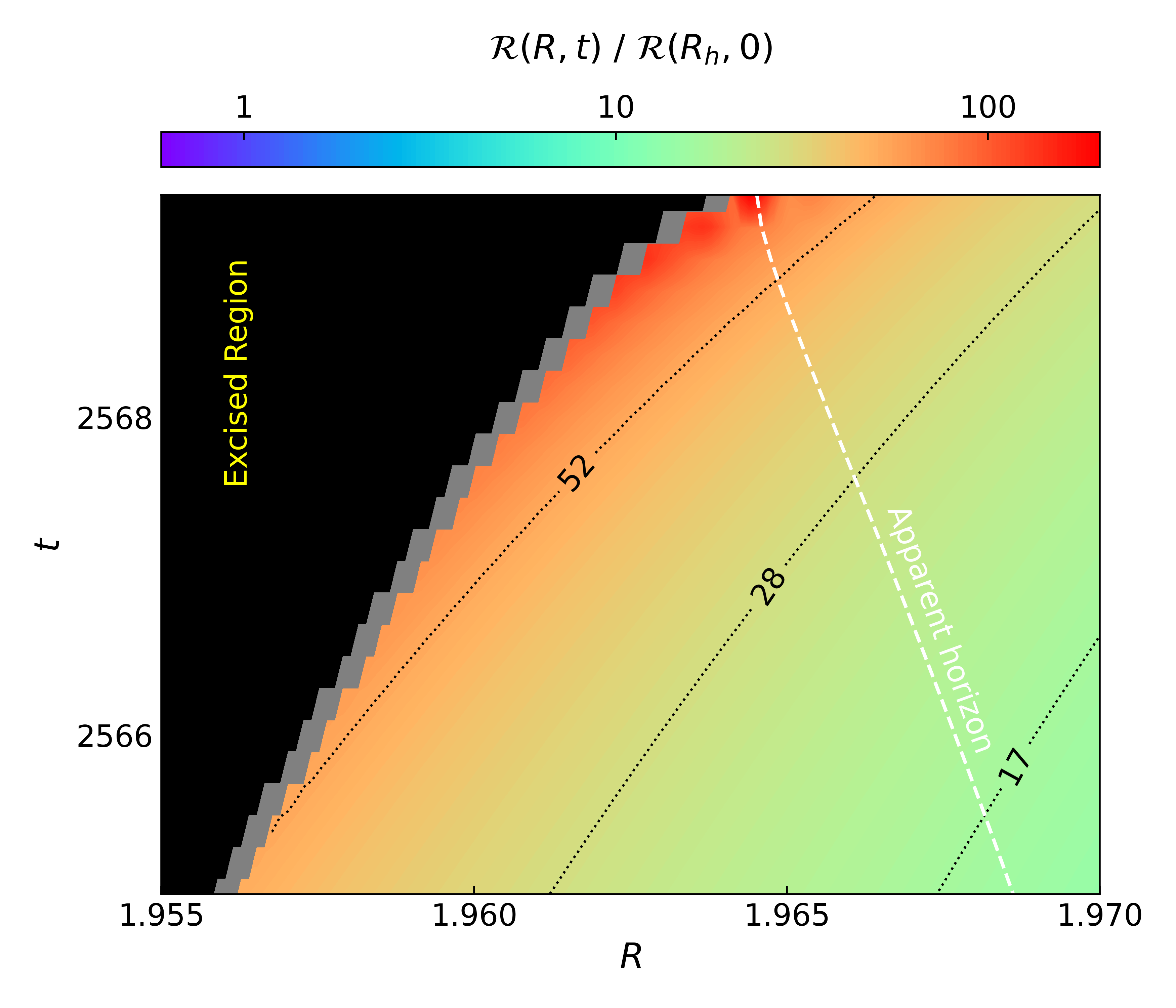}
	\caption{
	Contour plot of the Ricci scalar near the BH region for the simulation that passes the minimum BH mass. The black region is the excised region, and the gray region has been excluded from the computation to avoid inconsistencies due to the change of derivation and dissipation schemes near the excision boundary. The level curves of the $\RR$ follow the behavior of the excision boundary, and the region of high curvature expands. In the left panel we show the full evolution, while in the middle and right panels we focus on the region where the apparent horizon is about to cross the gray area. 
	}
	\label{fig:RicciColorPlot}
\end{figure*}

Since in the final time steps of the mass-loss evolution past the critical mass the apparent horizon is rapidly shrinking, it is interesting to understand whether it crosses the singularity, thus violating the weak cosmic censorship~\cite{1969NCimR...1..252P}.
Furthermore, as we have previously discussed, in the static case the curvature singularity is always inside the elliptic region, and thus it is natural to ask whether the expansion of the elliptic region\footnote{The elliptic region is always inside the excised region, and since the excised region cannot shrink, we do not know the real dynamics of the elliptic region. However, the evolution of the excision boundary is governed by the discriminants of the characteristic equation; therefore, if the radius of this boundary increases, then also the elliptic region is expanding.} is related to the curvature singularity moving outward. 

To address this point, in Fig.~\ref{fig:RicciColorPlot} we show the spacetime evolution of the Ricci scalar $\RR$ in this simulation. The black area is the excised region, while the gray area contains the first 3 grid points in the hyperbolic region. We decided to exclude this region from the computation of $\RR$ in order to avoid possible inconsistencies due to the change of the derivation and dissipation schemes.

\begin{figure}[th]
	\centering
	\includegraphics[width = 0.495\textwidth]{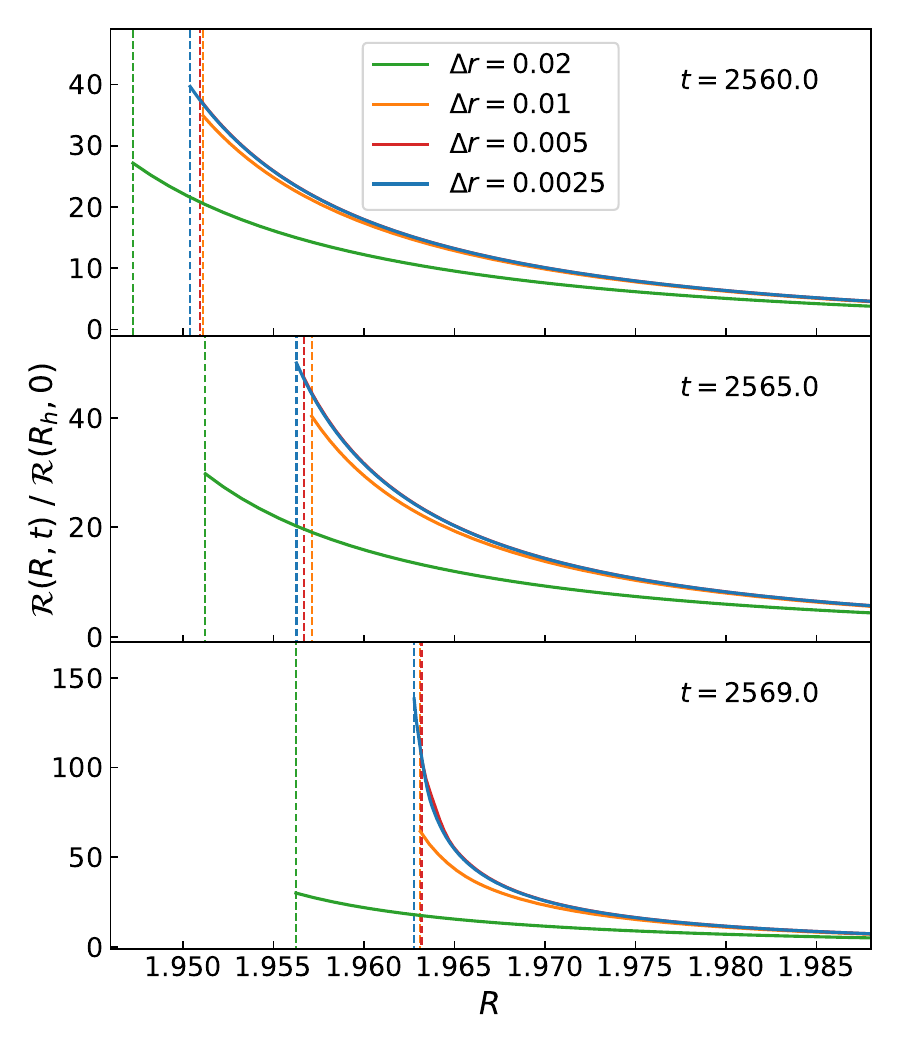}
	\caption{
	Radial profile of the Ricci curvature for the same simulation shown in Fig.~\ref{fig:RicciColorPlot} using different resolutions. Each panel shows a different time snapshot very close to the end of the simulation (as a reference, the apparent horizon has crossed the gray region in Fig.~\ref{fig:RicciColorPlot} at $t\approx2569.6$). Dashed vertical lines denote the outer boundary of the gray region in Fig.~\ref{fig:RicciColorPlot}, so the elliptic region starts close on their left.  Overall, as the apparent horizon approaches the elliptic region an increasingly higher resolution is required to make the curvature converge. Furthermore, the curvature dramatically grows before the simulation stops.
	}
	\label{fig:Ricciconvergence}
\end{figure}

The curvature at the horizon is modest at the beginning of the simulation ($\RR(r_h,t=0)\approx 0.0089 $). However, by the time the apparent horizon crosses the excision (in fact, already when it crosses the gray area in Fig.~\ref{fig:RicciColorPlot}), the Ricci scalar at the apparent horizon has grown by a factor $\approx 58$ compared to its initial value. 
Furthermore, we have performed this simulation with different spatial resolutions ($\Delta r = \{0.02, 0.01 , 0.005 , 0.0025\}$), finding that the curvature converges well until $t=2569.0$.
This is shown in Fig.~\ref{fig:Ricciconvergence}, in which we present the radial profile of the Ricci scalar at different time snapshots and for different resolutions. As a reference, at $t \approx 2569.6$ the apparent horizon has crossed the excision boundary, i.e. only $0.6$ after the last snapshot of the bottom panel\footnote{As a further check of our code, we have computed the Ricci scalar ${\cal R}$ by replacing the field equations in its definition both at the analytical and numerical level. The two computations give the same result.}.
Although the simulation becomes increasingly more demanding, our results suggest that the curvature when the apparent horizon crosses the gray region keeps growing as the grid step decreases.
This suggests that a large curvature region located just across the excision is emerging out of the apparent horizon.

\begin{figure}[th]
	\centering
	\includegraphics[width = 0.495\textwidth]{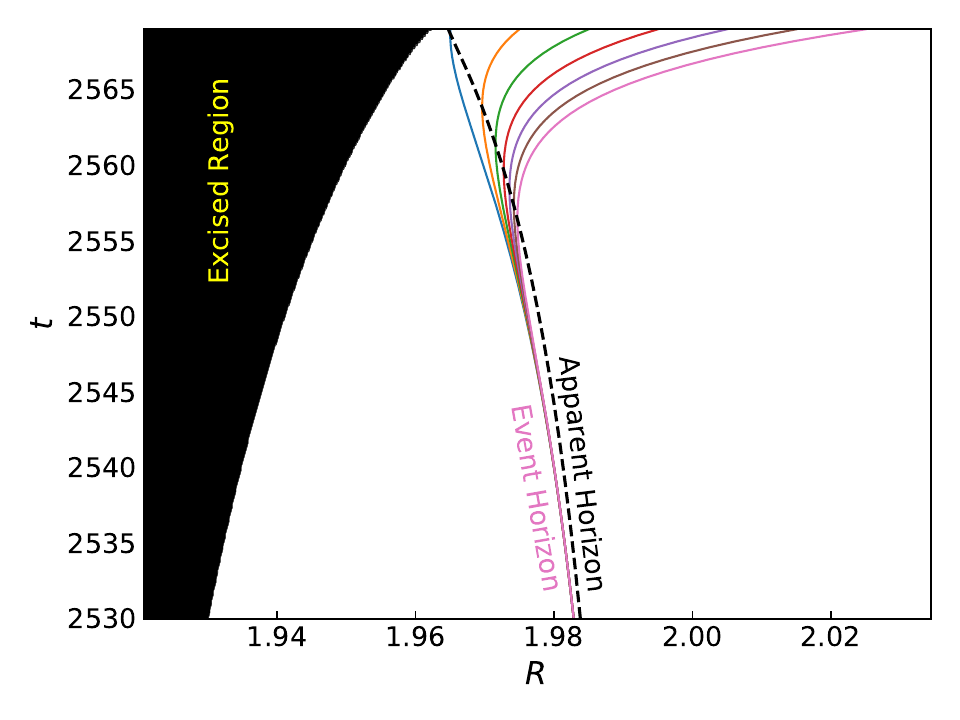}
	\includegraphics[width = 0.495\textwidth]{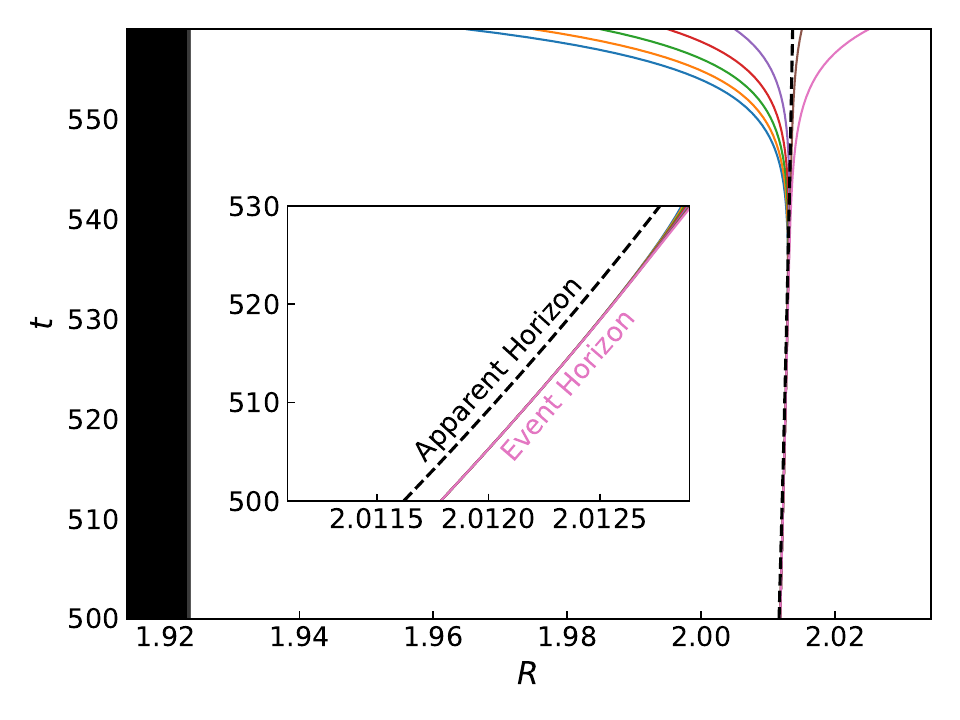}
	\caption{
	Upper panel: ray tracing for the simulation showed in Fig.~\ref{fig:RicciColorPlot} to find the event horizon, which corresponds to the surface where the geodesics converge. The event horizon tracks the apparent horizon and shrinks in time toward higher-curvature regions. The fact that the event horizon is within the apparent horizon is a feature of EdGB gravity but is not generic. This is shown in the lower panel for a transition from the unstable to the stable branch (rightmost simulation in Fig.~\ref{fig:LowerBranchDilatonParameterSpace}). In this case the event horizon is slightly outside the apparent horizon, as in GR.
	}
	\label{fig:rays}
\end{figure}

An important point is that the apparent horizon is foliation dependent and, in highly dynamical configurations, it does not generically coincides with the \emph{event} horizon. Furthermore, due to the violation of the null energy condition~\cite{Kanti:1995vq} in EdGB gravity the GR theorem~\cite{HawkingEllis} proving that the apparent horizon, if it exists, should always be enclosed by the event horizon does not necessarily apply. To explore the dynamics of the event horizon, we have studied the motion of null geodesics, tracing them backward in time and determining the surface where they converge (see, e.g., Ref.~\cite{Bosch:2017ccw} for a similar computation in a different context). In particular, for a given null tangent vector $n^\mu$, we compute the null geodesic equation by solving $n^\mu n^\nu g_{\mu\nu}=0$. In PG-like coordinates, this translates into
\begin{equation}
	\frac{d \, r(t)}{d \, t} =-\frac{\alpha(t,r)}{R'(r)}(\zeta(t,r) - 1) \,, \label{raytracing}
\end{equation} 
for outgoing rays described by the radial coordinate $r=r(t)$. We solve this equation backward in time with initial condition $r(t_F)=r_F$ where $t_F$ is near the final time of our simulation (which does not necessarily correspond to a stationary configuration) and $r_F$ is a free parameter. The result is presented in the upper panel of Fig.~\ref{fig:rays}. This shows two interesting features: i) in the last stages of the simulation the event horizon is \emph{inside} the apparent horizon; this effect is forbidden in GR and it is due to the GB coupling;\footnote{Note that the phantom field is tiny at late times, since it is initially already small and soon gets absorbed by the BH. Thus, the phantom perturbation cannot be responsible for the different dynamics of the horizons at late times.} ii) the event horizon shrinks in time following the same behavior as the apparent horizon, probing regions of increasing curvature.

Intrigued by the fact that the event horizon is located inside the apparent horizon, we have performed ray tracing also in other configurations. First of all, already for the same aforementioned simulation we note that the event horizon and the apparent horizon coincide at times earlier than those shown in the upper panel of Fig.~\ref{fig:rays}. This is because the dynamics is initially slow. Furthermore, when the dynamics is less extreme, the behavior of the event horizon is more similar to what is expected in GR. This is shown in the lower panel of Fig.~\ref{fig:rays}, in which we present the ray tracing for a transition from an unstable BH in the lower branch to a stable BH in the upper branch (rightmost simulation in Fig.~\ref{fig:LowerBranchDilatonParameterSpace}). The event horizon approximately tracks the apparent horizon also in this case, but it is (slightly) \emph{outside} of it, as in GR.

Since the curvature singularity is always located inside the excised region, our simulations cannot access the region where $\RR$ actually diverges\footnote{Note that for the minimum-mass solution the curvature singularity is initially already very close to the outer boundary of the elliptic region, see Fig.~\ref{fig:StaticCriticalExcisionSingularity}, so the high-curvature region is just across the boundary of the elliptic region.}. Nonetheless, it is important to note that the level curves in Fig.~\ref{fig:RicciColorPlot} follow the trajectory of the excision boundary, suggesting  that also the radius of the curvature singularity increases during the evolution.
Although our formalism is limited, these results might suggest that a naked singularity can form as the outcome of BH evaporation in EdGB gravity. We will come back to this point in the concluding discussion in Sec.~\ref{sec:conclusion}.

\begin{figure*}
	\centering
	\includegraphics[width = \columnwidth]{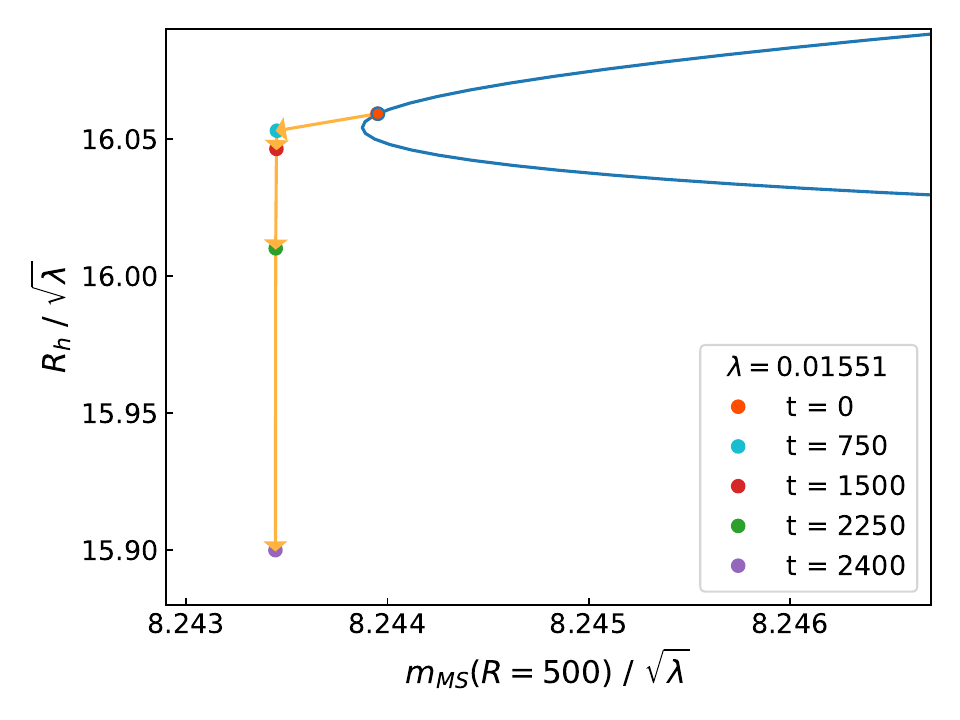}
	\includegraphics[width = \columnwidth]{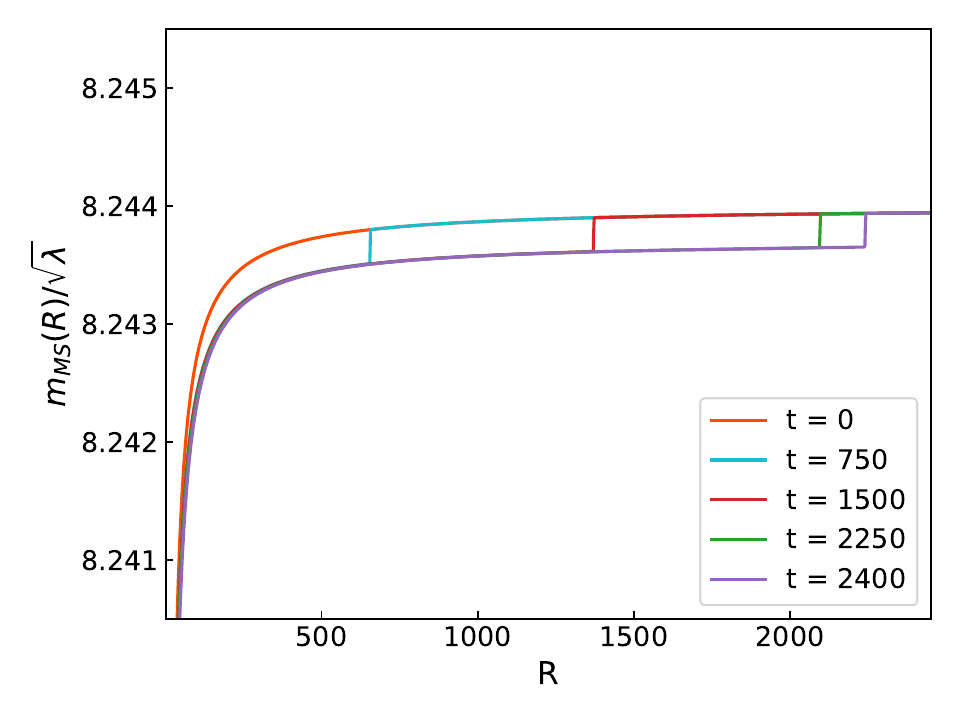}
	\caption{
	Left panel: analog of Fig.~\ref{fig:UpperBranchGhostLAMBDA} for a simulation of a pair of negative- and positive-energy wave packets onto a dilatonic BH near the critical mass. As in the leftmost evolution shown in Fig.~\ref{fig:UpperBranchGhostLAMBDA}, the BH mass decreases past criticality upon accreting the phantom perturbation, triggering a runaway instability on much longer time scales.
	Right panel: some time snapshots of the Misner-Sharp mass function, $m_{\rm MS}(R)$, for the same simulation. While the Misner-Sharp mass decreases near the BH due to the accretion of the phantom field, as the ordinary field $\chi$ moves outward it gives an outgoing positive contribution to the mass function.
	}
	\label{fig:pair}
\end{figure*}

\begin{figure}
	\centering
	\includegraphics[width = \columnwidth]{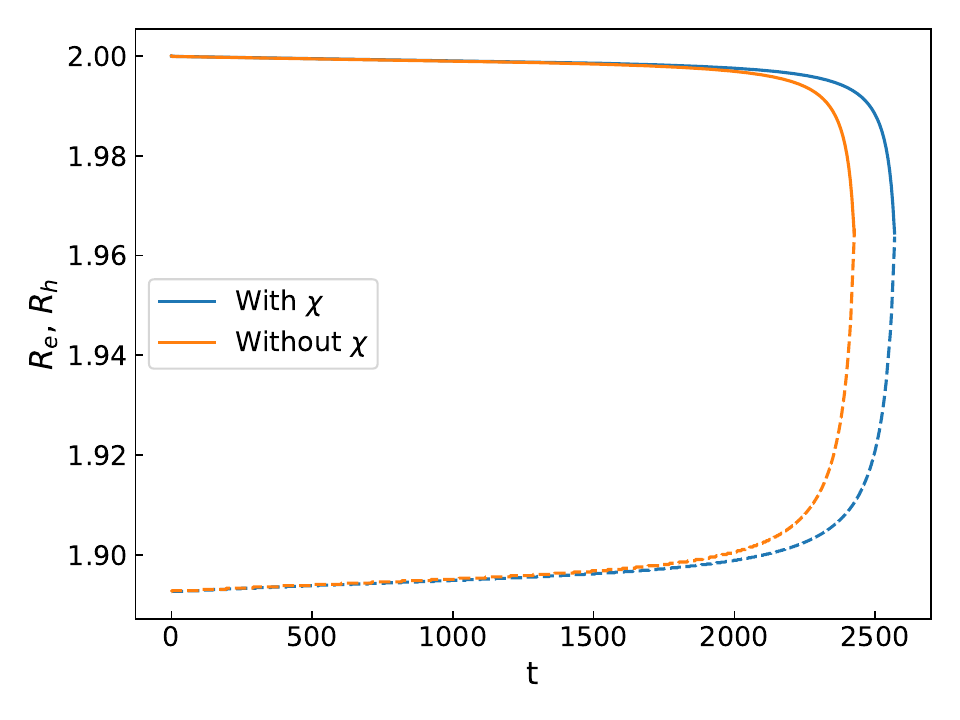}
	\caption{Analog of Fig.~\ref{fig:UpperGhostSubcriticalHorizon2} but for the case of a pair of phantom field $\xi$ and ordinary scalar field $\chi$. Solid and dashed curves correspond to the apparent horizons and excision boundaries, respectively. Regardless of the presence of $\chi$ and of the details of the initial phantom field, the dynamics is very similar and, on long time scales, leads to a shrink of the apparent horizon and to the formation of a naked elliptic region.
	}
	\label{fig:UpperGhostSubcriticalHorizon2}
\end{figure}

\subsection{Emulating Hawking pair production: negative- and positive-energy wave packets emitted near a dilatonic BH}  \label{subsec:pair}
So far we have emulated BH mass loss through the accretion of a phantom perturbation. This was a trick to mimic one of the salient features of Hawking evaporation at the classical level.
However, Hawking emission can be roughly interpreted as pair creation of entangled particles
near the horizon~\cite{Almheiri:2020cfm}, with one (``positive-energy'') particle escaping to infinity and the other (``negative-energy'') particle falling inside the BH and decreasing its mass. 
In order to emulate Hawking pair production more closely, in this section we consider an extended setup in which we evolve \emph{two} wave packets initially located near the horizon of a dilatonic BH. In particular, besides ``vacuum'' EdGB gravity, the matter content of the model is described by the action
%
\begin{equation}
	S_{\rm matter} = \frac{1}{16\pi} \gint \left( \bigl( \nabla \xi \bigr)^2-\bigl( \nabla \chi \bigr)^2\right)\,,
	\label{eq:MatterAction}
\end{equation}
where $\xi$ is again the phantom field (that will emulate the negative-energy Hawking quantum), while $\chi$ is a new minimally-coupled scalar field that will emulate the positive-energy Hawking quantum.

For concreteness, we shall present the simulation of a dilatonic BH near the critical configuration to which we add two Gaussian perturbations. For $\xi$ we have used the profile in Eq.~\eqref{eq:InitialXi}, while we initialized $\chi$ with the profile
\begin{align}
\chi(r, t=0) &= \delta \chi(r) = \frac{A_{0, \chi}}{R(r)} e^{-\frac{(R(r) - R_{0, \chi})^2}{\sigma_\chi^2}}, \notag \\
Y(r, t=0) &= \delta Y(r) = \rder \delta \chi(r), \notag \\
H(r, t=0) &= \delta H(r) = - \frac{\delta \chi(r)}{R(r)} - \partial_R \delta \chi(r) \notag \\
&= - \frac{\delta \chi(r)}{R(r)} - \frac{1}{R'} \delta Y(r),
\label{eq:InitialChi}
\end{align}
where $Y := \rder \chi$ and $H := \frac{1}{\alpha} \tder \chi - \frac{\zeta \, Y}{R'(r)}$ is the conjugate momentum of the scalar field $\chi$. With these choices the initial perturbation of the phantom field $\xi$ is (approximately) ingoing whereas the initial perturbation of the ordinary field $\chi$ is (approximately) outgoing. 

The parameters of the profiles \eqref{eq:InitialXi} and \eqref{eq:InitialChi} are set to
\begin{eqnarray}
A_\xi &= 8\times 10^{-4}\,,\quad R_{0, \xi} = 2.1\,,\quad \sigma_\xi = 0.02\,,\\
A_\chi &= 7\times 10^{-3} \,,\quad R_{0, \chi} = 2.1\,,\quad \sigma_\chi = 0.02\,.
\end{eqnarray} 
In this way the pulses are generated inside the BH photon-sphere (located at $R\approx 3.05$ for an almost critical configuration) and close to the horizon (initially located at $R_H = 2$), but the scalar perturbations approximately vanish on it. The amplitudes are chosen in such a way that the total Misner-Sharp mass is approximately the same as the one of the initial BH, but when the phantom field is absorbed the BH mass decreases below the critical value by an amount similar to those of the simulations presented in the previous section.
We also tried different choices for the wave-packet initial location (e.g., inside and outside the photon-sphere) and width, the latter parametrizing the frenquency content and hence --~within the Hawking pair emission analogy~-- the temperature scale of the evaporating BH. We used a grid step $\Delta r = 0.005$ since, as we can see from Fig.~\ref{fig:Ricciconvergence}, this is sufficient to obtain results accurate enough for our purposes. In all cases we obtained the same qualitative features as presented below.
%

Overall, we observe a very similar dynamics as that presented in Sec.~\ref{subsec:phantom} for a single phantom perturbation. As an example, in the left panel of Fig.~\ref{fig:pair} we show the equivalent of Fig.~\ref{fig:UpperBranchGhostLAMBDA} but for this setup with a pair of negative- and positive-energy wave packets. In this case the Misner-Sharp mass shown on the horizontal axis is evaluated at $R=500$ so for $t\gtrsim 500$ it represents the BH mass without the (positive) contribution of the outgoing field $\chi$. 
The behavior is qualitatively the same as previously reported: due to the absorption of the small phantom perturbation, the BH mass immediately goes slightly past criticality, where no static BH solutions exist. On much longer time scales, the horizon starts shrinking. The behavior of the Misner-Sharp mass function, $m_{\rm MS}(R)$, at different time snapshopts is shown in the right panel of Fig.~\ref{fig:pair}, from which it is evident that the BH mass shrinks upon accreting the phantom field $\xi$, whereas the (positive-energy) contribution of the ordinary field $\chi$ moves outward as this wave packet reaches infinity.

To further support the generality of this dynamics, in Fig.~\ref{fig:UpperGhostSubcriticalHorizon2} we compare the dynamics of the apparent horizon and excision boundary for two simulations with and without the initial perturbation of the ordinary field $\chi$, showing that the qualitative behavior already presented in Fig.~\ref{fig:UpperGhostSubcriticalHorizon} --~in particular the formation of a naked elliptic region~-- is the same. This is expected since, as discussed above, there exists a hierarchy of scales between the accretion of the phantom field (reducing the BH mass past criticality) and the formation of a naked elliptic region. The latter occurs when the small phantom field perturbation has been already accreted and cannot play any role in the late-time dynamics. Indeed, the shrinking of the horizon and the appearance of a naked elliptic region are entirely due to the intrinsic, nonperturbative, dynamics of the theory triggered by going past the critical BH solution.

\begin{figure*}[th]
	\centering
	\includegraphics[width = 0.32\textwidth]{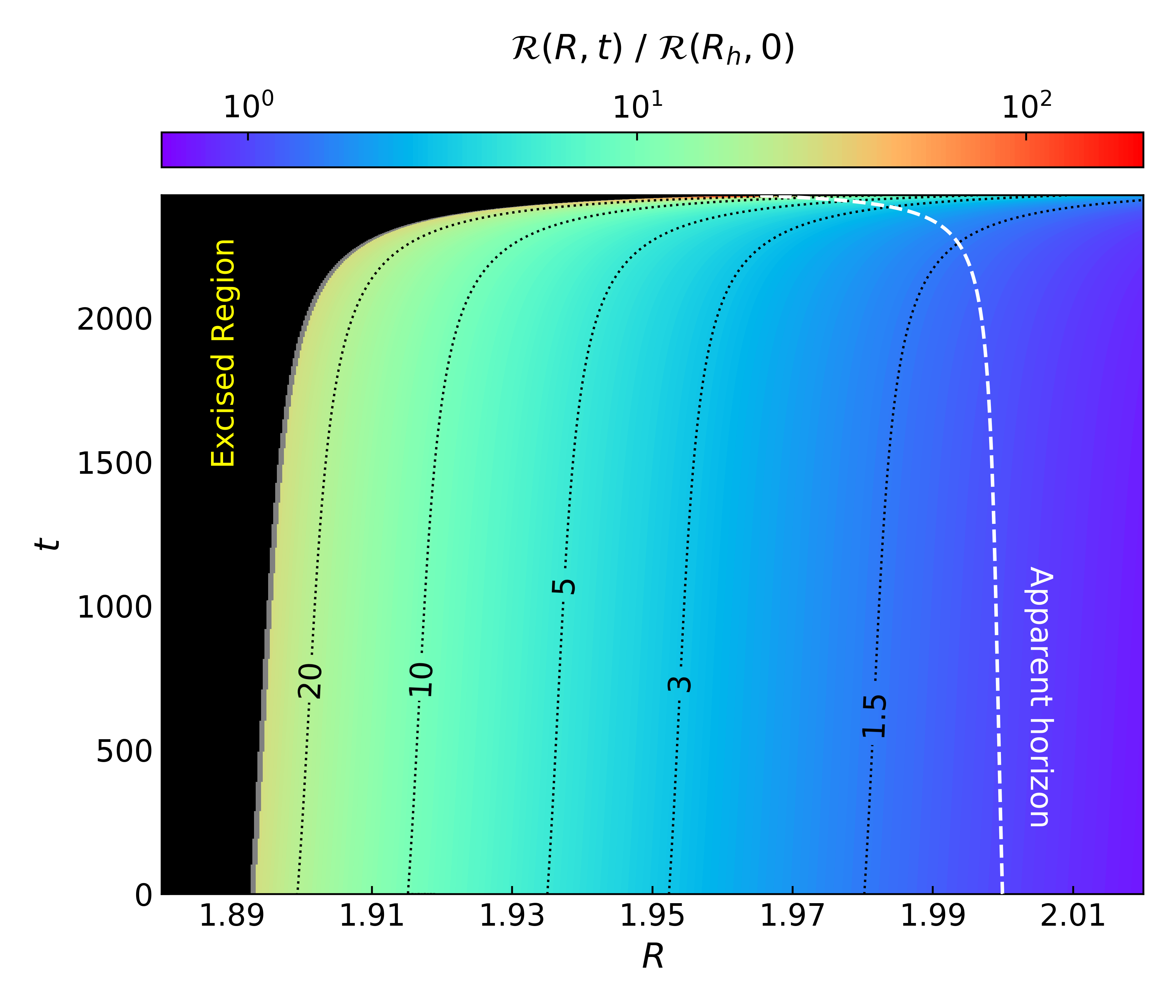}
	\includegraphics[width = 0.32\textwidth]{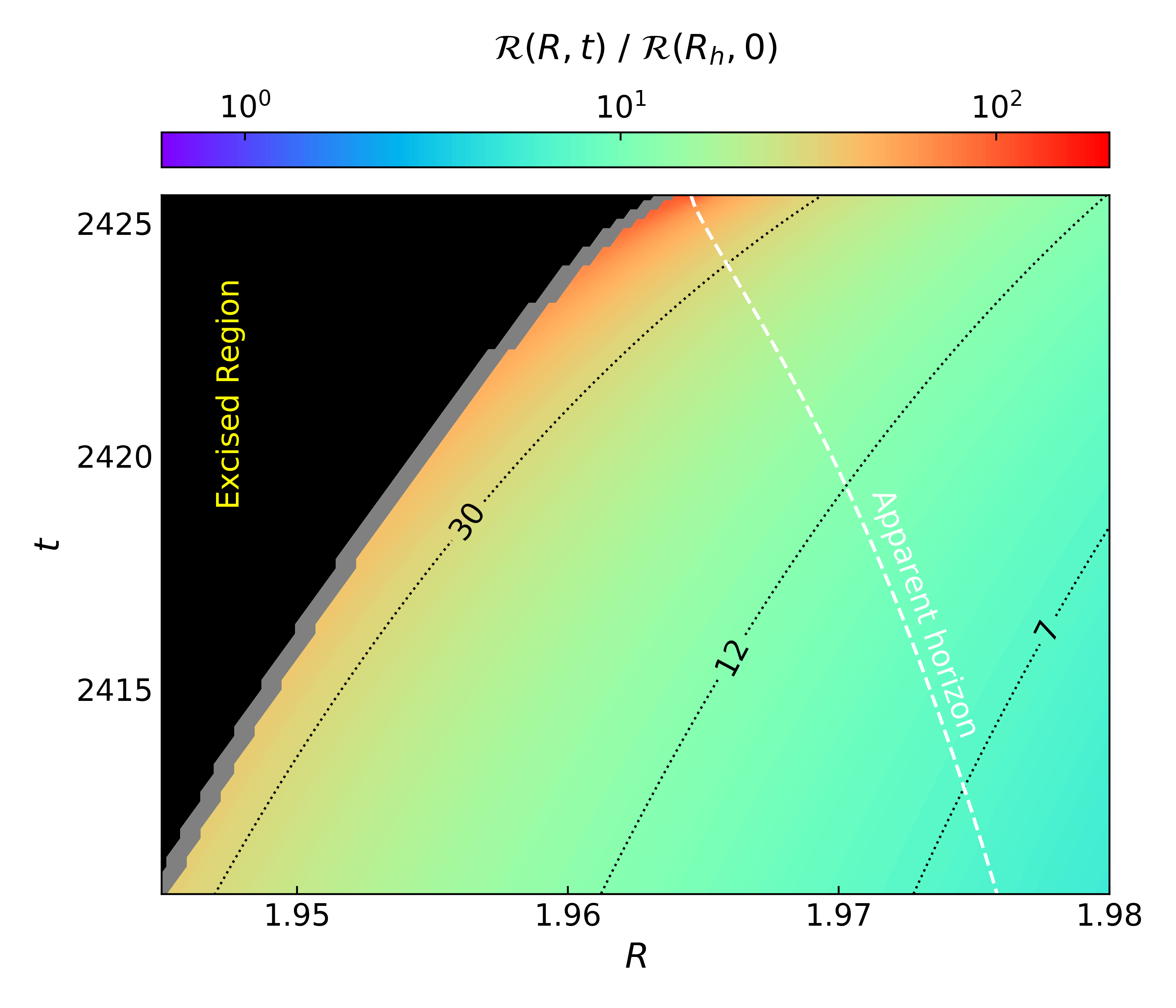}
	\includegraphics[width = 0.32\textwidth]{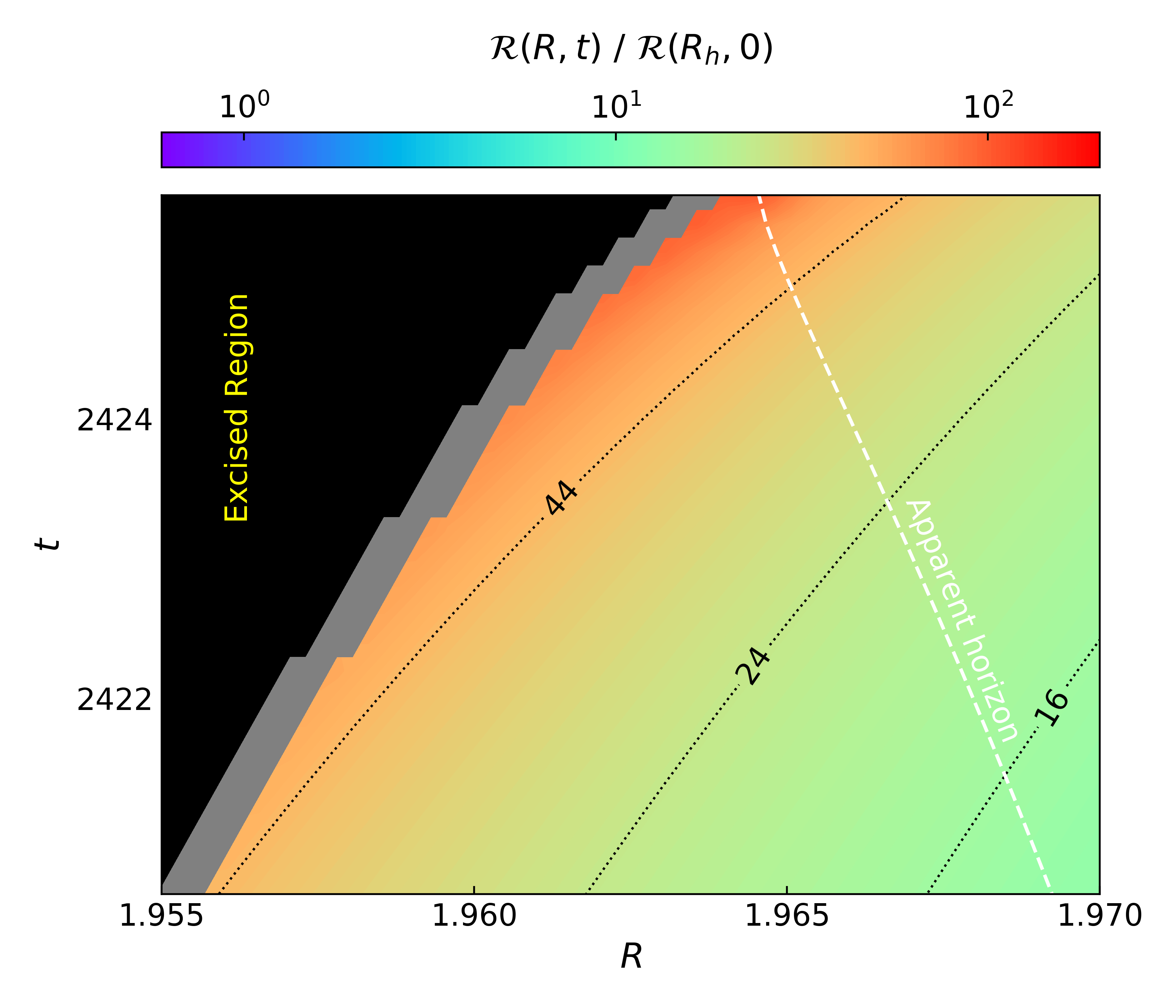}
	\caption{
	Analog of Fig.~\ref{fig:RicciColorPlot} but for an ingoing phantom perturbation and an outgoing standard field perturbation both starting near the horizon (which would more closely mimic the production of a Hawking quantum pair), see main text for details. 
	}
	\label{fig:RicciColorPlot2}
\end{figure*}

Finally, in Fig.~\ref{fig:RicciColorPlot2} we show the analog of Fig.~\ref{fig:RicciColorPlot} in this setup with a pair of negative- and positive-energy wave packets. The striking similarity between Figs.~\ref{fig:RicciColorPlot} and~\ref{fig:RicciColorPlot2} confirms that the late-time dynamics does not depend on the details of the BH mass loss past criticality.

\section{Concluding discussion} \label{sec:conclusion}
In this paper we have performed extensive numerical simulations of the spherical collapse onto dilatonic BHs in EdGB gravity, especially focusing on solutions near the minimum mass that emerges as a very special feature of gravity theories with higher-curvature terms.
We have also offered some broad motivations for this kind of studies, including the enigma related to the fate of Hawking evaporation in this theory.

While the current numerical formalism is insufficient to provide a definite answer to this puzzle, we wish to advance here some speculations supported by our results, and anticipate some interesting future directions.

First of all, the absence of BHs with mass smaller than the critical value makes it almost inevitable for Hawking evaporation in EdGB gravity to either violate the weak cosmic censorship (implying a breakdown of the theory and the need of a full quantum gravity completion) or to form (potentially classical) horizonless remnants.
Exploring the first option might require an evolution scheme (if it exists, see Ref.~\cite{R:2022hlf}) in which the system of equations remains hyperbolic at the singularity. Since the dynamics of the elliptic region depends on the gauge choice~\cite{Ripley:2019hxt,Ripley:2019irj,Bernard:2019fjb,letter}, a putative different formalism might be required to follow the evolution even if the weak cosmic censorship is preserved.
On the other hand, an intriguing result supporting the hypothesis of horizonless remnants is provided by the fact that the critical, minimum-mass solution actually corresponds to one of the double points in the phase space of the theory, wherein a BH and a wormhole solution co-exist.  One could therefore entertain the possibility that the minimum-mass BH can transit toward a (regular and horizonless) wormhole solution with slightly smaller mass, which cannot evaporate any further (see also~\cite{Alexeyev:2002tg,Alexeyev_model} for a model in which Hawking evaporation is halted). Here one interesting aspect for future investigation is the fact that the wormhole solution has matter at the throat~\cite{Kanti:2011jz,Kanti:2011yv}, whereas the BH is a vacuum solution of EdGB theory. However, we note that such transition requires mass loss and can therefore be triggered only by Hawking evaporation. It would be interesting to explore if Hawking particles around the minimum-mass BH can provide the correct matter content to support the wormhole throat.
The possibility of a transition toward the soliton seems more unlikely, given the fact that this solution connects to the unstable BH branch and has a singularity in the second radial derivatives of the scalar field~\cite{Kleihaus:2019rbg,Kleihaus:2020qwo}. However, given the special nature of this singularity, forming these solitons dynamically should be studied in more details.

Other possible outcomes of the evaporation might be simple dispersion of the fields or BH fragmentation, as argued in Ref.~\cite{Ahn:2014fwa} using thermodynamical arguments.
However, complete dispersion would require the disappearance of the apparent horizon, which seems incompatible with the existence of an elliptic region in the BH interior even at $t\sim0$. The only option here would be if the elliptic region shrinks together with the horizon but: i) this is the opposite to what our simulations show (the elliptic region actually expands), and ii) our formalism could not capture a shrinking of the elliptic region even in the case this occurs.
Concerning fragmentation~\cite{Ahn:2014fwa}, this is at least not an option in the spherically symmetric case discussed here, and should anyway occur above the minimum mass to allow for the fragmented BHs to exist.

Overall, beside formation of a naked singularity, it seems that, any other less pathological outcome would require a change of topology of the spacetime. An intriguing extension of our work concerns how to implement this dynamically in a consistent framework or at least to understand if the pathologies that dynamically emerge in this theory could be related to a change of topology.
It might also be interesting to revisit the causal structure of the theory (e.g.,~\cite{Izumi:2014loa,Reall:2014pwa}) in the regime we have identified.

Although we are admittedly providing more questions than answers, we hope that this intriguing problem will motivate further studies in several directions.
The possibility of forming horizonless remnants is particularly appealing, since these objects evade all the constraints on microscopic BHs~\cite{Carr:2020gox} which arise from Hawking evaporation and could form the entirety of the dark matter.
Indeed, the expectation that primordial BHs formed in the early Universe with masses below $M\sim 10^{15}\,{\rm g}$ should be completely evaporated by the present epoch and cannot therefore contribute to the dark matter is based on the assumption that GR is valid all the way down to full evaporation, which is most likely not the case. On the contrary, higher-curvature terms are bound to become dominant in the final stage of the evaporation. As we have discussed, in EdGB gravity this occurs at the length scale $M_{\rm crit}\propto\sqrt{\lambda}$ which might be much larger than the Planck length.
The scenario we have in mind here is a microscopic primordial BH formed in the early Universe with mass much larger than $M_{\rm crit}\propto\sqrt{\lambda}$, so that initially its dynamics is governed by GR. However, during Hawking evaporation the higher-curvature terms become stronger until the BH reaches the dynamical regime that we have explored here at the full nonperturbative level.

Given the nonperturbative nature of this phenomenon, an important extension of our work is to study possible higher-order terms and other corrections in the EdGB action. Some of these terms arise as naturally as the GB coupling in ultraviolet GR completions so they might play an important role in the nonlinear dynamics near the critical length scale.

Finally, our setup might provide a concrete first-principle model to form (stable?) horizonless remnants, which are interesting in the context of the information-loss paradox~\cite{Hawking:1975vcx,Mathur:2009hf,Polchinski:2016hrw}, see~\cite{Chen:2014jwq,Ong:2020xwv} for a review.


\begin{acknowledgments}
We are grateful to Daniela Doneva, Will East, Luis Lehner, Frans Pretorius, Justin Ripley, and Helvi Witek for useful comments on the draft.
We acknowledge financial support provided under the European Union's H2020 ERC, Starting Grant No.~DarkGRA--757480. 
Computations were performed at Sapienza University of Rome on the Vera cluster of the Amaldi Research Center.
This project has received funding from the European Union’s Horizon 2020 research and innovation programme under the Marie Skłodowska-Curie Grant No. 101007855.
We also acknowledge support under the MIUR PRIN and FARE programs (GW-NEXT, 
CUP:~B84I20000100001,  2020KR4KN2) and from the Amaldi Research Center funded by the MIUR program ``Dipartimento di Eccellenza'' (CUP:~B81I18001170001). This work is partially supported by the PRIN Grant No. 2020KR4KN2 ``String Theory as a Bridge between Gauge Theories and Quantum Gravity''.
\end{acknowledgments}

\appendix

\section{Equations} \label{app:Eqs}

In this appendix we provide the system of equations that we integrated numerically, both for the construction of static dilatonic solutions and for the simulations of the spherical collapse.

\subsection{Equations for constructing the static dilatonic BH solutions} \label{subsec:StaticEquations}

\subsubsection{Schwarzschild-like coordinates}

\begin{widetext}
\begin{eqnarray}
        &&\Lambda'\left(1+\frac{4}{\Sch{r}}F'[\phi]\phi'\right)-\frac{1-e^{\Lambda}}{\Sch{r}}-\frac{1}{2}\Sch{r}\phi'^2+\frac{4}{\Sch{r}}F'[\phi]\left[-3e^{-\Lambda}\Lambda'\phi'+2(e^{-\Lambda}-1)\phi''\right]+\frac{8}{\Sch{r}}F''[\phi]\phi'^2\left(e^{-\Lambda}-1\right)=0\,,  \label{eq:eqEinsteinttBHstaticCoordinates}\\
    &&\Gamma'\left(1+\frac{4}{\Sch{r}}F'[\phi]\phi'\right)+\frac{1-e^{\Lambda}}{\Sch{r}}-\frac{1}{2}\Sch{r}\phi'^2-\frac{12}{\Sch{r}}e^{-\Lambda}\Gamma'F'[\phi]\phi'=0\,,
    \label{eq:eqEinsteinrrBHstaticCoordinates}\\
        &&\Gamma''+\Gamma'\left(\frac{1}{\Sch{r}}+\frac{\Gamma'-\Lambda'}{2}\right)-\frac{\Lambda'}{\Sch{r}}+\phi'^2+\frac{4}{\Sch{r}}F'[\phi]\left[\Gamma'\phi'e^{-\Lambda}\left(3\Lambda'-\Gamma'\right)-2e^{-\Lambda}\left(\phi'\Gamma''+\Gamma'\phi''\right)\right]-\frac{8}{\Sch{r}}e^{-\Lambda}F''[\phi]\phi'^2\Gamma'=0\,,\nonumber\\
    \label{eq:eqEinsteinthetathetaBHstaticCoordinates}\\
       && \phi''+\frac{\Gamma'-\Lambda'}{2}\phi'+\frac{2\phi'}{\Sch{r}}+\frac{2}{\Sch{r}^2}F'[\phi]\left[\Gamma'\left(\Gamma'-\Lambda'+3e^{-\Lambda}\Lambda'-e^{-\Lambda}\Gamma'\right)+2\Gamma''\left(1-e^{-\Lambda}\right)\right]=0\,.
    \label{eq:eqDilatonBHstaticCoordinates}
\end{eqnarray}
 \end{widetext}

\subsubsection{PG-like coordinates}

In the static and spherically symmetric case $\phi$, $\alpha$ and $\zeta$ depend only on the coordinate radius $r$, and since we are interested in the dilatonic BH solutions, we set the phantom field to zero, and we consider only the equations for the metric and the dilaton (Eqs.~\eqref{eq:field_grav}-\eqref{eq:field_dilaton}).

After substituting the ansatz for the metric~\eqref{eq:PGLineElementRT} in the field equations, we perform some algebraic manipulation using Wolfram \textsc{Mathematica} obtaining the following system of ordinary differential equations:
\begin{widetext}
	\begin{align}
		\alpha' &= \frac{\alpha}{4 R' \left(R R'+4 \left(3 \zeta ^2-2\right) \phi ' F'[\phi]\right)}  \biggl\{48 \zeta ^3 R' \zeta ' \phi ' F'[\phi]+4 \zeta  R' \zeta ' \left(R R'-8 \phi ' F'[\phi]\right)+ \notag \\
			&\zeta ^2 \left(R^2 R' \phi '^2+2 R'^3\right)+R^2 R' \phi '^2+16 \zeta ^4 \left(R' \phi '^2 F''[\phi]+F'[\phi] \left(R' \phi ''-R'' \phi '\right)\right)\biggr\}, \label{eq:AlphaPrime} \\
		\zeta' &= \frac{1}{4 \zeta  R' \left(R R'+4 \left(3 \zeta ^2-2\right) \phi ' F'[\phi]\right)} \biggl\{R^2 R' \phi '^2-16 \zeta ^4 \Bigl[R' \phi '^2 F''[\phi]+F'[\phi] \left(R' \phi ''-R'' \phi '\right)\Bigr] \notag \\
			&-\zeta ^2 \left(16 R'' \phi ' F'[\phi]+R' \left(\phi '^2 \left(R^2-16 F''[\phi]\right)-16 \phi '' F'[\phi]\right)+2 R'^3\right)\biggr\}, \label{eq:ZetaPrime} \\
		\phi'' &= -\frac{1}{D_\phi}\biggl\{\left(\zeta ^2-1\right) R'^2 \phi ' \Bigl[\left(\zeta ^2-1\right) F'[\phi] \phi '^3-R' R''\Bigr] R^5+R' \phi ' \Bigl[\left(\zeta ^2-2\right) R'^4 \notag \\
			&-4 \zeta ^2 \left(\zeta ^2-1\right) \phi '^2 F''[\phi] R'^2-4 \left(7 \zeta ^4-13 \zeta ^2+6\right) F'[\phi] \phi ' R'' R'-8 \left(\zeta ^2-1\right)^2 F'[\phi]^2 \phi '^4\Bigr] R^4 \notag \\
			&+4 R' F'[\phi] \phi '^2 \Bigl[\left(12 \zeta ^4-25 \zeta ^2+12\right) R'^3-4 \zeta ^2 \left(3 \zeta ^4-5 \zeta ^2+2\right) \phi '^2 F''[\phi] R' \notag \\
			&-24 \left(\zeta ^2-1\right)^2 \left(3 \zeta ^2-2\right) F'[\phi] \phi ' R''\Bigr] R^3-32 \left(\zeta ^2-1\right) F'[\phi]^2 \phi '^3 \Bigl[6 \left(\zeta ^2-1\right) R' \phi '^2 F''[\phi] \zeta ^4 \notag \\
			&+\left(-21 \zeta ^4+32 \zeta ^2-12\right) R'^3+2 \left(15 \zeta ^6-39 \zeta ^4+32 \zeta ^2-8\right) F'[\phi] \phi ' R''\Bigr] R^2 \notag \\
			&+4 R'^2 F'[\phi] \Bigl[720 F'[\phi]^2 \phi '^4 \zeta ^8-24 \phi ' \left(94 F'[\phi]^2 \phi '^3+R'^2 F''[\phi] \phi '-R' F'[\phi] R''\right) \zeta ^6 \notag \\
			&+\left(2624 F'[\phi]^2 \phi '^4-3 R' \left(R'^3-8 \phi '^2 F''[\phi] R'+8 F'[\phi] \phi ' R''\right)\right) \zeta ^4-1344 F'[\phi]^2 \phi '^4 \zeta ^2 \notag \\
			&+256 F'[\phi]^2 \phi '^4\Bigr] R-96 \zeta ^4 \left(\zeta ^2-1\right) R'^2 F'[\phi]^2 \phi ' \Bigl[R'^3+8 \left(\zeta ^2-1\right) \phi '^2 F''[\phi] R' \notag \\
			&-8 \left(\zeta ^2-1\right) F'[\phi] \phi ' R''\Bigr]\biggr\}, \label{eq:PhiPrimePrime}
	\end{align}
	 where
	\begin{align}
		D_\phi &= \left(\zeta ^2-1\right) R' \Bigl[96 \left(3 \zeta ^4-5 \zeta ^2+2\right) R' F'[\phi]^2 \phi '^2 R^3+64 \left(15 \zeta ^6-39 \zeta ^4+32 \zeta ^2-8\right) F'[\phi]^3 \phi '^3 R^2 \notag \\
			&+R'^3 \left(R^4-96 \zeta ^4 F'[\phi]^2\right) R+4 R'^2 F'[\phi] \left(R^4 \left(7 \zeta ^2-6\right)-192 \zeta ^4 \left(\zeta ^2-1\right) F'[\phi]^2\right) \phi '\Bigr].
		\label{eq:PhiPrimePrimeDenominator}
	\end{align}
\end{widetext}
In these equations $F'[\phi] = \frac{\delta F[\phi]}{\delta \phi}$, while $\zeta'$, $\alpha'$, $\phi'$ and $R'$ are radial derivatives. 

The denominator $D_\phi$ vanishes at the horizon, since $\zeta = 1$. However the field equations are regular when the condition~\eqref{eq:HorizonPhiDer} is imposed. In this case $\zeta_h'$ and $\phi_h''$ are given by 
\begin{widetext}
	\begin{align}
		\phi_h'' &= -\frac{3}{R_h^4 R_h' \left(R_h^4-96 F'[\phi_h]^2\right)+4 R_h^7 \phi_h ' F'[\phi_h]} \Bigl[32 R_h^4 R_h'' \phi_h ' F'[\phi_h]^2+4 R_h^5 R_h' R_h'' F'[\phi_h] \label{eq:HorizonPhiDder} \\
			&+R_h^3 R_h'^2 \phi_h ' \left(4 R_h^2 F''[\phi_h]+R_h^4-32 F'[\phi_h]^2\right)+48 R_h'^3 \left(R_h^2 F'[\phi_h] F''[\phi_h]+4 F'[\phi_h]^3\right)\Bigr], \notag \\
		\zeta_h' &= -\frac{R_h'^2}{2 R_h R_h'+8 \phi_h ' F'[\phi_h]}. \label{eq:HorizonZetaDer}
	\end{align}
\end{widetext}

\subsection{System of equations for the simulation of the spherical collapse} \label{subsec:EvolutionEquations}

We now turn to discuss the system of equations used in the time evolution code. In this case the scalar fields and the metric functions depend on $(r,t)$. 

The evolution equations for the scalar fields have been obtained from the definition of the conjugate momenta (Eq.~\eqref{eq:PPiDef}) and are:
\begin{align}
	\tder \phi &= \alpha P + \frac{\alpha \zeta Q}{R'}, \label{eq:EqPhi} \\
	\tder \xi &= \alpha \Pi + \frac{\alpha \zeta \Theta}{R'}. \label{eq:EqXi} 
\end{align}
Note that we use a prime to indicate differentiation with respect to the single variable of a function, whereas we use $\partial_r$ and $\partial_t$ to denote partial differentiation of spacetime variables.
The evolution equations for $Q$ and $\Theta$ are obtained by performing the time derivative of their definitions, and substituting the radial derivatives of Eqs.~\eqref{eq:EqPhi},\eqref{eq:EqXi} in place of the mixed derivatives of the scalar fields. We get
\begin{align}
	\tder Q &= \rder \Bigl( \alpha P + \frac{\alpha \zeta Q}{R'} \Bigr), \label{eq:EqQ} \\
	\tder \Theta &= \rder \Bigl( \alpha \Pi + \frac{\alpha \zeta \Theta}{R'} \Bigr). \label{eq:EqTheta}
\end{align}
We finally used the field equations to obtain three evolution equations for $P$, $\Pi$, and $\zeta$ and two constraints for $\alpha$ and $\zeta$. The evolution equations are
\begin{widetext}
	\begin{align}
		\tder P &= \frac{1}{D_P} \biggl\{\alpha  R'^2 \Bigl[Q \left(R' (\rder \alpha)-\alpha  R''\right)+R' \Bigl(P \zeta  R' (\rder \alpha)+\alpha  \left((\rder Q)+R' \left(\zeta  (\rder P)+P (\rder \zeta)\right)\right)\Bigr)\Bigr] R^4 \notag \\
			&-2 \alpha  R' \Bigl[2 F'[\phi] \Bigl(\left(\zeta ^2+4\right) R' (\rder \alpha)+\alpha  \left(\zeta  R' (\rder \zeta)-4 R''\right)\Bigr) Q^2+R' \Bigl(20 P \zeta  R' F'[\phi] (\rder \alpha)  \notag \\
			&+\alpha  \bigl(-R'^2+4 F'[\phi] \left(2 \zeta  (\rder P)+3 P (\rder \zeta)\right) R'+8 F'[\phi] \left((\rder Q)-P \zeta  R''\right)\bigr)\Bigr) Q  \notag \\
			&+R' \Bigl(10 P^2 \zeta  F'[\phi] \bigl(\zeta  (\rder \alpha)+\alpha  (\rder \zeta)\bigr) R'^2+P \alpha  \zeta  \bigl(-R'^2+8 \zeta  F'[\phi] (\rder P) R'+8 F'[\phi] (\rder Q)\bigr) R'  \notag \\
			&-2 \bigl(2 \Theta  \Pi  R'+\zeta  \left(\Theta ^2+\Pi ^2 R'^2\right)\bigr) F'[\phi] \left(\zeta  (\rder \alpha)+\alpha  (\rder \zeta)\right)\Bigr)\Bigr] R^3-2 \alpha  F'[\phi] \Bigl[8 \alpha  \zeta ^2 R' F''[\phi] Q^4  \notag \\
			&-8 \Bigl(2 \left(\zeta ^2+2\right) R' F'[\phi] (\rder \alpha)+\alpha  \bigl(F'[\phi] \left(\left(\zeta ^2-4\right) R''+2 \zeta  R' (\rder \zeta)\right)-2 P \zeta  R'^2 F''[\phi]\bigr)\Bigr) Q^3  \notag \\
			&+R' \Bigl(\alpha  \bigl(\bigl(\left(8 \left(P^2-\Pi ^2\right) F''[\phi]-3\right) \zeta ^2+16\bigr) R'^2+8 \bigl(-2 \zeta  \left(\Theta  \Pi  F''[\phi]+2 F'[\phi] (\rder P)\right)  \notag \\
			&-P \left(\zeta ^2+8\right) F'[\phi] (\rder \zeta)\bigr) R'-8 \zeta  \bigl(\zeta  \Theta ^2 F''[\phi]-6 P F'[\phi] R''\bigr)+8 \left(\zeta ^2-4\right) F'[\phi] (\rder Q)\bigr)  \notag \\
			&-8 P \zeta  \left(\zeta ^2+16\right) R' F'[\phi] (\rder \alpha)\Bigr) Q^2+2 \Bigl(P \alpha  \zeta  \bigl(15 R'^2-32 \zeta  F'[\phi] (\rder P) R'-24 F'[\phi] (\rder Q)\bigr) R'^2  \notag \\
			&+4 \bigl(2 \Theta  \Pi  R'+\zeta  \left(\Theta ^2+\Pi ^2 R'^2\right)\bigr) F'[\phi] \left(2 \zeta  R' (\rder \alpha)+\alpha  \left(\zeta  R''+2 R' (\rder \zeta)\right)\right)  \notag \\
			&-4 P^2 \zeta  F'[\phi] \bigl(16 \zeta  R' (\rder \alpha)-3 \alpha  \left(\zeta  R''-4 R' (\rder \zeta)\right)\bigr) R'^2\Bigr) Q+\zeta  R' \Bigl(8 R' (\rder \alpha) \bigl(P \zeta  \bigl(2 \Theta  \Pi  R'  \notag \\
			&+\zeta  \left(\Theta ^2+\left(\Pi ^2-5 P^2\right) R'^2\right)\bigr) F'[\phi]+R' (\rder \zeta)\bigr)+\alpha  \bigl(-32 P^2 \zeta ^2 F'[\phi] (\rder P) R'^3  \notag \\
			&+2 \Theta  \Pi  \bigl(R'^2+8 P F'[\phi] (\rder \zeta) R'-8 F'[\phi] (\rder Q)\bigr) R'+\zeta  \bigl(17 P^2 R'^4-\Pi ^2 R'^4+3 \Theta ^2 R'^2  \notag \\
			&-24 P^2 F'[\phi] (\rder Q) R'^2-8 \Pi ^2 F'[\phi] (\rder Q) R'^2+8 P \bigl(\Theta ^2+\left(\Pi ^2-5 P^2\right) R'^2\bigr) F'[\phi] (\rder \zeta) R'  \notag \\
			&-8 \Theta ^2 F'[\phi] (\rder Q)\bigr)\bigr)\Bigr)\Bigr] R^2-8 R' F'[\phi] \Bigl[4 \alpha ^2 \left(\zeta ^2-4\right) R' F'[\phi] Q^3+4 \alpha  \zeta  R' \bigl(P \alpha  \left(\zeta ^2-12\right) R' F'[\phi]  \notag \\
			&+2 \zeta  F''[\phi] (\rder \alpha)\bigr) Q^2+4 \zeta  \Bigl(2 R' F'[\phi] (\rder \alpha)^2 \zeta ^3+\alpha ^2 R' \bigl(2 F'[\phi] (\rder \zeta)^2+2 P R' F''[\phi] (\rder \zeta)  \notag \\
			&-\left(\Theta ^2+\left(13 P^2-\Pi ^2\right) R'^2\right) F'[\phi]\bigr) \zeta +2 \alpha  (\rder \alpha) \bigl(\zeta  \left(3 P \zeta  R'^2 F''[\phi]-F'[\phi] R''\right)  \notag \\
			&+2 \left(\zeta ^2-2\right) R' F'[\phi] (\rder \zeta)\bigr)\Bigr) Q+\zeta ^2 R' \Bigl(-20 P^3 \alpha ^2 \zeta  F'[\phi] R'^3+16 P^2 \alpha  \zeta  F''[\phi] \left(\zeta  (\rder \alpha)+\alpha  (\rder \zeta)\right) R'^2  \notag \\
			&+4 P F'[\phi] \bigl(\zeta  \left(\left(\Pi ^2 R'^2-\Theta ^2\right) \alpha ^2+2 (\rder \alpha)^2\right)-6 \alpha  (\rder \alpha) (\rder \zeta)\bigr) R'+\alpha  \bigl(\bigl(R' \bigl(\left(2 \zeta ^2-1\right) R'  \notag \\
			&+8 \zeta  \left(3-2 \zeta ^2\right) F'[\phi] (\rder P)\bigr)+8 F'[\phi] (\rder Q)\bigr) (\rder \alpha)+2 \alpha  R' \bigl(\zeta  R'  \notag \\
			&+4 \left(1-2 \zeta ^2\right) F'[\phi] (\rder P)\bigr) (\rder \zeta)\bigr)\Bigr)\Bigr] R+4 \zeta  F'[\phi] \Bigl[\zeta  \Bigl(64 Q R' \left(2 Q+P \zeta  R'\right) F'[\phi]^2 (\rder \zeta)^2  \notag \\
			&+8 F'[\phi] \bigl(-8 \zeta  R' F''[\phi] Q^3+8 \left(2 P F''[\phi] R'^2+\zeta  F'[\phi] R''\right) Q^2+R' \bigl(R' \bigl(5 \zeta  R' \left(8 F''[\phi] P^2+1\right)  \notag \\
			&+16 \left(1-2 \zeta ^2\right) F'[\phi] (\rder P)\bigr)-8 \zeta  F'[\phi] (\rder Q)\bigr) Q+P \zeta  R'^3 \bigl(3 \zeta  R' \left(8 F''[\phi] P^2+1\right)  \notag \\
			&+8 \left(1-3 \zeta ^2\right) F'[\phi] (\rder P)\bigr)\bigr) (\rder \zeta)+\zeta  R' \bigl(8 \zeta  R' F''[\phi] P^2+8 Q F''[\phi] P+\zeta  R'  \notag \\
			&-8 \left(\zeta ^2-1\right) F'[\phi] (\rder P)\bigr) \left(R'^3-8 \left(F''[\phi] Q^2+F'[\phi] (\rder Q)\right) R'+8 Q F'[\phi] R''\right)\Bigr) \alpha ^2  \notag \\
			&+8 F'[\phi] (\rder \alpha) \Bigl(16 \zeta  R' F''[\phi] Q^3+16 \bigl(\zeta  \left(4 P \zeta  R'^2 F''[\phi]-F'[\phi] R''\right)+2 \left(\zeta ^2-1\right) R' F'[\phi] (\rder \zeta)\bigr) Q^2  \notag \\
			&+2 \zeta  R' \bigl(\left(\zeta ^2 \left(36 F''[\phi] P^2+2\right)-1\right) R'^2+4 F'[\phi] \left(2 \zeta  \left(3-2 \zeta ^2\right) (\rder P)+3 P \left(\zeta ^2-2\right) (\rder \zeta)\right) R'  \notag \\
			&+8 F'[\phi] \left((\rder Q)-P \zeta  R''\right)\bigr) Q+P \zeta ^2 R'^2 \bigl(\left(3 \zeta ^2 \left(8 F''[\phi] P^2+1\right)-2\right) R'^2  \notag \\
			&-8 F'[\phi] \left(\zeta  \left(3 \zeta ^2-5\right) (\rder P)+2 P (\rder \zeta)\right) R'+16 F'[\phi] (\rder Q)\bigr)\Bigr) \alpha  \notag \\
			&+128 \zeta ^2 R' \left(Q \zeta +P R'\right) \left(Q+P \zeta  R'\right) F'[\phi]^2 (\rder \alpha)^2\Bigr]\biggr\}, \label{eq:EqP} \\
		\tder \Pi &= \frac{R R' (\rder \alpha) \left(R' \zeta  \Pi +\Theta \right)+\alpha  \bigl[R' \bigl(R \left(R' (\rder \zeta) \Pi +(\rder \Theta)\right)+R' \zeta  \left(2 R' \Pi +R (\rder \Pi)\right)\bigr)+\Theta  \left(2 R'^2-R R''\right)\bigr]}{R R'^3}, \label{eq:EqPi} \\
		\tder \zeta &= \frac{1}{4 R' \zeta  \left(8 F'[\phi] \left(R' P \zeta +Q\right)-R R'\right)}\Bigl\{2 \zeta ^2 \Bigl[\alpha  \Bigl(8 (\rder \zeta) F'[\phi] \left(2 R' P \zeta +Q\right)+R' \bigl(-8 R' P^2 \zeta  F''[\phi] \notag \\
			&-R' \zeta -8 P Q F''[\phi]+8 (\rder P) \left(\zeta ^2-1\right) F'[\phi]\bigr)\Bigr)-8 R'^2 (\tder P) \zeta  F'[\phi]\Bigr] \notag \\
			&+R^2 \alpha  \left(2 R' (\Theta  \Pi -P Q)+R'^2 \zeta  \left(\Pi ^2-P^2\right)+\zeta  \left(\Theta ^2-Q^2\right)\right)-4 R R' \alpha  \zeta ^2 (\rder \zeta)\Bigr\}, \label{eq:EqZeta}
	\end{align}
	where
	\begin{align}
		D_P &= R'^2 \Bigl[\alpha  R'^3 R^4-16 \alpha  R'^2 \left(Q+P \zeta  R'\right) F'[\phi] R^3+64 \alpha  R' \left(Q+P \zeta  R'\right)^2 F'[\phi]^2 R^2  \notag \\
			&+128 \zeta ^3 R'^2 F'[\phi]^2 \left(\zeta  (\rder \alpha)+\alpha  (\rder \zeta)\right) R-32 \zeta ^3 F'[\phi]^2 \Bigl(8 \zeta  R' \left(4 Q+3 P \zeta  R'\right) F'[\phi] (\rder \alpha)  \notag \\
			&+\alpha  \bigl(\zeta  \left(R'^3-8 \left(F''[\phi] Q^2+F'[\phi] (\rder Q)\right) R'+8 Q F'[\phi] R''\right)  \notag \\
			&+8 R' \left(4 Q+3 P \zeta  R'\right) F'[\phi] (\rder \zeta)\bigr)\Bigr)\Bigr].
		\label{eq:DEqP}
	\end{align}
	The constraints are
	\begin{align}
		\rder \zeta &= -\frac{1}{4 R' \zeta  \left(R R'-8 F'[\phi] \left(R' P \zeta +Q\right)\right) \left(R R'-4 F'[\phi] \left(3 R' P \zeta +2 Q\right)\right)}\Bigl\{8 R'^2 \zeta ^2 F'[\phi] \bigl(R' P \notag \\
			&+Q \zeta \bigr) \Bigl[R^2 (P Q-\Theta  \Pi )+8 \zeta ^2 \left(P Q F''[\phi]+(\rder P) F'[\phi]\right)\Bigr]-\Bigl[4 F'[\phi] \left(Q \left(\zeta ^2-2\right)-2 R' P \zeta \right) \notag \\
			&+R R'\Bigr] \Bigr[R^2 R' \Bigl(2 R' \zeta  (P Q-\Theta  \Pi )+R'^2 \left(P^2-\Pi ^2\right)+Q^2-\Theta ^2\Bigr) \notag \\ 
			&+2 \zeta ^2 \Bigl(R' \left(8 R' \zeta  \left(P Q F''[\phi]+(\rder P) F'[\phi]\right)+8 Q^2 F''[\phi]+8 (\rder Q) F'[\phi]-R'^2\right)-8 R'' Q F'[\phi]\Bigr)\Bigr]\Bigr\}, \label{eq:ConstrZeta} \\
		\rder \alpha &= \frac{\alpha}{2 R' \zeta  \left(R R'-8 F'[\phi] \left(R' P \zeta +Q\right)\right) \left(R R'-4 F'[\phi] \left(3 R' P \zeta +2 Q\right)\right)}  \Bigl\{2 R^2 R' F'[\phi] \Bigl[R'^2 \zeta  \bigl(5 P^2 Q \notag \\
			&-6 P \Theta  \Pi +Q \Pi ^2\bigr)+4 R' Q (P Q-\Theta  \Pi )+Q \zeta  \left(\Theta ^2-Q^2\right)\Bigr]-8 R R'^3 \zeta ^2 \Bigl(P Q F''[\phi]+(\rder P) F'[\phi]\Bigr) \notag \\
			&+R^3 R'^3 (\Theta  \Pi -P Q)+4 \zeta ^2 F'[\phi] \Bigl[-8 R' Q^3 \zeta  F''[\phi]+24 R'^3 P (\rder P) \zeta  F'[\phi]+R' Q \Bigl(16 R' (\rder P) F'[\phi] \notag \\
			&+\zeta  \left(R'^2 \left(24 P^2 F''[\phi]+1\right)-8 (\rder Q) F'[\phi]\right)\Bigr)+8 Q^2 \left(R'' \zeta  F'[\phi]+2 R'^2 P F''[\phi]\right)\Bigr]\Bigr\}. \label{eq:ConstrAlpha}
	\end{align}
\end{widetext}
These equations have been derived using Wolfram \textsc{Mathematica}.

\section{Phase diagram in EdGB gravity: BHs, wormholes, and solitons} \label{app:Phases}

Here we present the methods used to find the static wormholes and the solitonic solutions with cusp singularities discussed in Sec.~\ref{sec:phase}, and we discuss some of their properties~\cite{Kanti:2011jz,Kanti:2011yv,Kleihaus:2019rbg,Kleihaus:2020qwo}. We start with the wormhole solution. In spherical symmetry, the starting point is to consider the ansatz~\eqref{eq:Sch coord} in Schwarzschild coordinates. However, in this ansatz wormholes have a coordinate singularity, that can be removed defining the new coordinate $l^2=\Sch{r}^2-r_0^2$, where $l>0$ and $r_0$ is the wormhole throat~\cite{Kanti:2011jz}. In terms of the coordinates $(t,l,\theta,\varphi)$, we use the following ansatz for the metric:
\begin{equation}
    ds^2=-e^{2\nu(l)}dt^2+f(l)dl^2+\left(l^2+r_0^2\right)\left(d\theta^2+d\varphi^2\right)\,.
    \label{eq:AnsatzMetricWormholeCoordinates}
\end{equation}
Substituting into the modified Einstein equations~\eqref{eq:field_grav}-\eqref{eq:field_dilaton}, this yields 

\begin{widetext}
\begin{eqnarray}
        f'+\frac{f}{l\Sch{r}^2}\left(f\Sch{r}^2-l^2-2\Sch{r}_0^2 \right)-\frac{\Sch{r}^2}{2l}f\left(\phi'\right)^2-
        \frac{4\gamma\lambda}{l\Sch{r}^2}e^{-\gamma \phi}\left[\frac{4l\Sch{r}_0^2}{r^2}\phi'+\left(\Sch{r}^2-\frac{3l^2}{f} \right)f'\phi'+2\gamma\phi'^2\left(\Sch{r}^2f-l^2\right)+2\phi''\left(l^2-\Sch{r}^2f\right)\right]&=&0\,,\nonumber\\
    \label{eq:eqEinsteinttWormholeCoordinates}\\
    \nu'-\nu'\frac{4\gamma\lambda}{l }e^{-\gamma\phi}\left(1-\frac{3l^2}{\Sch{r}^2f}\right)\phi'+ \frac{l}{2\Sch{r}^2}-\frac{f}{2l}-\frac{\Sch{r}^2}{4l}\phi'^2&=&0\,,\nonumber\\
    \label{eq:eqEinsteinrrWormholeCoordinates}\\
        \nu''+\nu'^2+\nu'\left(\frac{l}{\Sch{r}^2}-\frac{f'}{2f}\right)-\frac{lf'}{2\Sch{r}^2f}+
        \frac{\Sch{r}_0^2}{\Sch{r}^4}+\frac{1}{2}\phi'^2+\frac{4\gamma\lambda}{\Sch{r}^2f}e^{-\gamma\phi}\left[\left(\frac{2 \Sch{r}_0^2}{\Sch{r}^2}-3\frac{lf'}{f}+2 l\nu'-2l\gamma\phi'\right)\nu'\phi'+\right.
        \left.2l\phi'\nu''+2l\nu'\phi''\right]&=&0\,,\nonumber\\
    \label{eq:eqEinsteinthetathetaWormholeCoordinates}\\
        \phi''+\nu'\phi'+\left(\frac{2l}{\Sch{r}^2}-\frac{f'}{2f}\right)\phi'+
        \frac{4e^{-\gamma\phi}\gamma\lambda}{\Sch{r}^2f}\left[-4\frac{l \Sch{r}_0^2}{\Sch{r}^4}\nu'+2\left(f-\frac{l^2}{\Sch{r}^2}\right)\nu''+\left(\frac{3l^2}{\Sch{r}^2f}-1\right)f'\nu'+\left(2f-\frac{2l^2}{\Sch{r}^2}\right)\nu'^2\right]&=&0\,.\nonumber \\
    \label{eq:eqDilatonWormholeCoordinates}
\end{eqnarray}
 \end{widetext}

To impose the boundary conditions, we first expand the dilaton and the metric functions at the throat (i.e., near $l\sim0$):
\begin{equation}
    \begin{split}
        &f(l)=f_0+f_1 l+\mathcal{O}(l^2)\,,\\
        &e^{2\nu(l)}=e^{2\nu_0}\left(1+\nu_1 l\right)+\mathcal{O}(l^2)\,, \\
        &\phi(l)=\phi_0+\phi_1 l+\mathcal{O}(l^2)\,,
    \end{split}
    \label{eq:InitialConditionsWormhole}
\end{equation}
where the parameters $(f_1,\nu_1,\phi_1)$ are functions of $(f_0,\nu_0,\phi_0)$~\cite{Kanti:2011jz,Kanti:2011yv}.
At spatial infinity, we require:
\begin{equation}
    \begin{split}
        &f(l)=1+\frac{2M}{l}+\mathcal{O}(l^{-2})\\
        &\nu(l)=-\frac{M}{l}+\mathcal{O}(l^{-2})\\
        &\phi(l)=-\frac{D}{l}+\mathcal{O}(l^{-2})
    \end{split}
    \label{eq:BoundaryConditionsInfWormhole}
\end{equation}
where $M$ and $D$ are the mass and scalar charge of the wormhole as measured by an observer at infinity.
To obtain the wormhole solutions, we integrate Eqs.~\eqref{eq:eqEinsteinttWormholeCoordinates}--\eqref{eq:eqDilatonWormholeCoordinates} from the throat at $l=0$ outward, imposing Eqs.~\eqref{eq:InitialConditionsWormhole} as initial conditions. The parameter $\nu_0$ is fixed though a rescaling by requiring asymptotic flatness of the metric, while $\phi_0$ is fixed through a shooting procedure such that the dilaton field at infinity vanishes as in Eq.~\eqref{eq:BoundaryConditionsInfWormhole}. 
We use units such that $r_0=2$. In this case the dimensionality of the parameter space is larger than for BHs: for each value of $\lambda$ there exists a one-parameter family identified by $f_0$. This yields a two-dimensional domain of existence, see Fig.~\ref{fig:SolutionsEdGB}.

In particular, for $\lambda<0.015228$ in the limit $f_0\rightarrow 1$ we obtain wormhole solutions that coexist with BH solutions, as can be seen from the inset of Fig.~\ref{fig:SolutionsEdGB}. For $\lambda=0.015228$ and $f_0\rightarrow 1$ the wormhole solutions coexist with the singular BH at the end of the unstable branch. For $\lambda > 0.015228$ we find $f_0>1$ for all wormhole solutions. In particular, at the minimum value of $f_0$ allowed for these families, the wormhole solutions coexist with asymptotically flat and horizonless solutions, characterized by a singularity in the second radial derivative of the dilaton field. These ``cusp'' solutions also bound the domain of existence of horizonless, particle-like solutions whose scalar field diverges at the origin~\cite{Kleihaus:2019rbg,Kleihaus:2020qwo}.

The coordinates in Eq.~\eqref{eq:AnsatzMetricWormholeCoordinates} cover only part of the spacetime. If we try to extend them to values $l<0$, we find a curvature singularity~\cite{Kanti:2011yv}. An interesting feature of these wormhole solutions is that this singularity disappears if we consider the existence of matter at the throat, as discussed extensively in \cite{Kanti:2011yv}. 

\section{Code details and convergence tests} \label{app:Tests}

\subsection{Nonuniform grid in areal radius coordinate}
As discussed in the main text, in order to increase resolution in the high-curvature regions we introduce a radial coordinate $r$ such that the areal radius is given by $R = R(r)$, where
\begin{equation} 
	\begin{cases}
		R &=  \eta_2 r + \frac{1 - \eta_1}{\Delta} \ln \Bigl( \frac{1 + e^{-\Delta (r - r_1)}}{1 + e^{\Delta r_1}} \Bigr) + \\
			 &+ \frac{\eta_2 - 1}{\Delta} \ln \Bigl( \frac{1 + e^{-\Delta (r - r_2)}}{1 + e^{\Delta r_2}} \Bigr) \\
		\frac{\partial R}{\partial r} &=  \eta_1 + \frac{1 - \eta_1}{1 + e^{-\Delta (r - r_1)}} + \frac{\eta_2 - 1}{1 + e^{-\Delta (r - r_2)}}
	\end{cases}
	\label{eq:RTildeDoubleExponential},
\end{equation}
and $\Delta$, $\eta_1 < 1$, $\eta_2 > 1$, $r_1 \le r_2$ are real parameters.

\begin{figure}
	\centering
	\includegraphics[width = \columnwidth]{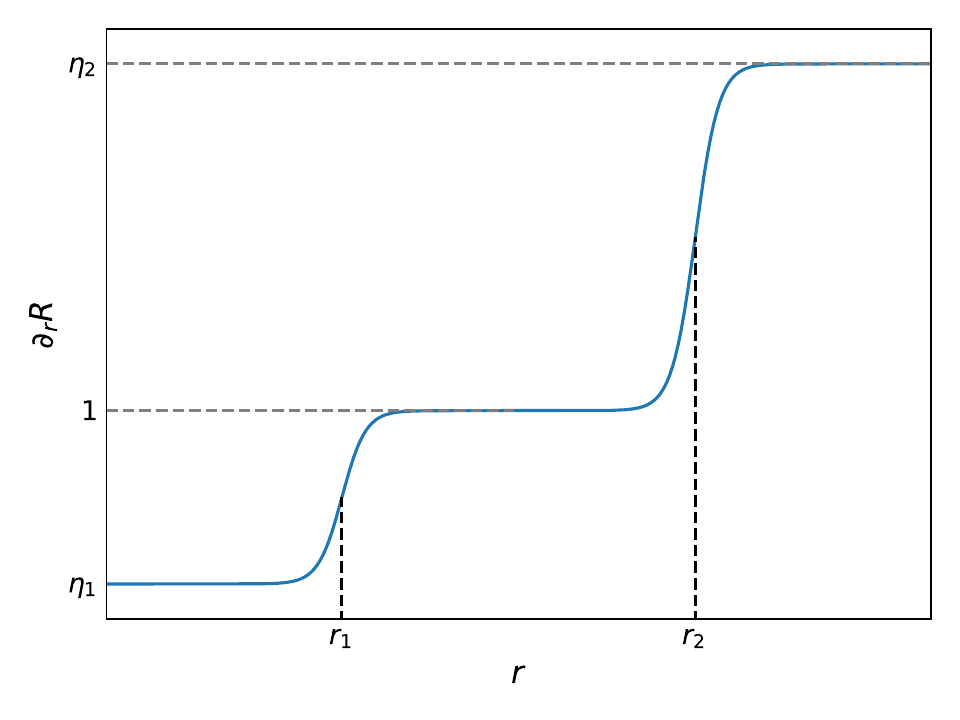}
	\caption{Schematic behavior of the derivative of the transformation function $R(r)$ between coordinate and areal radius. The resolution in areal radius is higher in the inner region, where $R' \to \eta_1 < 1$, and lower in the outer region, where $R' \to \eta_2 > 1$. $r_{1, 2}$ identify the position of the buffer regions between two different resolutions.}
	\label{fig:RadialTransformation}
\end{figure}

In Fig.~\ref{fig:RadialTransformation}, we show a representative plot of $R'(r)$. As we can see, in the inner region $R' \sim \eta_1 < 1$, and in the outer region $R' \sim \eta_2 > 1$; therefore, if we discretize the radial coordinate $r$ using a uniform grid step, we will obtain a higher resolution in $R$ in the inner region and a lower resolution in the outer region. In particular, $\eta_1$ and $\eta_2$ represent the ratio between the grid steps in $R$ and in $r$ in the inner and in the outer regions, respectively, while $\Delta^{-1}$ and $r_{1,2}$ respectively represent the width and the positions of the buffer regions where the resolution in the areal radius changes. 

In this way we can reduce the computational cost of the simulations by avoiding the use of high resolution in all the spatial domain of integration, thus restricting the use of a small grid step only near the BH region, where the singularity and the horizon are situated. 

Throughout the paper the parameters in the transformation \eqref{eq:RTildeDoubleExponential} are set to
\begin{gather}
    \Delta = 1, \quad \eta_1 = 0.05, \quad \eta_2 = 15, \notag \\
	\quad r_1 = 51 , \quad r_2 = 61.
	\label{eq:RTParameters}
\end{gather}

\subsection{Code testing and convergence}
Here we discuss the simulations we performed to test the accuracy of the integration algorithm. 

We first evolved a static BH in the upper branch ($\gamma = 4$, $\lambda = 0.01536$) in absence of perturbations of both scalar fields ($A_\phi = A_\xi = 0$). The outer boundary is placed at $R_\infty = 520 $, the final time is $T = 500 $, and the CFL factor was set to ${\rm CFL} = 0.025$. 

\begin{figure}
	\centering
	\includegraphics[width = \columnwidth]{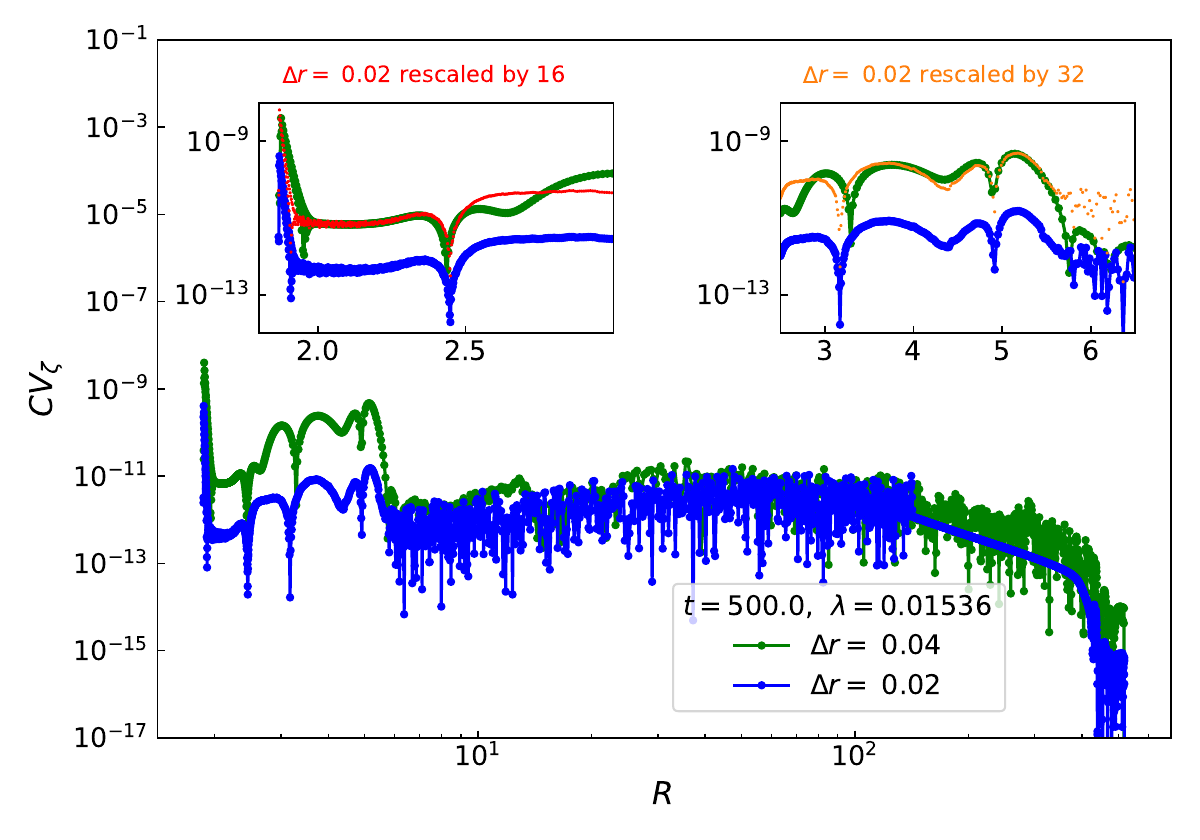}
	\caption{Scaling of the violation $CV_{\zeta}$ of the constraint~\eqref{eq:ConstrZeta}, at the end of the evolution of a stable static dilatonic BH configuration. We can see a fourth-order scaling near the horizon and a fifth order scaling for $3  \lesssim R \lesssim 6 $.}
	\label{fig:TestStaticEvolutionConvergence}
\end{figure}

In Fig.~\ref{fig:TestStaticEvolutionConvergence} we show how the violation $CV_{\zeta}$ of the constraint~\eqref{eq:ConstrZeta} at $t = T$ scales with the resolution.
As we can see the fourth-order scaling is not satisfied in all the radial domain. This can be due to the fact that $CV_{\zeta}$ assumes small values and is dominated by noise. However, as we can see from the insets, the constraint violation scales as a fourth-order term in $\Delta r$ in the horizon region, and as a fifth-order term in the region $3 \lesssim R \lesssim 6$. While the behavior near the horizon is consistent with the accuracy of the evolution algorithm, the fifth-order scaling might be due to the Kreiss-Oliger dissipation term, which is of order 5 in $\Delta r$.

Moreover, we observe that the profile of the dilaton field remains constant in time, which is consistent with the fact that our starting configuration is a static solution to the field equations.

\begin{figure}
	\centering
	\includegraphics[width = \columnwidth]{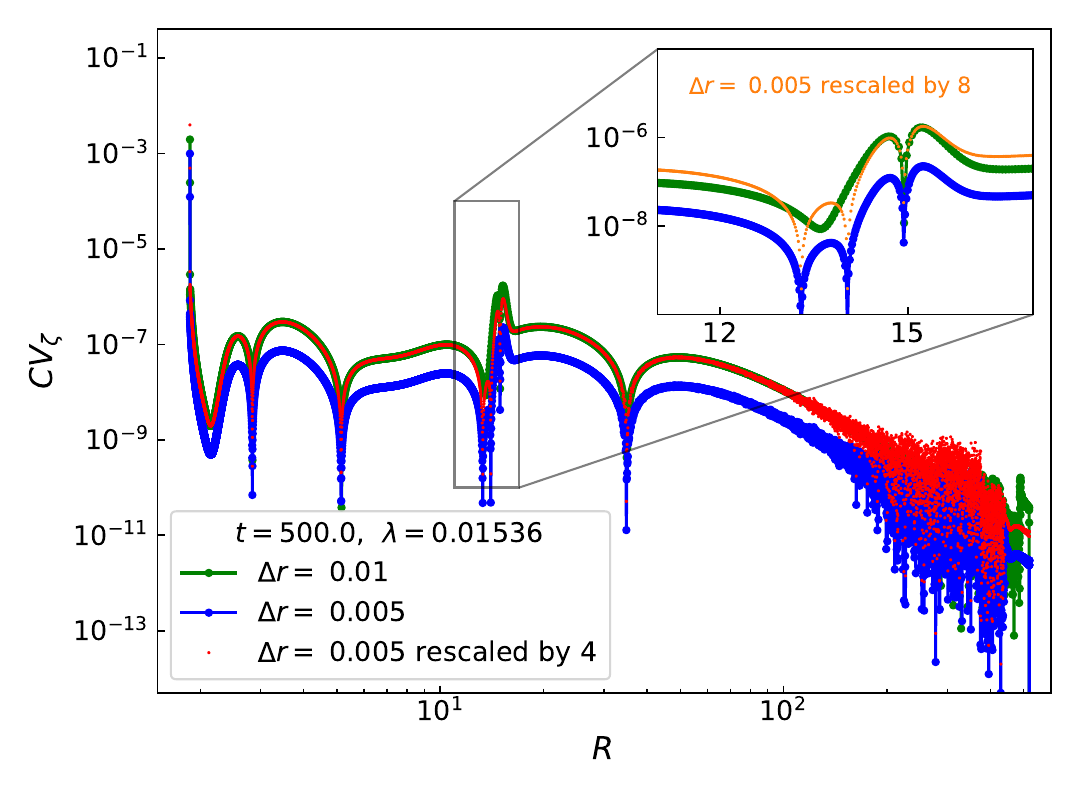}
	\caption{Scaling of the constraint violation at the last time step of the evolution of a static stable dilatonic BH with $\lambda = 0.01536$ using the second-order accurate code. In this case $CV_\zeta$ scales as a second-order term, except in a region around $R \sim 15 $, where there is third order scaling. This behavior may be due to the presence of the Kreiss-Oliger dissipation term.}
	\label{fig:SecondOrderStaticEvolutionConvergence}
\end{figure}

In order to corroborate the reliability of the integration algorithm in the region in which the constraint violation is dominated by noise, we used a second-order accurate version of the code. In this way $CV_{\zeta}$ is typically larger, allowing us to check its scaling properties above the noise floor. The modifications introduced alter as little as possible the structure of the integration algorithm and they can be summarized as follows:
\begin{itemize}
	\item We use the second-order Runge-Kutta method for the time integration;
	\item We use the second-order accurate finite differences method for the radial derivatives; we continue using the (second-order) upwind scheme for the first 2 grid points (instead of 1);
	\item We perform the integration of the constraint for $\alpha$ using only the trapezoidal rule;
	\item We perform the integration in the shooting procedure and in the initialization part with the second-order accurate Runge-Kutta method;
	\item We compute the numerical derivatives in the right-hand side of the constraints during the initialization part with second-order accuracy; however the resolution of the shooting procedure is still the double of the resolution in the evolution (half of the grid points are discarded after initialization);
	\item We use the third order Kreiss-Oliger dissipation term
		\begin{equation}
			Q \, u = - \frac{\eta_{\rm KO}}{16 \, \Delta t} (\Delta r)^4 (D_{+}^2) \rho (D_{-}^2) u,  
			\label{eq:ThirdOrderKO}
		\end{equation}
		where $u$ is a generic field, $\eta_{\rm KO} = 0.05$, and  
		\begin{equation}
			\rho = \frac{1}{1 + e^{5(R - 15)}};
			\label{eq:ThirdOrderKODampingFunction}
		\end{equation}
		we continue excluding the innermost and outermost 3 grid points from the computation of the dissipation term (instead of 2);
	\item We use ${\rm CFL} = 0.01$ since we observed that when using the second-order accurate code a lower CFL is needed.
\end{itemize}
We performed the numerical evolution of the same initial configuration as before with this version of the code. In Fig.~\ref{fig:SecondOrderStaticEvolutionConvergence} we show the scaling of the constraint violation at the end of the simulations with resolutions $\Delta r = 0.005 $ and $\Delta r = 0.01 $. In this case we obtain the expected second-order scaling in all the radial domain except in a small region around $R \sim 15 $ where $CV_\zeta$ seems to scale as a third order term. This may be due to the Kreiss-Oliger dissipation term, which is of order 3. 

We then moved to consider some collapsing scenarios in order to test the behavior of our second- and fourth-order accurate codes in the dynamical setups of our interest.
We first considered the collapse of a wave packet of the dilaton on a Schwarzschild BH in GR ($\lambda = 0$). In this case we can estimate the BH mass at the horizon as $M_h = \frac{R_h}{2}$, and we can compare it with the Misner-Sharp mass at infinity to check that the results of the numerical evolution are in agreement with physical expectations. We obtained that initially $M_h < M_{\rm MS}$ since part of the total mass in the spacetime is stored in the profile of the dilaton outside the horizon, while at the end $M_h =M_{\rm MS}$ with excellent accuracy. This is consistent with the fact that the pulse of the dilaton has been absorbed by the BH.

We then considered a wave packet of the phantom field instead of the dilaton. In this case $M_h > M_{\rm MS}$ at the beginning of the simulation since the profile of the phantom field outside the BH adds a negative contribution to the total Misner-Sharp mass. At the end of the simulation instead, $M_h = M_{\rm MS}$. Also in this case the results of the simulations are consistent with physical expectations since the BH mass decreases upon absorbing the phantom perturbation.

\begin{figure}
	\centering
	\includegraphics[width = \columnwidth]{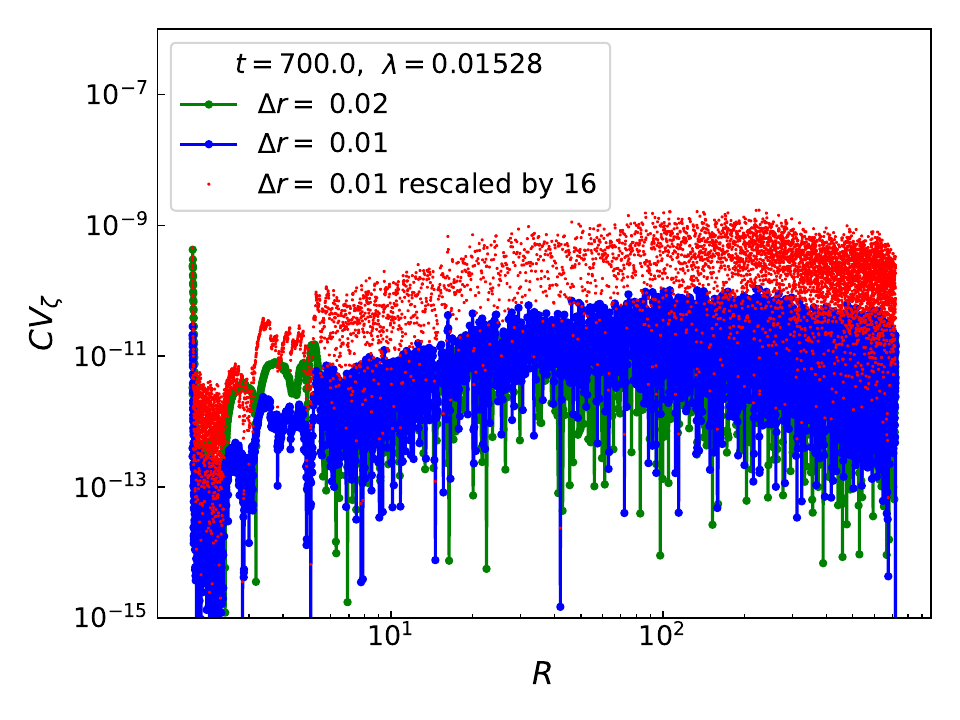} \\
	\includegraphics[width = \columnwidth]{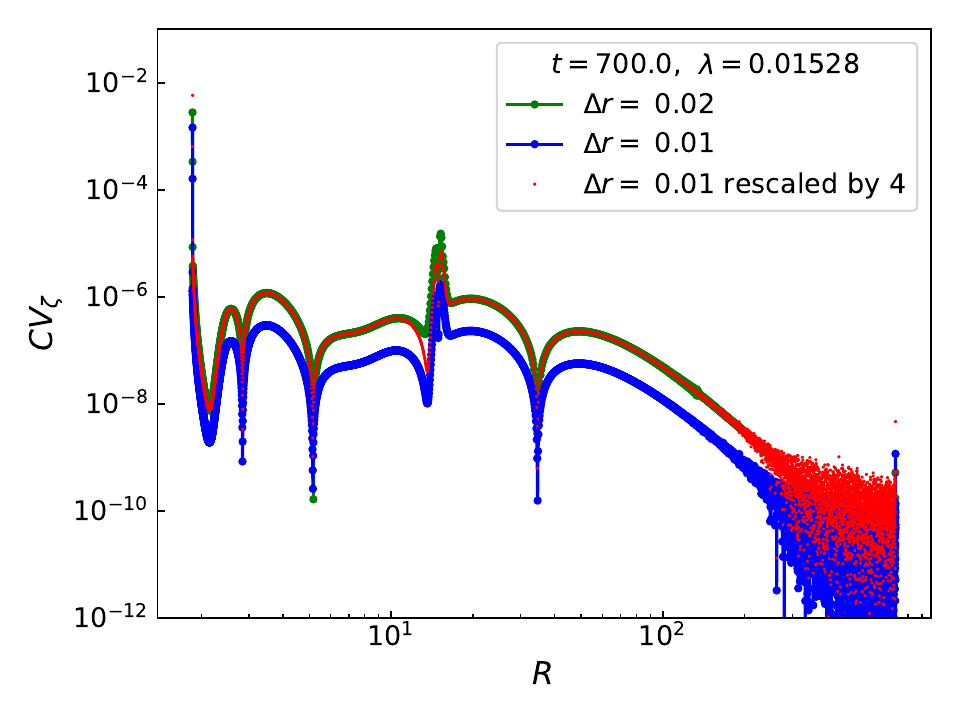}
	\caption{Scaling of the constraint violation at the end of the simulation of the collapse of a wave packet of the phantom field on a dilatonic BH in the upper branch. The upper and the lower panels refer respectively to the fourth and the second-order accurate versions of the code. 
	}
	\label{fig:TestUpperUpperGhostConvergence}
\end{figure}

We finally studied the convergence in some collapsing scenarios when $\lambda \neq 0$. We discuss here a test simulation of the collapse of a wave packet of the phantom field on a static dilatonic BH in the case $\lambda = 0.01528$. The outer boundary is at $R_\infty = 720 $, the final time of integration is $T = 700 $, and the parameters of the initial wave packet are
\begin{equation}
	A_{0, \xi} = 0.02 , \qquad R_{0, \xi} = 15 , \qquad \sigma_\xi = 0.5 .
	\label{eq:TestGhostPerturbationParameters}
\end{equation}
In the upper panel of Fig.~\ref{fig:TestUpperUpperGhostConvergence} we show the scaling of the constraint violation at the end of the numerical evolution. As we can see it is not possible to evaluate the convergence of the code since $CV_{\zeta}$ is very small and dominated by noise. However we repeated the simulation with the second-order accurate version of the code and we obtained the expected scaling properties (see the lower panel of Fig.~\ref{fig:TestUpperUpperGhostConvergence})

In summary, even though it is not possible to evaluate properly the convergence of the code, the constraint violation appears to be very small and dominated by noise. The results of the test simulations are consistent with physical expectations, and the good scaling properties of the second-order accurate version of the code corroborate the reliability of our implementation of the integration algorithm that we used.

Finally, for some selected simulations we have also checked that the time evolution is in agreement between the second- and fourth-order accurate codes.

\bibliographystyle{apsrev4}
\bibliography{References}

\begin{thebibliography}{71}%
\makeatletter
\providecommand \@ifxundefined [1]{%
 \@ifx{#1\undefined}
}%
\providecommand \@ifnum [1]{%
 \ifnum #1\expandafter \@firstoftwo
 \else \expandafter \@secondoftwo
 \fi
}%
\providecommand \@ifx [1]{%
 \ifx #1\expandafter \@firstoftwo
 \else \expandafter \@secondoftwo
 \fi
}%
\providecommand \natexlab [1]{#1}%
\providecommand \enquote  [1]{``#1''}%
\providecommand \bibnamefont  [1]{#1}%
\providecommand \bibfnamefont [1]{#1}%
\providecommand \citenamefont [1]{#1}%
\providecommand \href@noop [0]{\@secondoftwo}%
\providecommand \href [0]{\begingroup \@sanitize@url \@href}%
\providecommand \@href[1]{\@@startlink{#1}\@@href}%
\providecommand \@@href[1]{\endgroup#1\@@endlink}%
\providecommand \@sanitize@url [0]{\catcode `\\12\catcode `\$12\catcode
  `\&12\catcode `\#12\catcode `\^12\catcode `\_12\catcode `\%12\relax}%
\providecommand \@@startlink[1]{}%
\providecommand \@@endlink[0]{}%
\providecommand \url  [0]{\begingroup\@sanitize@url \@url }%
\providecommand \@url [1]{\endgroup\@href {#1}{\urlprefix }}%
\providecommand \urlprefix  [0]{URL }%
\providecommand \Eprint [0]{\href }%
\providecommand \doibase [0]{http://dx.doi.org/}%
\providecommand \selectlanguage [0]{\@gobble}%
\providecommand \bibinfo  [0]{\@secondoftwo}%
\providecommand \bibfield  [0]{\@secondoftwo}%
\providecommand \translation [1]{[#1]}%
\providecommand \BibitemOpen [0]{}%
\providecommand \bibitemStop [0]{}%
\providecommand \bibitemNoStop [0]{.\EOS\space}%
\providecommand \EOS [0]{\spacefactor3000\relax}%
\providecommand \BibitemShut  [1]{\csname bibitem#1\endcsname}%
\let\auto@bib@innerbib\@empty
\bibitem [{\citenamefont {{Penrose}}(1969)}]{1969NCimR...1..252P}%
  \BibitemOpen
  \bibfield  {author} {\bibinfo {author} {\bibfnamefont {R.}~\bibnamefont
  {{Penrose}}}, }\href@noop {} {\bibfield  {journal} {\bibinfo  {journal}
  {\emph {Nuovo Cimento Rivista Serie}} }\textbf {\bibinfo {volume} {1}}
  (\bibinfo {year} {1969})}\BibitemShut {NoStop}%
\bibitem [{\citenamefont {Wald}(1999)}]{Wald:1997wa}%
  \BibitemOpen
  \bibfield  {author} {\bibinfo {author} {\bibfnamefont {R.M.} \bibnamefont
  {Wald}}, }\enquote {\bibinfo {title} {Gravitational collapse and cosmic
  censorship},} in \href {\doibase 10.1007/978-94-017-0934-7_5} {\emph
  {\bibinfo {booktitle} {Black Holes, Gravitational Radiation and the Universe:
  Essays in Honor of C.V. Vishveshwara}}} (\bibinfo  {publisher} {Springer
  Netherlands}, \bibinfo {address} {Dordrecht}, \bibinfo {year} {1999}) pp.
  \bibinfo {pages} {69--86}\BibitemShut {NoStop}%
\bibitem [{\citenamefont {Witt-Hansen}(1976)}]{gedankenexperiments}%
  \BibitemOpen
  \bibfield  {author} {\bibinfo {author} {\bibfnamefont {J.}~\bibnamefont
  {Witt-Hansen}}, }\href {\doibase https://doi.org/10.1163/24689300-01301004}
  {\bibfield  {journal} {\bibinfo  {journal} {\emph {Danish Yearbook of
  Philosophy}} }\textbf {\bibinfo {volume} {13}}, \bibinfo {pages} {48 }
  (\bibinfo {year} {1976})}\BibitemShut {NoStop}%
\bibitem [{\citenamefont {{Wald}}(1974)}]{1974AnPhy..82..548W}%
  \BibitemOpen
  \bibfield  {author} {\bibinfo {author} {\bibfnamefont {R.}~\bibnamefont
  {{Wald}}}, }\href {\doibase 10.1016/0003-4916(74)90125-0} {\bibfield
  {journal} {\bibinfo  {journal} {\emph {Annals of Physics}} }\textbf {\bibinfo
  {volume} {82}}, \bibinfo {pages} {548} (\bibinfo {year} {1974})}\BibitemShut
  {NoStop}%
\bibitem [{\citenamefont {Hubeny}(1999)}]{Hubeny:1998ga}%
  \BibitemOpen
  \bibfield  {author} {\bibinfo {author} {\bibfnamefont {V.E.} \bibnamefont
  {Hubeny}}, }\href {\doibase 10.1103/PhysRevD.59.064013} {\bibfield  {journal}
  {\bibinfo  {journal} {\emph {Phys. Rev. D}} }\textbf {\bibinfo {volume}
  {59}}, \bibinfo {pages} {064013} (\bibinfo {year} {1999})}, \Eprint
  {http://arxiv.org/abs/gr-qc/9808043} {arXiv:gr-qc/9808043}\BibitemShut
  {NoStop}%
\bibitem [{\citenamefont {Jacobson} and \citenamefont
  {Sotiriou}(2009)}]{Jacobson:2009kt}%
  \BibitemOpen
  \bibfield  {author} {\bibinfo {author} {\bibfnamefont {T.}~\bibnamefont
  {Jacobson}} and \bibinfo {author} {\bibfnamefont {T.P.} \bibnamefont
  {Sotiriou}}, }\href {\doibase 10.1103/PhysRevLett.103.141101} {\bibfield
  {journal} {\bibinfo  {journal} {\emph {Phys. Rev. Lett.}} }\textbf {\bibinfo
  {volume} {103}}, \bibinfo {pages} {141101} (\bibinfo {year} {2009})},
  \bibinfo {note} {[Erratum: Phys.Rev.Lett. 103, 209903 (2009)]}, \Eprint
  {http://arxiv.org/abs/0907.4146} {arXiv:0907.4146}\BibitemShut {NoStop}%
\bibitem [{\citenamefont {Saa} and \citenamefont
  {Santarelli}(2011)}]{Saa:2011wq}%
  \BibitemOpen
  \bibfield  {author} {\bibinfo {author} {\bibfnamefont {A.}~\bibnamefont
  {Saa}} and \bibinfo {author} {\bibfnamefont {R.}~\bibnamefont {Santarelli}},
  }\href {\doibase 10.1103/PhysRevD.84.027501} {\bibfield  {journal} {\bibinfo
  {journal} {\emph {Phys. Rev. D}} }\textbf {\bibinfo {volume} {84}}, \bibinfo
  {pages} {027501} (\bibinfo {year} {2011})}, \Eprint
  {http://arxiv.org/abs/1105.3950} {arXiv:1105.3950}\BibitemShut {NoStop}%
\bibitem [{\citenamefont {Isoyama} \emph {et~al.}(2011)\citenamefont {Isoyama},
  \citenamefont {Sago}, and \citenamefont {Tanaka}}]{Isoyama:2011ea}%
  \BibitemOpen
  \bibfield  {author} {\bibinfo {author} {\bibfnamefont {S.}~\bibnamefont
  {Isoyama}}, \bibinfo {author} {\bibfnamefont {N.}~\bibnamefont {Sago}},  and
  \bibinfo {author} {\bibfnamefont {T.}~\bibnamefont {Tanaka}}, }\href
  {\doibase 10.1103/PhysRevD.84.124024} {\bibfield  {journal} {\bibinfo
  {journal} {\emph {Phys. Rev. D}} }\textbf {\bibinfo {volume} {84}}, \bibinfo
  {pages} {124024} (\bibinfo {year} {2011})}, \Eprint
  {http://arxiv.org/abs/1108.6207} {arXiv:1108.6207}\BibitemShut {NoStop}%
\bibitem [{\citenamefont {Nat\'ario} \emph {et~al.}(2016)\citenamefont
  {Nat\'ario}, \citenamefont {Queimada}, and \citenamefont
  {Vicente}}]{Natario:2016bay}%
  \BibitemOpen
  \bibfield  {author} {\bibinfo {author} {\bibfnamefont {J.}~\bibnamefont
  {Nat\'ario}}, \bibinfo {author} {\bibfnamefont {L.}~\bibnamefont {Queimada}},
   and \bibinfo {author} {\bibfnamefont {R.}~\bibnamefont {Vicente}}, }\href
  {\doibase 10.1088/0264-9381/33/17/175002} {\bibfield  {journal} {\bibinfo
  {journal} {\emph {Class. Quant. Grav.}} }\textbf {\bibinfo {volume} {33}},
  \bibinfo {pages} {175002} (\bibinfo {year} {2016})}, \Eprint
  {http://arxiv.org/abs/1601.06809} {arXiv:1601.06809}\BibitemShut {NoStop}%
\bibitem [{\citenamefont {Siahaan} and \citenamefont
  {Tjiang}(2021)}]{Siahaan:2021bzc}%
  \BibitemOpen
  \bibfield  {author} {\bibinfo {author} {\bibfnamefont {H.M.} \bibnamefont
  {Siahaan}} and \bibinfo {author} {\bibfnamefont {P.C.} \bibnamefont
  {Tjiang}}, }\href@noop {} {  (\bibinfo {year} {2021})}, \Eprint
  {http://arxiv.org/abs/2108.06523} {arXiv:2108.06523}\BibitemShut {NoStop}%
\bibitem [{\citenamefont {Aniceto} \emph {et~al.}(2016)\citenamefont {Aniceto},
  \citenamefont {Pani}, and \citenamefont {Rocha}}]{Aniceto:2015klq}%
  \BibitemOpen
  \bibfield  {author} {\bibinfo {author} {\bibfnamefont {P.}~\bibnamefont
  {Aniceto}}, \bibinfo {author} {\bibfnamefont {P.}~\bibnamefont {Pani}},  and
  \bibinfo {author} {\bibfnamefont {J.V.} \bibnamefont {Rocha}}, }\href
  {\doibase 10.1007/JHEP05(2016)115} {\bibfield  {journal} {\bibinfo  {journal}
  {\emph {JHEP}} }\textbf {\bibinfo {volume} {05}}, \bibinfo {pages} {115}
  (\bibinfo {year} {2016})}, \Eprint {http://arxiv.org/abs/1512.08550}
  {arXiv:1512.08550}\BibitemShut {NoStop}%
\bibitem [{\citenamefont {Semiz}(2011)}]{Semiz:2005gs}%
  \BibitemOpen
  \bibfield  {author} {\bibinfo {author} {\bibfnamefont {{\.I}.}~\bibnamefont
  {Semiz}}, }\href {\doibase 10.1007/s10714-010-1108-z} {\bibfield  {journal}
  {\bibinfo  {journal} {\emph {Gen. Rel. Grav.}} }\textbf {\bibinfo {volume}
  {43}}, \bibinfo {pages} {833} (\bibinfo {year} {2011})}, \Eprint
  {http://arxiv.org/abs/gr-qc/0508011} {arXiv:gr-qc/0508011}\BibitemShut
  {NoStop}%
\bibitem [{\citenamefont {D{\"u}zta\c{s}} and \citenamefont
  {Semiz}(2013)}]{Duztas:2013wua}%
  \BibitemOpen
  \bibfield  {author} {\bibinfo {author} {\bibfnamefont {K.}~\bibnamefont
  {D{\"u}zta\c{s}}} and \bibinfo {author} {\bibfnamefont {{\.I}.}~\bibnamefont
  {Semiz}}, }\href {\doibase 10.1103/PhysRevD.88.064043} {\bibfield  {journal}
  {\bibinfo  {journal} {\emph {Phys. Rev. D}} }\textbf {\bibinfo {volume}
  {88}}, \bibinfo {pages} {064043} (\bibinfo {year} {2013})}, \Eprint
  {http://arxiv.org/abs/1307.1481} {arXiv:1307.1481}\BibitemShut {NoStop}%
\bibitem [{\citenamefont {D\"uzta\c{s}}(2021)}]{Duztas:2021kuj}%
  \BibitemOpen
  \bibfield  {author} {\bibinfo {author} {\bibfnamefont {K.}~\bibnamefont
  {D\"uzta\c{s}}}, }\href {\doibase 10.1140/epjc/s10052-021-09937-5} {\bibfield
   {journal} {\bibinfo  {journal} {\emph {Eur. Phys. J. C}} }\textbf {\bibinfo
  {volume} {81}}, \bibinfo {pages} {1131} (\bibinfo {year} {2021})}, \Eprint
  {http://arxiv.org/abs/2107.05345} {arXiv:2107.05345}\BibitemShut {NoStop}%
\bibitem [{\citenamefont {Barausse} \emph {et~al.}(2010)\citenamefont
  {Barausse}, \citenamefont {Cardoso}, and \citenamefont
  {Khanna}}]{Barausse:2010ka}%
  \BibitemOpen
  \bibfield  {author} {\bibinfo {author} {\bibfnamefont {E.}~\bibnamefont
  {Barausse}}, \bibinfo {author} {\bibfnamefont {V.}~\bibnamefont {Cardoso}},
  and \bibinfo {author} {\bibfnamefont {G.}~\bibnamefont {Khanna}}, }\href
  {\doibase 10.1103/PhysRevLett.105.261102} {\bibfield  {journal} {\bibinfo
  {journal} {\emph {Phys. Rev. Lett.}} }\textbf {\bibinfo {volume} {105}},
  \bibinfo {pages} {261102} (\bibinfo {year} {2010})}, \Eprint
  {http://arxiv.org/abs/1008.5159} {arXiv:1008.5159}\BibitemShut {NoStop}%
\bibitem [{\citenamefont {Barausse} \emph {et~al.}(2011)\citenamefont
  {Barausse}, \citenamefont {Cardoso}, and \citenamefont
  {Khanna}}]{Barausse:2011vx}%
  \BibitemOpen
  \bibfield  {author} {\bibinfo {author} {\bibfnamefont {E.}~\bibnamefont
  {Barausse}}, \bibinfo {author} {\bibfnamefont {V.}~\bibnamefont {Cardoso}},
  and \bibinfo {author} {\bibfnamefont {G.}~\bibnamefont {Khanna}}, }\href
  {\doibase 10.1103/PhysRevD.84.104006} {\bibfield  {journal} {\bibinfo
  {journal} {\emph {Phys. Rev. D}} }\textbf {\bibinfo {volume} {84}}, \bibinfo
  {pages} {104006} (\bibinfo {year} {2011})}, \Eprint
  {http://arxiv.org/abs/1106.1692} {arXiv:1106.1692}\BibitemShut {NoStop}%
\bibitem [{\citenamefont {Corelli} \emph {et~al.}(2021)\citenamefont {Corelli},
  \citenamefont {Ikeda}, and \citenamefont {Pani}}]{Corelli:2021ikv}%
  \BibitemOpen
  \bibfield  {author} {\bibinfo {author} {\bibfnamefont {F.}~\bibnamefont
  {Corelli}}, \bibinfo {author} {\bibfnamefont {T.}~\bibnamefont {Ikeda}},  and
  \bibinfo {author} {\bibfnamefont {P.}~\bibnamefont {Pani}}, }\href {\doibase
  10.1103/PhysRevD.104.084069} {\bibfield  {journal} {\bibinfo  {journal}
  {\emph {Phys. Rev. D}} }\textbf {\bibinfo {volume} {104}}, \bibinfo {pages}
  {084069} (\bibinfo {year} {2021})}, \Eprint {http://arxiv.org/abs/2108.08328}
  {arXiv:2108.08328}\BibitemShut {NoStop}%
\bibitem [{\citenamefont {Corelli} \emph {et~al.}(2022)\citenamefont {Corelli},
  \citenamefont {De~Amicis}, \citenamefont {Ikeda}, and \citenamefont
  {Pani}}]{letter}%
  \BibitemOpen
  \bibfield  {author} {\bibinfo {author} {\bibfnamefont {F.}~\bibnamefont
  {Corelli}}, \bibinfo {author} {\bibfnamefont {M.}~\bibnamefont {De~Amicis}},
  \bibinfo {author} {\bibfnamefont {T.}~\bibnamefont {Ikeda}},  and \bibinfo
  {author} {\bibfnamefont {P.}~\bibnamefont {Pani}}, }\href@noop {} {
  (\bibinfo {year} {2022})}, \Eprint {http://arxiv.org/abs/2205.13006}
  {arXiv:2205.13006}\BibitemShut {NoStop}%
\bibitem [{\citenamefont {Kanti} \emph {et~al.}(1996)\citenamefont {Kanti},
  \citenamefont {Mavromatos}, \citenamefont {Rizos}, \citenamefont {Tamvakis},
  and \citenamefont {Winstanley}}]{Kanti:1995vq}%
  \BibitemOpen
  \bibfield  {author} {\bibinfo {author} {\bibfnamefont {P.}~\bibnamefont
  {Kanti}}, \bibinfo {author} {\bibfnamefont {N.E.} \bibnamefont {Mavromatos}},
  \bibinfo {author} {\bibfnamefont {J.}~\bibnamefont {Rizos}}, \bibinfo
  {author} {\bibfnamefont {K.}~\bibnamefont {Tamvakis}},  and \bibinfo {author}
  {\bibfnamefont {E.}~\bibnamefont {Winstanley}}, }\href {\doibase
  10.1103/PhysRevD.54.5049} {\bibfield  {journal} {\bibinfo  {journal} {\emph
  {Phys. Rev. D}} }\textbf {\bibinfo {volume} {54}}, \bibinfo {pages} {5049}
  (\bibinfo {year} {1996})}, \Eprint {http://arxiv.org/abs/hep-th/9511071}
  {arXiv:hep-th/9511071}\BibitemShut {NoStop}%
\bibitem [{\citenamefont {Woodard}(2007)}]{Woodard:2006nt}%
  \BibitemOpen
  \bibfield  {author} {\bibinfo {author} {\bibfnamefont {R.P.} \bibnamefont
  {Woodard}}, }\href {\doibase 10.1007/978-3-540-71013-4_14} {\bibfield
  {journal} {\bibinfo  {journal} {\emph {Lect. Notes Phys.}} }\textbf {\bibinfo
  {volume} {720}}, \bibinfo {pages} {403} (\bibinfo {year} {2007})}, \Eprint
  {http://arxiv.org/abs/astro-ph/0601672} {arXiv:astro-ph/0601672}\BibitemShut
  {NoStop}%
\bibitem [{\citenamefont {Ripley} and \citenamefont
  {Pretorius}(2019{\natexlab{a}})}]{Ripley:2019hxt}%
  \BibitemOpen
  \bibfield  {author} {\bibinfo {author} {\bibfnamefont {J.L.} \bibnamefont
  {Ripley}} and \bibinfo {author} {\bibfnamefont {F.}~\bibnamefont
  {Pretorius}}, }\href {\doibase 10.1103/PhysRevD.99.084014} {\bibfield
  {journal} {\bibinfo  {journal} {\emph {Phys. Rev. D}} }\textbf {\bibinfo
  {volume} {99}}, \bibinfo {pages} {084014} (\bibinfo {year}
  {2019}{\natexlab{a}})}, \Eprint {http://arxiv.org/abs/1902.01468}
  {arXiv:1902.01468}\BibitemShut {NoStop}%
\bibitem [{\citenamefont {Ripley} and \citenamefont
  {Pretorius}(2019{\natexlab{b}})}]{Ripley:2019irj}%
  \BibitemOpen
  \bibfield  {author} {\bibinfo {author} {\bibfnamefont {J.L.} \bibnamefont
  {Ripley}} and \bibinfo {author} {\bibfnamefont {F.}~\bibnamefont
  {Pretorius}}, }\href {\doibase 10.1088/1361-6382/ab2416} {\bibfield
  {journal} {\bibinfo  {journal} {\emph {Class. Quant. Grav.}} }\textbf
  {\bibinfo {volume} {36}}, \bibinfo {pages} {134001} (\bibinfo {year}
  {2019}{\natexlab{b}})}, \Eprint {http://arxiv.org/abs/1903.07543}
  {arXiv:1903.07543}\BibitemShut {NoStop}%
\bibitem [{\citenamefont {Kov\'acs} and \citenamefont
  {Reall}(2020{\natexlab{a}})}]{Kovacs:2020pns}%
  \BibitemOpen
  \bibfield  {author} {\bibinfo {author} {\bibfnamefont {A.D.} \bibnamefont
  {Kov\'acs}} and \bibinfo {author} {\bibfnamefont {H.S.} \bibnamefont
  {Reall}}, }\href {\doibase 10.1103/PhysRevLett.124.221101} {\bibfield
  {journal} {\bibinfo  {journal} {\emph {Phys. Rev. Lett.}} }\textbf {\bibinfo
  {volume} {124}}, \bibinfo {pages} {221101} (\bibinfo {year}
  {2020}{\natexlab{a}})}, \Eprint {http://arxiv.org/abs/2003.04327}
  {arXiv:2003.04327}\BibitemShut {NoStop}%
\bibitem [{\citenamefont {Kov\'acs} and \citenamefont
  {Reall}(2020{\natexlab{b}})}]{Kovacs:2020ywu}%
  \BibitemOpen
  \bibfield  {author} {\bibinfo {author} {\bibfnamefont {A.D.} \bibnamefont
  {Kov\'acs}} and \bibinfo {author} {\bibfnamefont {H.S.} \bibnamefont
  {Reall}}, }\href {\doibase 10.1103/PhysRevD.101.124003} {\bibfield  {journal}
  {\bibinfo  {journal} {\emph {Phys. Rev. D}} }\textbf {\bibinfo {volume}
  {101}}, \bibinfo {pages} {124003} (\bibinfo {year} {2020}{\natexlab{b}})},
  \Eprint {http://arxiv.org/abs/2003.08398} {arXiv:2003.08398}\BibitemShut
  {NoStop}%
\bibitem [{\citenamefont {East} and \citenamefont
  {Ripley}(2021{\natexlab{a}})}]{East:2020hgw}%
  \BibitemOpen
  \bibfield  {author} {\bibinfo {author} {\bibfnamefont {W.E.} \bibnamefont
  {East}} and \bibinfo {author} {\bibfnamefont {J.L.} \bibnamefont {Ripley}},
  }\href {\doibase 10.1103/PhysRevD.103.044040} {\bibfield  {journal} {\bibinfo
   {journal} {\emph {Phys. Rev. D}} }\textbf {\bibinfo {volume} {103}},
  \bibinfo {pages} {044040} (\bibinfo {year} {2021}{\natexlab{a}})}, \Eprint
  {http://arxiv.org/abs/2011.03547} {arXiv:2011.03547}\BibitemShut {NoStop}%
\bibitem [{\citenamefont {East} and \citenamefont
  {Ripley}(2021{\natexlab{b}})}]{East:2021bqk}%
  \BibitemOpen
  \bibfield  {author} {\bibinfo {author} {\bibfnamefont {W.E.} \bibnamefont
  {East}} and \bibinfo {author} {\bibfnamefont {J.L.} \bibnamefont {Ripley}},
  }\href {\doibase 10.1103/PhysRevLett.127.101102} {\bibfield  {journal}
  {\bibinfo  {journal} {\emph {Phys. Rev. Lett.}} }\textbf {\bibinfo {volume}
  {127}}, \bibinfo {pages} {101102} (\bibinfo {year} {2021}{\natexlab{b}})},
  \Eprint {http://arxiv.org/abs/2105.08571} {arXiv:2105.08571}\BibitemShut
  {NoStop}%
\bibitem [{\citenamefont {Kuan} \emph {et~al.}(2021{\natexlab{a}})\citenamefont
  {Kuan}, \citenamefont {Doneva}, and \citenamefont
  {Yazadjiev}}]{Kuan:2021lol}%
  \BibitemOpen
  \bibfield  {author} {\bibinfo {author} {\bibfnamefont {H.J.} \bibnamefont
  {Kuan}}, \bibinfo {author} {\bibfnamefont {D.D.} \bibnamefont {Doneva}},  and
  \bibinfo {author} {\bibfnamefont {S.S.} \bibnamefont {Yazadjiev}}, }\href
  {\doibase 10.1103/PhysRevLett.127.161103} {\bibfield  {journal} {\bibinfo
  {journal} {\emph {Phys. Rev. Lett.}} }\textbf {\bibinfo {volume} {127}},
  \bibinfo {pages} {161103} (\bibinfo {year} {2021}{\natexlab{a}})}, \Eprint
  {http://arxiv.org/abs/2103.11999} {arXiv:2103.11999}\BibitemShut {NoStop}%
\bibitem [{\citenamefont {Kuan} \emph {et~al.}(2021{\natexlab{b}})\citenamefont
  {Kuan}, \citenamefont {Singh}, \citenamefont {Doneva}, \citenamefont
  {Yazadjiev}, and \citenamefont {Kokkotas}}]{Kuan:2021yih}%
  \BibitemOpen
  \bibfield  {author} {\bibinfo {author} {\bibfnamefont {H.J.} \bibnamefont
  {Kuan}}, \bibinfo {author} {\bibfnamefont {J.}~\bibnamefont {Singh}},
  \bibinfo {author} {\bibfnamefont {D.D.} \bibnamefont {Doneva}}, \bibinfo
  {author} {\bibfnamefont {S.S.} \bibnamefont {Yazadjiev}},  and \bibinfo
  {author} {\bibfnamefont {K.D.} \bibnamefont {Kokkotas}}, }\href {\doibase
  10.1103/PhysRevD.104.124013} {\bibfield  {journal} {\bibinfo  {journal}
  {\emph {Phys. Rev. D}} }\textbf {\bibinfo {volume} {104}}, \bibinfo {pages}
  {124013} (\bibinfo {year} {2021}{\natexlab{b}})}, \Eprint
  {http://arxiv.org/abs/2105.08543} {arXiv:2105.08543}\BibitemShut {NoStop}%
\bibitem [{\citenamefont {Witek} \emph {et~al.}(2019)\citenamefont {Witek},
  \citenamefont {Gualtieri}, \citenamefont {Pani}, and \citenamefont
  {Sotiriou}}]{Witek:2018dmd}%
  \BibitemOpen
  \bibfield  {author} {\bibinfo {author} {\bibfnamefont {H.}~\bibnamefont
  {Witek}}, \bibinfo {author} {\bibfnamefont {L.}~\bibnamefont {Gualtieri}},
  \bibinfo {author} {\bibfnamefont {P.}~\bibnamefont {Pani}},  and \bibinfo
  {author} {\bibfnamefont {T.P.} \bibnamefont {Sotiriou}}, }\href {\doibase
  10.1103/PhysRevD.99.064035} {\bibfield  {journal} {\bibinfo  {journal} {\emph
  {Phys. Rev. D}} }\textbf {\bibinfo {volume} {99}}, \bibinfo {pages} {064035}
  (\bibinfo {year} {2019})}, \Eprint {http://arxiv.org/abs/1810.05177}
  {arXiv:1810.05177}\BibitemShut {NoStop}%
\bibitem [{\citenamefont {Okounkova} \emph {et~al.}(2020)\citenamefont
  {Okounkova}, \citenamefont {Stein}, \citenamefont {Moxon}, \citenamefont
  {Scheel}, and \citenamefont {Teukolsky}}]{Okounkova:2019zjf}%
  \BibitemOpen
  \bibfield  {author} {\bibinfo {author} {\bibfnamefont {M.}~\bibnamefont
  {Okounkova}}, \bibinfo {author} {\bibfnamefont {L.C.} \bibnamefont {Stein}},
  \bibinfo {author} {\bibfnamefont {J.}~\bibnamefont {Moxon}}, \bibinfo
  {author} {\bibfnamefont {M.A.} \bibnamefont {Scheel}},  and \bibinfo {author}
  {\bibfnamefont {S.A.} \bibnamefont {Teukolsky}}, }\href {\doibase
  10.1103/PhysRevD.101.104016} {\bibfield  {journal} {\bibinfo  {journal}
  {\emph {Phys. Rev. D}} }\textbf {\bibinfo {volume} {101}}, \bibinfo {pages}
  {104016} (\bibinfo {year} {2020})}, \Eprint {http://arxiv.org/abs/1911.02588}
  {arXiv:1911.02588}\BibitemShut {NoStop}%
\bibitem [{\citenamefont {Okounkova}(2020)}]{Okounkova:2020rqw}%
  \BibitemOpen
  \bibfield  {author} {\bibinfo {author} {\bibfnamefont {M.}~\bibnamefont
  {Okounkova}}, }\href {\doibase 10.1103/PhysRevD.102.084046} {\bibfield
  {journal} {\bibinfo  {journal} {\emph {Phys. Rev. D}} }\textbf {\bibinfo
  {volume} {102}}, \bibinfo {pages} {084046} (\bibinfo {year} {2020})}, \Eprint
  {http://arxiv.org/abs/2001.03571} {arXiv:2001.03571}\BibitemShut {NoStop}%
\bibitem [{\citenamefont {Silva} \emph {et~al.}(2021)\citenamefont {Silva},
  \citenamefont {Witek}, \citenamefont {Elley}, and \citenamefont
  {Yunes}}]{Silva:2020omi}%
  \BibitemOpen
  \bibfield  {author} {\bibinfo {author} {\bibfnamefont {H.O.} \bibnamefont
  {Silva}}, \bibinfo {author} {\bibfnamefont {H.}~\bibnamefont {Witek}},
  \bibinfo {author} {\bibfnamefont {M.}~\bibnamefont {Elley}},  and \bibinfo
  {author} {\bibfnamefont {N.}~\bibnamefont {Yunes}}, }\href {\doibase
  10.1103/PhysRevLett.127.031101} {\bibfield  {journal} {\bibinfo  {journal}
  {\emph {Phys. Rev. Lett.}} }\textbf {\bibinfo {volume} {127}}, \bibinfo
  {pages} {031101} (\bibinfo {year} {2021})}, \Eprint
  {http://arxiv.org/abs/2012.10436} {arXiv:2012.10436}\BibitemShut {NoStop}%
\bibitem [{\citenamefont {Doneva} \emph {et~al.}(2022)\citenamefont {Doneva},
  \citenamefont {Va{\~n}{\'o}{-}Vi{\~n}uales}, and \citenamefont
  {Yazadjiev}}]{Doneva:2022byd}%
  \BibitemOpen
  \bibfield  {author} {\bibinfo {author} {\bibfnamefont {D.D.} \bibnamefont
  {Doneva}}, \bibinfo {author} {\bibfnamefont {A.}~\bibnamefont
  {Va{\~n}{\'o}{-}Vi{\~n}uales}},  and \bibinfo {author} {\bibfnamefont {S.S.}
  \bibnamefont {Yazadjiev}}, }\href {\doibase 10.1103/PhysRevD.106.L061502}
  {\bibfield  {journal} {\bibinfo  {journal} {\emph {Phys. Rev. D}} }\textbf
  {\bibinfo {volume} {106}}, \bibinfo {pages} {L061502} (\bibinfo {year}
  {2022})}, \Eprint {http://arxiv.org/abs/2204.05333}
  {arXiv:2204.05333}\BibitemShut {NoStop}%
\bibitem [{\citenamefont {Elley} \emph {et~al.}(2022)\citenamefont {Elley},
  \citenamefont {Silva}, \citenamefont {Witek}, and \citenamefont
  {Yunes}}]{Elley:2022ept}%
  \BibitemOpen
  \bibfield  {author} {\bibinfo {author} {\bibfnamefont {M.}~\bibnamefont
  {Elley}}, \bibinfo {author} {\bibfnamefont {H.O.} \bibnamefont {Silva}},
  \bibinfo {author} {\bibfnamefont {H.}~\bibnamefont {Witek}},  and \bibinfo
  {author} {\bibfnamefont {N.}~\bibnamefont {Yunes}}, }\href {\doibase
  10.1103/PhysRevD.106.044018} {\bibfield  {journal} {\bibinfo  {journal}
  {\emph {Phys. Rev. D}} }\textbf {\bibinfo {volume} {106}}, \bibinfo {pages}
  {044018} (\bibinfo {year} {2022})}, \Eprint {http://arxiv.org/abs/2205.06240}
  {arXiv:2205.06240}\BibitemShut {NoStop}%
\bibitem [{\citenamefont {Torii} \emph {et~al.}(1997)\citenamefont {Torii},
  \citenamefont {Yajima}, and \citenamefont {Maeda}}]{Torii:1996yi}%
  \BibitemOpen
  \bibfield  {author} {\bibinfo {author} {\bibfnamefont {T.}~\bibnamefont
  {Torii}}, \bibinfo {author} {\bibfnamefont {H.}~\bibnamefont {Yajima}},  and
  \bibinfo {author} {\bibfnamefont {K.i.} \bibnamefont {Maeda}}, }\href
  {\doibase 10.1103/PhysRevD.55.739} {\bibfield  {journal} {\bibinfo  {journal}
  {\emph {Phys. Rev. D}} }\textbf {\bibinfo {volume} {55}}, \bibinfo {pages}
  {739} (\bibinfo {year} {1997})}, \Eprint {http://arxiv.org/abs/gr-qc/9606034}
  {arXiv:gr-qc/9606034}\BibitemShut {NoStop}%
\bibitem [{\citenamefont {Alexeyev} and \citenamefont
  {Pomazanov}(1997)}]{Alexeev:1996vs}%
  \BibitemOpen
  \bibfield  {author} {\bibinfo {author} {\bibfnamefont {S.O.} \bibnamefont
  {Alexeyev}} and \bibinfo {author} {\bibfnamefont {M.V.} \bibnamefont
  {Pomazanov}}, }\href {\doibase 10.1103/PhysRevD.55.2110} {\bibfield
  {journal} {\bibinfo  {journal} {\emph {Phys. Rev. D}} }\textbf {\bibinfo
  {volume} {55}}, \bibinfo {pages} {2110} (\bibinfo {year} {1997})}, \Eprint
  {http://arxiv.org/abs/hep-th/9605106} {arXiv:hep-th/9605106}\BibitemShut
  {NoStop}%
\bibitem [{\citenamefont {Pani} and \citenamefont
  {Cardoso}(2009)}]{Pani:2009wy}%
  \BibitemOpen
  \bibfield  {author} {\bibinfo {author} {\bibfnamefont {P.}~\bibnamefont
  {Pani}} and \bibinfo {author} {\bibfnamefont {V.}~\bibnamefont {Cardoso}},
  }\href {\doibase 10.1103/PhysRevD.79.084031} {\bibfield  {journal} {\bibinfo
  {journal} {\emph {Phys. Rev. D}} }\textbf {\bibinfo {volume} {79}}, \bibinfo
  {pages} {084031} (\bibinfo {year} {2009})}, \Eprint
  {http://arxiv.org/abs/0902.1569} {arXiv:0902.1569}\BibitemShut {NoStop}%
\bibitem [{\citenamefont {Guo} \emph {et~al.}(2008)\citenamefont {Guo},
  \citenamefont {Ohta}, and \citenamefont {Torii}}]{Guo:2008hf}%
  \BibitemOpen
  \bibfield  {author} {\bibinfo {author} {\bibfnamefont {Z.K.} \bibnamefont
  {Guo}}, \bibinfo {author} {\bibfnamefont {N.}~\bibnamefont {Ohta}},  and
  \bibinfo {author} {\bibfnamefont {T.}~\bibnamefont {Torii}}, }\href {\doibase
  10.1143/PTP.120.581} {\bibfield  {journal} {\bibinfo  {journal} {\emph {Prog.
  Theor. Phys.}} }\textbf {\bibinfo {volume} {120}}, \bibinfo {pages} {581}
  (\bibinfo {year} {2008})}, \Eprint {http://arxiv.org/abs/0806.2481}
  {arXiv:0806.2481}\BibitemShut {NoStop}%
\bibitem [{\citenamefont {De~Amicis}(2021)}]{MarinaThesis}%
  \BibitemOpen
  \bibfield  {author} {\bibinfo {author} {\bibfnamefont {M.}~\bibnamefont
  {De~Amicis}}, }\href@noop {} {\bibfield  {journal} {\bibinfo  {journal}
  {\emph {Master Thesis discussed at Sapienza University of Rome}} } (\bibinfo
  {year} {2021})}\BibitemShut {NoStop}%
\bibitem [{\citenamefont {Bl\'azquez-Salcedo} \emph {et~al.}(2017)\citenamefont
  {Bl\'azquez-Salcedo}, \citenamefont {Khoo}, and \citenamefont
  {Kunz}}]{Blazquez-Salcedo:2017txk}%
  \BibitemOpen
  \bibfield  {author} {\bibinfo {author} {\bibfnamefont {J.L.} \bibnamefont
  {Bl\'azquez-Salcedo}}, \bibinfo {author} {\bibfnamefont {F.S.} \bibnamefont
  {Khoo}},  and \bibinfo {author} {\bibfnamefont {J.}~\bibnamefont {Kunz}},
  }\href {\doibase 10.1103/PhysRevD.96.064008} {\bibfield  {journal} {\bibinfo
  {journal} {\emph {Phys. Rev. D}} }\textbf {\bibinfo {volume} {96}}, \bibinfo
  {pages} {064008} (\bibinfo {year} {2017})}, \Eprint
  {http://arxiv.org/abs/1706.03262} {arXiv:1706.03262}\BibitemShut {NoStop}%
\bibitem [{\citenamefont {Kanti} \emph {et~al.}(2011)\citenamefont {Kanti},
  \citenamefont {Kleihaus}, and \citenamefont {Kunz}}]{Kanti:2011jz}%
  \BibitemOpen
  \bibfield  {author} {\bibinfo {author} {\bibfnamefont {P.}~\bibnamefont
  {Kanti}}, \bibinfo {author} {\bibfnamefont {B.}~\bibnamefont {Kleihaus}},
  and \bibinfo {author} {\bibfnamefont {J.}~\bibnamefont {Kunz}}, }\href
  {\doibase 10.1103/PhysRevLett.107.271101} {\bibfield  {journal} {\bibinfo
  {journal} {\emph {Phys. Rev. Lett.}} }\textbf {\bibinfo {volume} {107}},
  \bibinfo {pages} {271101} (\bibinfo {year} {2011})}, \Eprint
  {http://arxiv.org/abs/1108.3003} {arXiv:1108.3003}\BibitemShut {NoStop}%
\bibitem [{\citenamefont {Hawking}(1975)}]{Hawking:1975vcx}%
  \BibitemOpen
  \bibfield  {author} {\bibinfo {author} {\bibfnamefont {S.W.} \bibnamefont
  {Hawking}}, }\href {\doibase 10.1007/BF02345020} {\bibfield  {journal}
  {\bibinfo  {journal} {\emph {Commun. Math. Phys.}} }\textbf {\bibinfo
  {volume} {43}}, \bibinfo {pages} {199} (\bibinfo {year} {1975})}, \bibinfo
  {note} {[Erratum: Commun.Math.Phys. 46, 206 (1976)]}\BibitemShut {NoStop}%
\bibitem [{\citenamefont {Konoplya} \emph {et~al.}(2019)\citenamefont
  {Konoplya}, \citenamefont {Zinhailo}, and \citenamefont
  {Stuchl\'\i{}k}}]{Konoplya:2019hml}%
  \BibitemOpen
  \bibfield  {author} {\bibinfo {author} {\bibfnamefont {R.A.} \bibnamefont
  {Konoplya}}, \bibinfo {author} {\bibfnamefont {A.F.} \bibnamefont
  {Zinhailo}},  and \bibinfo {author} {\bibfnamefont {Z.}~\bibnamefont
  {Stuchl\'\i{}k}}, }\href {\doibase 10.1103/PhysRevD.99.124042} {\bibfield
  {journal} {\bibinfo  {journal} {\emph {Phys. Rev. D}} }\textbf {\bibinfo
  {volume} {99}}, \bibinfo {pages} {124042} (\bibinfo {year} {2019})}, \Eprint
  {http://arxiv.org/abs/1903.03483} {arXiv:1903.03483}\BibitemShut {NoStop}%
\bibitem [{\citenamefont {Alexeyev} \emph {et~al.}(2002)\citenamefont
  {Alexeyev}, \citenamefont {Barrau}, \citenamefont {Boudoul}, \citenamefont
  {Khovanskaya}, and \citenamefont {Sazhin}}]{Alexeyev:2002tg}%
  \BibitemOpen
  \bibfield  {author} {\bibinfo {author} {\bibfnamefont {S.}~\bibnamefont
  {Alexeyev}}, \bibinfo {author} {\bibfnamefont {A.}~\bibnamefont {Barrau}},
  \bibinfo {author} {\bibfnamefont {G.}~\bibnamefont {Boudoul}}, \bibinfo
  {author} {\bibfnamefont {O.}~\bibnamefont {Khovanskaya}},  and \bibinfo
  {author} {\bibfnamefont {M.}~\bibnamefont {Sazhin}}, }\href {\doibase
  10.1088/0264-9381/19/16/314} {\bibfield  {journal} {\bibinfo  {journal}
  {\emph {Class. Quant. Grav.}} }\textbf {\bibinfo {volume} {19}}, \bibinfo
  {pages} {4431} (\bibinfo {year} {2002})}, \Eprint
  {http://arxiv.org/abs/gr-qc/0201069} {arXiv:gr-qc/0201069}\BibitemShut
  {NoStop}%
\bibitem [{\citenamefont {Adler} \emph {et~al.}(2001)\citenamefont {Adler},
  \citenamefont {Chen}, and \citenamefont {Santiago}}]{Adler:2001vs}%
  \BibitemOpen
  \bibfield  {author} {\bibinfo {author} {\bibfnamefont {R.J.} \bibnamefont
  {Adler}}, \bibinfo {author} {\bibfnamefont {P.}~\bibnamefont {Chen}},  and
  \bibinfo {author} {\bibfnamefont {D.I.} \bibnamefont {Santiago}}, }\href
  {\doibase 10.1023/A:1015281430411} {\bibfield  {journal} {\bibinfo  {journal}
  {\emph {Gen. Rel. Grav.}} }\textbf {\bibinfo {volume} {33}}, \bibinfo {pages}
  {2101} (\bibinfo {year} {2001})}, \Eprint
  {http://arxiv.org/abs/gr-qc/0106080} {arXiv:gr-qc/0106080}\BibitemShut
  {NoStop}%
\bibitem [{\citenamefont {Gross} and \citenamefont
  {Sloan}(1987)}]{Gross:1986mw}%
  \BibitemOpen
  \bibfield  {author} {\bibinfo {author} {\bibfnamefont {D.J.} \bibnamefont
  {Gross}} and \bibinfo {author} {\bibfnamefont {J.H.} \bibnamefont {Sloan}},
  }\href {\doibase 10.1016/0550-3213(87)90465-2} {\bibfield  {journal}
  {\bibinfo  {journal} {\emph {Nucl. Phys. B}} }\textbf {\bibinfo {volume}
  {291}}, \bibinfo {pages} {41} (\bibinfo {year} {1987})}\BibitemShut {NoStop}%
\bibitem [{\citenamefont {Torii} and \citenamefont
  {Maeda}(1998)}]{Torii:1998gm}%
  \BibitemOpen
  \bibfield  {author} {\bibinfo {author} {\bibfnamefont {T.}~\bibnamefont
  {Torii}} and \bibinfo {author} {\bibfnamefont {K.i.} \bibnamefont {Maeda}},
  }\href {\doibase 10.1103/PhysRevD.58.084004} {\bibfield  {journal} {\bibinfo
  {journal} {\emph {Phys. Rev. D}} }\textbf {\bibinfo {volume} {58}}, \bibinfo
  {pages} {084004} (\bibinfo {year} {1998})}\BibitemShut {NoStop}%
\bibitem [{\citenamefont {Sotiriou} and \citenamefont
  {Zhou}(2014{\natexlab{a}})}]{Sotiriou:2013qea}%
  \BibitemOpen
  \bibfield  {author} {\bibinfo {author} {\bibfnamefont {T.P.} \bibnamefont
  {Sotiriou}} and \bibinfo {author} {\bibfnamefont {S.Y.} \bibnamefont {Zhou}},
  }\href {\doibase 10.1103/PhysRevLett.112.251102} {\bibfield  {journal}
  {\bibinfo  {journal} {\emph {Phys. Rev. Lett.}} }\textbf {\bibinfo {volume}
  {112}}, \bibinfo {pages} {251102} (\bibinfo {year} {2014}{\natexlab{a}})},
  \Eprint {http://arxiv.org/abs/1312.3622} {arXiv:1312.3622}\BibitemShut
  {NoStop}%
\bibitem [{\citenamefont {Sotiriou} and \citenamefont
  {Zhou}(2014{\natexlab{b}})}]{Sotiriou:2014pfa}%
  \BibitemOpen
  \bibfield  {author} {\bibinfo {author} {\bibfnamefont {T.P.} \bibnamefont
  {Sotiriou}} and \bibinfo {author} {\bibfnamefont {S.Y.} \bibnamefont {Zhou}},
  }\href {\doibase 10.1103/PhysRevD.90.124063} {\bibfield  {journal} {\bibinfo
  {journal} {\emph {Phys. Rev. D}} }\textbf {\bibinfo {volume} {90}}, \bibinfo
  {pages} {124063} (\bibinfo {year} {2014}{\natexlab{b}})}, \Eprint
  {http://arxiv.org/abs/1408.1698} {arXiv:1408.1698}\BibitemShut {NoStop}%
\bibitem [{\citenamefont {Gibbons} and \citenamefont
  {Hawking}(1993)}]{gibbons1993action}%
  \BibitemOpen
  \bibfield  {author} {\bibinfo {author} {\bibfnamefont {G.W.} \bibnamefont
  {Gibbons}} and \bibinfo {author} {\bibfnamefont {S.W.} \bibnamefont
  {Hawking}}, }in \href@noop {} {\emph {\bibinfo {booktitle} {Euclidean Quantum
  Gravity}}} (\bibinfo  {publisher} {World Scientific}, \bibinfo {year} {1993})
  pp. \bibinfo {pages} {233--237}\BibitemShut {NoStop}%
\bibitem [{\citenamefont {Regge} and \citenamefont
  {Wheeler}(1957)}]{Regge:1957td}%
  \BibitemOpen
  \bibfield  {author} {\bibinfo {author} {\bibfnamefont {T.}~\bibnamefont
  {Regge}} and \bibinfo {author} {\bibfnamefont {J.A.} \bibnamefont {Wheeler}},
  }\href {\doibase 10.1103/PhysRev.108.1063} {\bibfield  {journal} {\bibinfo
  {journal} {\emph {Phys. Rev.}} }\textbf {\bibinfo {volume} {108}}, \bibinfo
  {pages} {1063} (\bibinfo {year} {1957})}\BibitemShut {NoStop}%
\bibitem [{\citenamefont {Kanti} \emph {et~al.}(2012)\citenamefont {Kanti},
  \citenamefont {Kleihaus}, and \citenamefont {Kunz}}]{Kanti:2011yv}%
  \BibitemOpen
  \bibfield  {author} {\bibinfo {author} {\bibfnamefont {P.}~\bibnamefont
  {Kanti}}, \bibinfo {author} {\bibfnamefont {B.}~\bibnamefont {Kleihaus}},
  and \bibinfo {author} {\bibfnamefont {J.}~\bibnamefont {Kunz}}, }\href
  {\doibase 10.1103/PhysRevD.85.044007} {\bibfield  {journal} {\bibinfo
  {journal} {\emph {Phys. Rev. D}} }\textbf {\bibinfo {volume} {85}}, \bibinfo
  {pages} {044007} (\bibinfo {year} {2012})}, \Eprint
  {http://arxiv.org/abs/1111.4049} {arXiv:1111.4049}\BibitemShut {NoStop}%
\bibitem [{\citenamefont {Kleihaus} \emph
  {et~al.}(2020{\natexlab{a}})\citenamefont {Kleihaus}, \citenamefont {Kunz},
  and \citenamefont {Kanti}}]{Kleihaus:2019rbg}%
  \BibitemOpen
  \bibfield  {author} {\bibinfo {author} {\bibfnamefont {B.}~\bibnamefont
  {Kleihaus}}, \bibinfo {author} {\bibfnamefont {J.}~\bibnamefont {Kunz}},  and
  \bibinfo {author} {\bibfnamefont {P.}~\bibnamefont {Kanti}}, }\href {\doibase
  10.1016/j.physletb.2020.135401} {\bibfield  {journal} {\bibinfo  {journal}
  {\emph {Phys. Lett. B}} }\textbf {\bibinfo {volume} {804}}, \bibinfo {pages}
  {135401} (\bibinfo {year} {2020}{\natexlab{a}})}, \Eprint
  {http://arxiv.org/abs/1910.02121} {arXiv:1910.02121}\BibitemShut {NoStop}%
\bibitem [{\citenamefont {Kleihaus} \emph
  {et~al.}(2020{\natexlab{b}})\citenamefont {Kleihaus}, \citenamefont {Kunz},
  and \citenamefont {Kanti}}]{Kleihaus:2020qwo}%
  \BibitemOpen
  \bibfield  {author} {\bibinfo {author} {\bibfnamefont {B.}~\bibnamefont
  {Kleihaus}}, \bibinfo {author} {\bibfnamefont {J.}~\bibnamefont {Kunz}},  and
  \bibinfo {author} {\bibfnamefont {P.}~\bibnamefont {Kanti}}, }\href {\doibase
  10.1103/PhysRevD.102.024070} {\bibfield  {journal} {\bibinfo  {journal}
  {\emph {Phys. Rev. D}} }\textbf {\bibinfo {volume} {102}}, \bibinfo {pages}
  {024070} (\bibinfo {year} {2020}{\natexlab{b}})}, \Eprint
  {http://arxiv.org/abs/2005.07650} {arXiv:2005.07650}\BibitemShut {NoStop}%
\bibitem [{\citenamefont {Ripley} and \citenamefont
  {Pretorius}(2020)}]{Ripley:2019aqj}%
  \BibitemOpen
  \bibfield  {author} {\bibinfo {author} {\bibfnamefont {J.L.} \bibnamefont
  {Ripley}} and \bibinfo {author} {\bibfnamefont {F.}~\bibnamefont
  {Pretorius}}, }\href {\doibase 10.1103/PhysRevD.101.044015} {\bibfield
  {journal} {\bibinfo  {journal} {\emph {Phys. Rev. D}} }\textbf {\bibinfo
  {volume} {101}}, \bibinfo {pages} {044015} (\bibinfo {year} {2020})}, \Eprint
  {http://arxiv.org/abs/1911.11027} {arXiv:1911.11027}\BibitemShut {NoStop}%
\bibitem [{\citenamefont {Kokkotas} \emph {et~al.}(2017)\citenamefont
  {Kokkotas}, \citenamefont {Konoplya}, and \citenamefont
  {Zhidenko}}]{Kokkotas:2017ymc}%
  \BibitemOpen
  \bibfield  {author} {\bibinfo {author} {\bibfnamefont {K.D.} \bibnamefont
  {Kokkotas}}, \bibinfo {author} {\bibfnamefont {R.A.} \bibnamefont
  {Konoplya}},  and \bibinfo {author} {\bibfnamefont {A.}~\bibnamefont
  {Zhidenko}}, }\href {\doibase 10.1103/PhysRevD.96.064004} {\bibfield
  {journal} {\bibinfo  {journal} {\emph {Phys. Rev. D}} }\textbf {\bibinfo
  {volume} {96}}, \bibinfo {pages} {064004} (\bibinfo {year} {2017})}, \Eprint
  {http://arxiv.org/abs/1706.07460} {arXiv:1706.07460}\BibitemShut {NoStop}%
\bibitem [{\citenamefont {Babiuc} \emph {et~al.}(2008)}]{Babiuc:2007vr}%
  \BibitemOpen
  \bibfield  {author} {\bibinfo {author} {\bibfnamefont {M.C.} \bibnamefont
  {Babiuc}} \emph {et~al.}, }\href {\doibase 10.1088/0264-9381/25/12/125012}
  {\bibfield  {journal} {\bibinfo  {journal} {\emph {Class. Quant. Grav.}}
  }\textbf {\bibinfo {volume} {25}}, \bibinfo {pages} {125012} (\bibinfo {year}
  {2008})}, \Eprint {http://arxiv.org/abs/0709.3559}
  {arXiv:0709.3559}\BibitemShut {NoStop}%
\bibitem [{\citenamefont {Hawking} and \citenamefont
  {Ellis}(1973)}]{HawkingEllis}%
  \BibitemOpen
  \bibfield  {author} {\bibinfo {author} {\bibfnamefont {S.}~\bibnamefont
  {Hawking}} and \bibinfo {author} {\bibfnamefont {G.F.R.} \bibnamefont
  {Ellis}}, }\href@noop {} {\emph {\bibinfo {title} {The Large Scale Structure
  of Space-Time}}} (\bibinfo  {publisher} {Cambridge University Press},
  \bibinfo {year} {1973})\BibitemShut {NoStop}%
\bibitem [{\citenamefont {Bosch} \emph {et~al.}(2017)\citenamefont {Bosch},
  \citenamefont {Buchel}, and \citenamefont {Lehner}}]{Bosch:2017ccw}%
  \BibitemOpen
  \bibfield  {author} {\bibinfo {author} {\bibfnamefont {P.}~\bibnamefont
  {Bosch}}, \bibinfo {author} {\bibfnamefont {A.}~\bibnamefont {Buchel}},  and
  \bibinfo {author} {\bibfnamefont {L.}~\bibnamefont {Lehner}}, }\href
  {\doibase 10.1007/JHEP07(2017)135} {\bibfield  {journal} {\bibinfo  {journal}
  {\emph {Journal of High Energy Physics}} }\textbf {\bibinfo {volume} {2017}},
  \bibinfo {pages} {135} (\bibinfo {year} {2017})}, \Eprint
  {http://arxiv.org/abs/1704.05454} {arXiv:1704.05454}\BibitemShut {NoStop}%
\bibitem [{\citenamefont {Almheiri} \emph {et~al.}(2021)\citenamefont
  {Almheiri}, \citenamefont {Hartman}, \citenamefont {Maldacena}, \citenamefont
  {Shaghoulian}, and \citenamefont {Tajdini}}]{Almheiri:2020cfm}%
  \BibitemOpen
  \bibfield  {author} {\bibinfo {author} {\bibfnamefont {A.}~\bibnamefont
  {Almheiri}}, \bibinfo {author} {\bibfnamefont {T.}~\bibnamefont {Hartman}},
  \bibinfo {author} {\bibfnamefont {J.}~\bibnamefont {Maldacena}}, \bibinfo
  {author} {\bibfnamefont {E.}~\bibnamefont {Shaghoulian}},  and \bibinfo
  {author} {\bibfnamefont {A.}~\bibnamefont {Tajdini}}, }\href {\doibase
  10.1103/RevModPhys.93.035002} {\bibfield  {journal} {\bibinfo  {journal}
  {\emph {Rev. Mod. Phys.}} }\textbf {\bibinfo {volume} {93}}, \bibinfo {pages}
  {035002} (\bibinfo {year} {2021})}, \Eprint {http://arxiv.org/abs/2006.06872}
  {arXiv:2006.06872}\BibitemShut {NoStop}%
\bibitem [{\citenamefont {R.} \emph {et~al.}(2022)\citenamefont {R.},
  \citenamefont {Ripley}, and \citenamefont {Yunes}}]{R:2022hlf}%
  \BibitemOpen
  \bibfield  {author} {\bibinfo {author} {\bibfnamefont {A.H.K.} \bibnamefont
  {R.}}, \bibinfo {author} {\bibfnamefont {J.L.} \bibnamefont {Ripley}},  and
  \bibinfo {author} {\bibfnamefont {N.}~\bibnamefont {Yunes}}, }\href@noop {} {
   (\bibinfo {year} {2022})}, \Eprint {http://arxiv.org/abs/2211.08477}
  {arXiv:2211.08477}\BibitemShut {NoStop}%
\bibitem [{\citenamefont {Bernard} \emph {et~al.}(2019)\citenamefont {Bernard},
  \citenamefont {Lehner}, and \citenamefont {Luna}}]{Bernard:2019fjb}%
  \BibitemOpen
  \bibfield  {author} {\bibinfo {author} {\bibfnamefont {L.}~\bibnamefont
  {Bernard}}, \bibinfo {author} {\bibfnamefont {L.}~\bibnamefont {Lehner}},
  and \bibinfo {author} {\bibfnamefont {R.}~\bibnamefont {Luna}}, }\href
  {\doibase 10.1103/PhysRevD.100.024011} {\bibfield  {journal} {\bibinfo
  {journal} {\emph {Phys. Rev. D}} }\textbf {\bibinfo {volume} {100}}, \bibinfo
  {pages} {024011} (\bibinfo {year} {2019})}, \Eprint
  {http://arxiv.org/abs/1904.12866} {arXiv:1904.12866}\BibitemShut {NoStop}%
\bibitem [{\citenamefont {{Alexeyev}} \emph {et~al.}(2002)\citenamefont
  {{Alexeyev}}, \citenamefont {{Barrow}}, \citenamefont {{Bowdole}},
  \citenamefont {{Sazhin}}, and \citenamefont
  {{Khovanskaya}}}]{Alexeyev_model}%
  \BibitemOpen
  \bibfield  {author} {\bibinfo {author} {\bibfnamefont {S.O.} \bibnamefont
  {{Alexeyev}}}, \bibinfo {author} {\bibfnamefont {A.}~\bibnamefont
  {{Barrow}}}, \bibinfo {author} {\bibfnamefont {G.}~\bibnamefont {{Bowdole}}},
  \bibinfo {author} {\bibfnamefont {M.V.} \bibnamefont {{Sazhin}}},  and
  \bibinfo {author} {\bibfnamefont {O.S.} \bibnamefont {{Khovanskaya}}}, }\href
  {\doibase 10.1134/1.1491964} {\bibfield  {journal} {\bibinfo  {journal}
  {\emph {Astronomy Letters}} }\textbf {\bibinfo {volume} {28}}, \bibinfo
  {pages} {428} (\bibinfo {year} {2002})}\BibitemShut {NoStop}%
\bibitem [{\citenamefont {Ahn} \emph {et~al.}(2015)\citenamefont {Ahn},
  \citenamefont {Gwak}, \citenamefont {Lee}, and \citenamefont
  {Lee}}]{Ahn:2014fwa}%
  \BibitemOpen
  \bibfield  {author} {\bibinfo {author} {\bibfnamefont {W.K.} \bibnamefont
  {Ahn}}, \bibinfo {author} {\bibfnamefont {B.}~\bibnamefont {Gwak}}, \bibinfo
  {author} {\bibfnamefont {B.H.} \bibnamefont {Lee}},  and \bibinfo {author}
  {\bibfnamefont {W.}~\bibnamefont {Lee}}, }\href {\doibase
  10.1140/epjc/s10052-015-3568-5} {\bibfield  {journal} {\bibinfo  {journal}
  {\emph {Eur. Phys. J. C}} }\textbf {\bibinfo {volume} {75}}, \bibinfo {pages}
  {372} (\bibinfo {year} {2015})}, \Eprint {http://arxiv.org/abs/1412.4189}
  {arXiv:1412.4189}\BibitemShut {NoStop}%
\bibitem [{\citenamefont {Izumi}(2014)}]{Izumi:2014loa}%
  \BibitemOpen
  \bibfield  {author} {\bibinfo {author} {\bibfnamefont {K.}~\bibnamefont
  {Izumi}}, }\href {\doibase 10.1103/PhysRevD.90.044037} {\bibfield  {journal}
  {\bibinfo  {journal} {\emph {Phys. Rev. D}} }\textbf {\bibinfo {volume}
  {90}}, \bibinfo {pages} {044037} (\bibinfo {year} {2014})}, \Eprint
  {http://arxiv.org/abs/1406.0677} {arXiv:1406.0677}\BibitemShut {NoStop}%
\bibitem [{\citenamefont {Reall} \emph {et~al.}(2014)\citenamefont {Reall},
  \citenamefont {Tanahashi}, and \citenamefont {Way}}]{Reall:2014pwa}%
  \BibitemOpen
  \bibfield  {author} {\bibinfo {author} {\bibfnamefont {H.S.} \bibnamefont
  {Reall}}, \bibinfo {author} {\bibfnamefont {N.}~\bibnamefont {Tanahashi}},
  and \bibinfo {author} {\bibfnamefont {B.}~\bibnamefont {Way}}, }\href
  {\doibase 10.1088/0264-9381/31/20/205005} {\bibfield  {journal} {\bibinfo
  {journal} {\emph {Class. Quant. Grav.}} }\textbf {\bibinfo {volume} {31}},
  \bibinfo {pages} {205005} (\bibinfo {year} {2014})}, \Eprint
  {http://arxiv.org/abs/1406.3379} {arXiv:1406.3379}\BibitemShut {NoStop}%
\bibitem [{\citenamefont {Carr} \emph {et~al.}(2021)\citenamefont {Carr},
  \citenamefont {Kohri}, \citenamefont {Sendouda}, and \citenamefont
  {Yokoyama}}]{Carr:2020gox}%
  \BibitemOpen
  \bibfield  {author} {\bibinfo {author} {\bibfnamefont {B.}~\bibnamefont
  {Carr}}, \bibinfo {author} {\bibfnamefont {K.}~\bibnamefont {Kohri}},
  \bibinfo {author} {\bibfnamefont {Y.}~\bibnamefont {Sendouda}},  and \bibinfo
  {author} {\bibfnamefont {J.}~\bibnamefont {Yokoyama}}, }\href {\doibase
  10.1088/1361-6633/ac1e31} {\bibfield  {journal} {\bibinfo  {journal} {\emph
  {Rept. Prog. Phys.}} }\textbf {\bibinfo {volume} {84}}, \bibinfo {pages}
  {116902} (\bibinfo {year} {2021})}, \Eprint {http://arxiv.org/abs/2002.12778}
  {arXiv:2002.12778}\BibitemShut {NoStop}%
\bibitem [{\citenamefont {Mathur}(2009)}]{Mathur:2009hf}%
  \BibitemOpen
  \bibfield  {author} {\bibinfo {author} {\bibfnamefont {S.D.} \bibnamefont
  {Mathur}}, }\href {\doibase 10.1088/0264-9381/26/22/224001} {\bibfield
  {journal} {\bibinfo  {journal} {\emph {Class. Quant. Grav.}} }\textbf
  {\bibinfo {volume} {26}}, \bibinfo {pages} {224001} (\bibinfo {year}
  {2009})}, \Eprint {http://arxiv.org/abs/0909.1038}
  {arXiv:0909.1038}\BibitemShut {NoStop}%
\bibitem [{\citenamefont {Polchinski}(2017)}]{Polchinski:2016hrw}%
  \BibitemOpen
  \bibfield  {author} {\bibinfo {author} {\bibfnamefont {J.}~\bibnamefont
  {Polchinski}}, }\enquote {\bibinfo {title} {The black hole information
  problem},} in \href {\doibase 10.1142/9789813149441_0006} {\emph {\bibinfo
  {booktitle} {New Frontiers in Fields and Strings}}} (\bibinfo  {publisher}
  {World Scientific}, \bibinfo {address} {Singapore}, \bibinfo {year} {2017})
  Chap.~\bibinfo {chapter} {6}, pp. \bibinfo {pages} {353--397}\BibitemShut
  {NoStop}%
\bibitem [{\citenamefont {Chen} \emph {et~al.}(2015)\citenamefont {Chen},
  \citenamefont {Ong}, and \citenamefont {Yeom}}]{Chen:2014jwq}%
  \BibitemOpen
  \bibfield  {author} {\bibinfo {author} {\bibfnamefont {P.}~\bibnamefont
  {Chen}}, \bibinfo {author} {\bibfnamefont {Y.C.} \bibnamefont {Ong}},  and
  \bibinfo {author} {\bibfnamefont {D.h.} \bibnamefont {Yeom}}, }\href
  {\doibase 10.1016/j.physrep.2015.10.007} {\bibfield  {journal} {\bibinfo
  {journal} {\emph {Phys. Rept.}} }\textbf {\bibinfo {volume} {603}}, \bibinfo
  {pages} {1} (\bibinfo {year} {2015})}, \Eprint
  {http://arxiv.org/abs/1412.8366} {arXiv:1412.8366}\BibitemShut {NoStop}%
\bibitem [{\citenamefont {Ong}(2020)}]{Ong:2020xwv}%
  \BibitemOpen
  \bibfield  {author} {\bibinfo {author} {\bibfnamefont {Y.C.} \bibnamefont
  {Ong}}, }\href {\doibase 10.1142/S0217751X20300070} {\bibfield  {journal}
  {\bibinfo  {journal} {\emph {Int. J. Mod. Phys. A}} }\textbf {\bibinfo
  {volume} {35}}, \bibinfo {pages} {2030007} (\bibinfo {year} {2020})}, \Eprint
  {http://arxiv.org/abs/2005.07032} {arXiv:2005.07032}\BibitemShut {NoStop}%
\end{thebibliography}%

\end{document}